\documentclass[letterpaper,twocolumn,10pt]{article}
\usepackage{usenix-2020-09_optimize}

\usepackage{tikz}
\usepackage{amsmath}
\usepackage[english]{babel}
\usepackage{blindtext}
\usepackage{textcomp}
\usepackage[utf8]{inputenc}
\usepackage{colortbl}
\usepackage{xcolor}
\usepackage{multirow}
\usepackage{hhline}
\usepackage{booktabs} 
\usepackage{subcaption}
\usepackage{changepage}
\usepackage{stfloats}
\usepackage[numbers,sort&compress]{natbib}

\setcitestyle{maxcitenames=5,mincitenames=1}

\usepackage{caption}
\usepackage{colortbl}  
\usepackage{xcolor}    

\definecolor{myred}{HTML}{FFCDD2}    
\definecolor{myyellow}{HTML}{FFE0B2} 
\definecolor{mygreen}{HTML}{C8E6C9}  
\usepackage{makecell}

\usepackage{titlesec}
\titlespacing*{\section}{0pt}{7pt plus 3pt minus 3pt}{3pt plus 3pt minus 2pt}
\titlespacing*{\subsection}{0pt}{4pt plus 3pt minus 2pt}{1pt plus 3pt minus 1pt}
\titlespacing*{\subsubsection}{0pt}{4pt plus 3pt minus 2pt}{0pt plus 3pt minus 1pt}
\titleformat{\section}{\large\bfseries}{\thesection}{1em}{}
\titleformat{\subsection}{\normalfont\bfseries}{\thesubsection}{1em}{}
\titleformat{\subsubsection}{\normalfont\bfseries}{\thesubsubsection}{1em}{}

\renewcommand\smallskip{\vspace{2pt}}

\begin{document}

\def\commentsDisplay{} 



\date{}


\title{TrainMover: An Interruption-Resilient Runtime for ML Training}
\newcommand{\sysname}{TrainMover\xspace}

\newcommand{\system}{\sysname\xspace}

\author{
{\rm ChonLam Lao}$^{1,2}$, {\rm Jiaqi Gao}$^{2,*}$, {\rm Jiamin Cao}$^{2}$, {\rm Zhipeng Zhang}$^{2}$, {\rm Pengcheng Zhang}$^{2}$,
{\rm Jiangfei Duan}$^{2}$,\\ {\rm Zhilong Zheng}$^{2}$, {\rm Yu Guan}$^{2}$, {\rm Yichi Xu}$^{2}$, {\rm Yong Li}$^{2}$,{\rm Zhengping Qian}$^{2}$, {\rm Aditya Akella}$^{3}$, \\{\rm Minlan Yu}$^{1}$, {\rm Ennan Zhai}$^{2}$, {\rm Dennis Cai}$^{2}$, {\rm Jingren Zhou}$^{2}$\\
[0.35em]
$^{1}$Harvard University, $^{2}$Alibaba Group, $^{3}$UT Austin
}
\ifdefined\commentsDisplay
    \newcommand{\lam}[1]{\textbf{\color{blue}Lam: #1}}
    \newcommand{\minlan}[1]{\textbf{\color{red}Minlan: #1}}
    \newcommand{\aditya}[1]{\textbf{\color{green}AA: #1}}
    \newcommand{\jiaqi}[1]{\textbf{\color{brown}Jiaqi: #1}}
    \newcommand{\jiamin}[1]{\textbf{\color{purple}Jiamin: #1}}
    \newcommand{\zhipeng}[1]{\textbf{\color{purple}Zhipeng: #1}}
    \newcommand{\fixed}[1]{}
\else
    \newcommand{\lam}[1]{}
    \newcommand{\minlan}[1]{}
    \newcommand{\aditya}[1]{}
    \newcommand{\jiaqi}[1]{}
    \newcommand{\jiamin}[1]{}
    \newcommand{\fixed}[1]{}
\fi

\newcommand{\para}[1]{\smallskip\noindent {\bf #1}}
\newcommand{\parait}[1]{\smallskip\noindent \textit{ #1}}

\newcommand{\squishlist}{
   \begin{list}{$\bullet$}
    { \setlength{\itemsep}{0pt}      \setlength{\parsep}{3pt}
      \setlength{\topsep}{3pt}       \setlength{\partopsep}{0pt}
      \setlength{\leftmargin}{3.5mm} \setlength{\labelwidth}{1em}
      \setlength{\labelsep}{0.5em} } }

\newcommand{\squishend}{
    \end{list}  }

\newcounter{boxlblcounter}
\newcommand{\squishnumlist}{
   \begin{list}{\arabic{boxlblcounter}.}
    { \usecounter{boxlblcounter}
      \setlength{\itemsep}{0pt}      \setlength{\parsep}{3pt}
      \setlength{\topsep}{3pt}       \setlength{\partopsep}{0pt}
      \setlength{\leftmargin}{3.5mm} \setlength{\labelwidth}{1em}
      \setlength{\labelsep}{0.5em} } }

\newcommand{\squishnumend}{
    \end{list}  }

\ifdefined\revisionMark
    \newcommand{\revision}[1]{\textcolor{blue}{#1}}
\else
    \newcommand{\revision}[1]{#1}
\fi
\newcommand{\mypar}[1]{\vspace{0.00cm} \noindent \textbf{#1}}
\newcommand{\secref}[1]{\S\ref{#1}}
\newcommand*\circled[1]{\tikz[baseline=(char.base)]{
            \node[shape=circle,draw,inner sep=2pt] (char) {#1};}}
\newcommand{\codesm}[1]{\texttt{\small #1}}

\maketitle
\begingroup
\renewcommand{\thefootnote}{*}
\footnotetext[1]{Jiaqi Gao is the corresponding author.}
\endgroup
\begin{abstract}
Large-scale ML training jobs are frequently interrupted by hardware and software anomalies, failures, and management events. Existing solutions like checkpoint-restart or runtime reconfiguration suffer from long downtimes and degraded performance. We present TrainMover, a resilient LLM training runtime that leverages elastic and standby machines to handle interruptions with minimal downtime and zero memory overhead. To achieve these goals, TrainMover introduces three key techniques: two-phase, delta-based communication group setup; communication-free sandboxed warmup; and general standby design that enables failure recovery from any role. Our evaluation shows that TrainMover consistently achieves around 20 seconds of downtime when handling various interruptions at the 1024-GPU scale. TrainMover is projected to reduce wasted GPU hours by 55\% compared to the best alternative, saving 1.4 million GPU-hours per week at the 64K-GPU scale.

\end{abstract}

\section{Introduction}
\label{sec:introduction}
Large Language Models (LLMs) have gained significant attention in recent years~\cite{palm, dlrm, Llama, Llama2, gpt1, gpt2, gpt3}. Scaling laws continue to guide the design and training of increasingly large models. LLM training jobs are typically deployed on parallel training frameworks (\emph{e.g.,} Megatron-LM~\cite{megatron-lm} and NeMo~\cite{nemo}), require tight coordination across thousands of GPUs, and run for weeks to months. For example, training GPT-3 with 175 billion parameters on 1,024 GPUs required approximately 34 days~\cite{megatron-lm-in-scale}, while Llama 3 with 405 billion parameters was trained over 54 days using up to 16,000 H100 GPUs~\cite{llama3}. Meta~\cite{si2025collectivecommunication100kgpus} and xAI~\cite{xai_blog_2024} are further pushing the training scale to 100K+ GPUs.

This large-scale, long-running, and tightly coupled distributed training frequently encounters \textit{interruptions}, including hardware anomalies, software contentions, network failures, and various management events.
First, failures and anomalies become common as scale increases, and even a single component fault can slow down or halt the entire job. Alibaba reports that 60\% of large-scale training jobs experience slowness from such issues, causing a 35\% increase in average JCT~\cite{falcon_ali}, and the Llama 3 training job observed a mean-time-to-failure (MTTF) of only 2.7 hours~\cite{llama3}.
Second, because training runs for weeks to months, maintenance tasks (repairs, security upgrades, patches, etc.) cannot be deferred; they often require rebooting servers or switches and thus interrupt training~\cite{meta_maintainence_blog, meta2_realibility}.
Third, in shared clusters, operators need to frequently reschedule jobs or rebalance resources when high-priority jobs arrive~\cite{parcae, cant_be_late} or when fragmented resources become consolidated~\cite{borg}.

To preserve high ETTR (Effective Training Time Ratio), operators always keep a small pool of backup or elastic GPU machines (e.g., 6\% in Alibaba~\cite{alihpn}, Bytedance~\cite{robust_bytedance}, Google~\cite{gemini-google}, Meta~\cite{llama3}) to replace affected machines as quickly as possible when interruptions occur. This is because large-scale training jobs are highly tailored for peak performance and GPU memory utilization. For example, Meta~\cite{llama3} and Deepseek~\cite{deepseekv3} tailor their model layouts to balance memory, computation, and communication for peak throughput (see \secref{mov:thegoal}). Therefore, even small layout changes can degrade performance, trigger out-of-memory errors, or leave GPUs idle~\cite{recycle}.


Existing solutions for replacing affected machines fall into two broad categories. The first, and the current industry best practice, is the \textit{stop–reschedule–reinitialize} approach~\cite{bytecheckpoint,just-in-time-checkpoint,check_n_run,bytedance10000nsdi24}: the job is stopped, faulty nodes are replaced with healthy ones from a backup pool, and the training framework re-initializes from the latest checkpoint. For a 16,000-GPU LLM job, the interruptions result in more than an hour of downtime and \$86K of daily waste. The second category is \textit{reconfiguration systems}, such as ReCycle~\cite{recycle}, Oobleck~\cite{oobleck}, and Parcae~\cite{parcae}, which can be retargeted to replace a machine by tolerating interruption drop (–1) and adding back healthy machine (+1) elastically without halting the job.
However, a fundamental limitation persists across both categories: bringing back a new machine (\textit{joiner}) online after each interruption requires re-initialization, which remains slow and forces the entire job to stall until the joiner resumes, leaving the critical path unchanged (see \secref{sec:related_limitation}).

In this paper, we introduce \sysname, an \textbf{interruption-resilient} LLM training runtime that allows a new joiner to prepare initialization in advance and overlap it with ongoing training, minimizing the interruption impact to the rest.
\sysname is carefully designed to avoid interfering with ongoing training and to incur zero additional GPU memory overhead during handling.

Achieving this is nontrivial. Modern LLM training stacks tightly couple initialization with globally synchronized communication setup, making it impossible to prepare joiners in advance without dedicated handling. Decoupling joiner initialization from this global path is essential. \sysname leverages the two-phase communication-group setup (\secref{sec:nccl}) and sandboxed shadow iterations (\secref{sec:warmup}) to prepare communication and computation state in the background. The general-standby design (\secref{subsec:standby}) further enables role-agnostic preparation, allowing recovery regardless of which machine fails.

\mypar{Executing shadow iteration on sandbox (\secref{sec:warmup}):}
\sysname introduces a communication-free sandbox that allows joiners to trigger initialization independently by running shadow training iterations in isolation before entering the main training loop, replacing actual communication with pre-recorded tensors. This decouples joiners from existing machines and significantly reduces transitional downtime.

\mypar{Two-phase delta-based communication group setup (\secref{sec:nccl}):}
\sysname extends a two-phase delta-based communication group setup to allow all non–critical-path steps to be performed upfront (the first-phase) and overlapped with ongoing training. In the second phase, \sysname applies the membership changes in a delta manner, minimizing disruption to users and reducing transition downtime.

\mypar{General standby to handle unexpected failures (\secref{subsec:standby}):}
\sysname leverages a general standby that can be immediately promoted to a joiner when a failure occurs, exploiting the high symmetry of distributed training. This symmetry removes the need for any prior role knowledge, allowing a single pre-warmed standby to recover any failed machine. The resulting performance is nearly identical to handling interruptions at a known, specific role.
For expected interruptions, \sysname prepares the joiners in a sandbox (\secref{sec:warmup}) and two-phase communication setup (\secref{sec:nccl}), allowing the ongoing training job to continue uninterrupted and preserving performance during migration. Once the joiners are ready, \sysname replaces the affected communication group members and synchronizes the latest model state from the leavers to the joiners before resuming training. For unexpected interruptions, \sysname uses the general standby design (\secref{subsec:standby}) to pre-warm the standby machine with the same techniques in the background, enabling rapid replacement of the affected machine when a failure occurs. 




Our experiments show that \sysname achieves a consistent downtime of around 20 seconds at the 1,024-GPU scale and is projected to reduce wasted GPU hours by 55\% compared to other systems at the 64K-GPU scale. It also enables other use cases, such as resource load balancing at 10-minute intervals, with less than 3\% training throughput loss.

\section{Background and Motivation}
\label{sec:motivation}

\subsection{Interruptions Prevail in LLM Training}

LLM training jobs commonly span tens to hundreds of thousands of GPUs and run for weeks to months. Such clusters are expected to operate at peak efficiency to amortize their substantial infrastructure cost. However, sustaining this performance at scale is difficult, as numerous infrastructure and operational events can hinder or slow progress. We refer to these events collectively as \textbf{interruptions}, which include:
\begin{figure*}[t]
\centering
\footnotesize

\begin{minipage}[t]{0.35\textwidth}   
    \centering
    \vspace{10pt}
\begin{tabular}{l|rr}
\hline
\textbf{Restart Stages}                 & \textbf{Avg. Time}    & \textbf{Percentage} \\
\hline
\noalign{\vskip 1pt}
\hline  
Job Stop \& Cleanup            & 0.52 min     & 7.69\%  \\
\cline{1-3}  
Job Reschedule                 & 1.5  min     & 23.08\% \\
\cline{1-3}  
Job Initialization             & 4.45 min     & 69.23\% \\
{\scriptsize \quad• Checkpoint time} & \multicolumn{1}{r@{\hspace{0.5em}}}{\scriptsize\color{gray} 1.56 min} & {\scriptsize\color{gray} 35.1\%} \\
{\scriptsize \quad• NCCL instantiation}       & \multicolumn{1}{r@{\hspace{0.5em}}}{\scriptsize\color{gray} 1.09 min}  & {\scriptsize\color{gray} 24.5\%} \\
{\scriptsize \quad• Cold warmup}     & \multicolumn{1}{r@{\hspace{0.5em}}}{\scriptsize\color{gray} 1.80 min}  & {\scriptsize\color{gray} 40.4\%} \\
\hline
Total                          & 6.47 min     & 100\%   \\
\hline
\end{tabular}
    \captionof{table}{Average restart time breakdown for 8192-GPU training jobs.}
    \label{tab:restart_time_8192}
    \label{tab:scale_comparison}
\end{minipage}
\hfill
\begin{minipage}[t]{0.31\textwidth}
    \centering
    \vspace{0pt}
    \includegraphics[width=\linewidth]{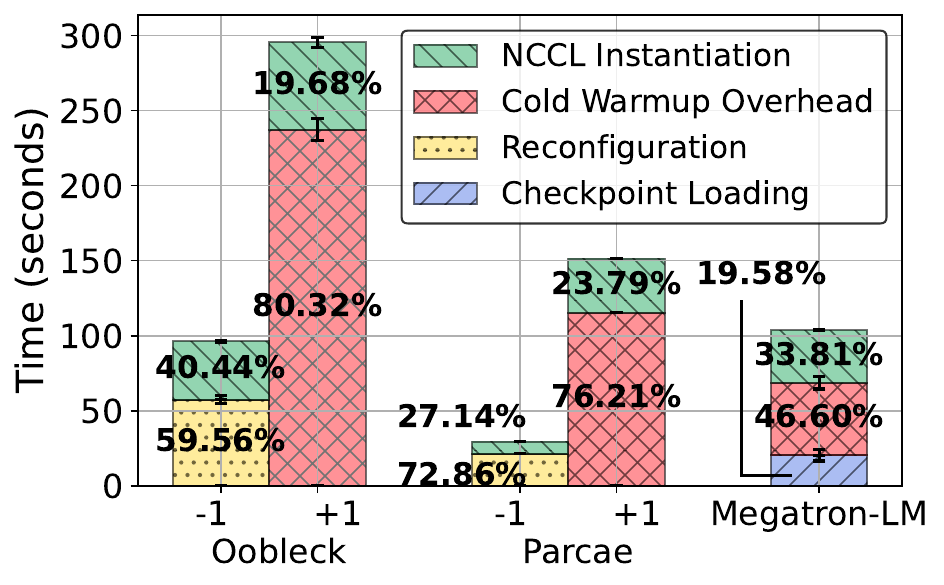}
    \caption{Reconfiguring and booting from scratch are both costly.}
    \label{fig:boot_breakdown}
\end{minipage}
\hfill
\begin{minipage}[t]{0.33\textwidth}
    \centering
    \vspace{0pt}
    \includegraphics[width=\linewidth]{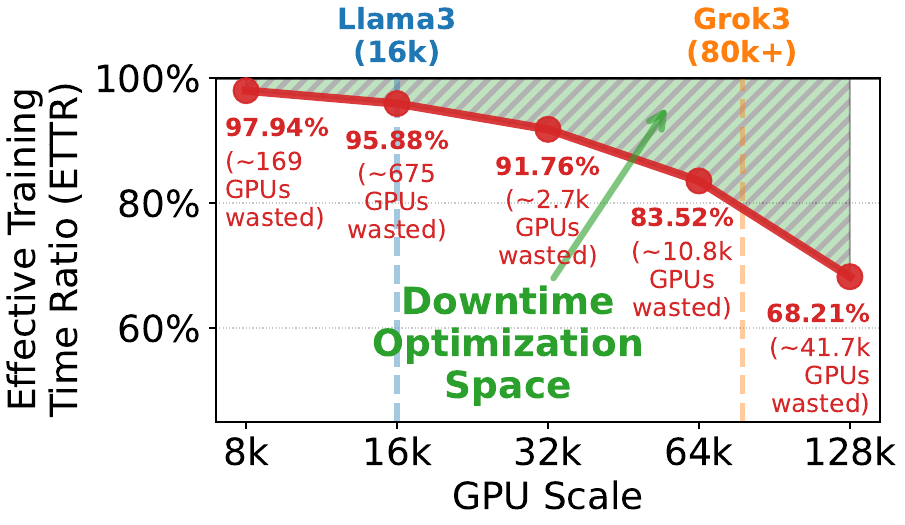}
    \caption{Impact of downtime on ETTR at different scales.}
    \label{fig:8192_gpu}
\end{minipage}

\end{figure*}
\squishlist
\item \textbf{Hardware anomalies} such as GPU failures or down-voltage from overheating or power limits~\cite{llama3}. Aegis~\cite{aegis} reports that GPU servers exhibit failure rates orders of magnitude higher than general cloud nodes, and even within accelerator families, the H100’s MTTF is twice that of the A100~\cite{aegis}. Meta further reports that the MTTF of a 1,024-GPU job is 7.9 hours~\cite{meta2_realibility}\revision{, and the Llama~3 16,000-GPU training job is 2.7 hours~\cite{llama3}}.
\item \textbf{Software contentions} such as co-located programs running on the same CPU~\cite{falcon_ali}, OS or garbage collections in the Python training framework~\cite{bytedance_straggler}. Falcon~\cite{falcon_ali} shows that such contentions from background tasks and network anomalies can slow LLM training by 34.59\% on average.
\item \textbf{Network failures} including optical module or switch failures and network congestion. Although network architects provision sufficient redundancy to maintain connectivity~\cite{alihpn}, Falcon~\cite{falcon_ali} reports that 40–50\% of large-scale LLM training jobs still experience network-induced fail-slows, with average slowdowns of 15–35\% due to congestion or switch failures.
\item \textbf{Management events} including driver updates~\cite{llama3}, GPU resource balancing, and power or thermal planning~\cite{astral}.
First, bespoke training hardware adopts the latest techniques and therefore requires frequent maintenance, including bug fixes, urgent security patches, and firmware or driver updates~\cite{meta_maintainence_blog}; Llama-3 reported at least one such interruption per day~\cite{llama3}.
Second, in shared clusters, operators often reassign GPU servers to improve locality and avoid network congestion~\cite{alihpn, llama3}.
Third, to reduce power and thermal fluctuations and better balance resources, operators regularly redistribute workloads across the cluster~\cite{llama3, astral, google-blog}.
\squishend

Depending on their impact on the training job, these interruptions can be classified as two types: (1) \textit{expected/planned interruptions}, such as scheduled maintenance where the system is notified in advance; software contentions and network failures where the training job is still alive but not performant and waits for the operator to adjust. (2) \textit{unexpected interruptions}, such as GPU and CPU failures that crash the training framework instantly without notice\footnote{This paper does not consider failures resulting from user-level bugs; ensuring application correctness in long-running jobs lies with the user.}.

\subsection{The Goal: Sustaining High ETTR}
\label{mov:thegoal}
These interruptions severely impact existing LLM training frameworks (\emph{e.g.}, Megatron-LM~\cite{megatron-lm} and DeepSpeed~\cite{deepspeed}) that rely on tight synchronization across all GPUs. Because each GPU must exchange intermediate results before every iteration\footnote{We do not handle asynchronous training frameworks because, to the best of our knowledge, no such framework is deployed in production at scale yet.}, any slowdown propagates to the entire cluster, and a single failure terminates the job~\cite{falcon_ali}.
Alibaba~\cite{falcon_ali} reports that slowness can delay large-scale LLM training completion time by up to 90\%. ByteDance reports that 42.5\% of training jobs are at least 10\% slower due to stragglers~\cite{bytedance_straggler}. Meta~\cite{meta2_realibility} reports the ETTR drops to 0.6 when a job spans across
8192 GPUs due to high hardware failure rates. Such a longer training time directly translates to the cost of the training infrastructure. For example, based on AWS’s public pricing, training with 16k GPUs costs \$1.44M per day~\cite{ec2-princing}. An ETTR of 0.6 translates to roughly \$0.58M of cost wasted per day.
Therefore, when an interruption happens, the highest priority is to restore the training job to its peak performance.

In practice, this goal translates to \textit{quickly replacing the interrupted GPUs with
the healthy ones while maintaining the training layout}.
\revision{Because the training layout is specially optimized from the ground up to achieve
peak performance on the specific scale.
For example, before training starts, Meta~\cite{llama3} manually fine-tunes
the model partitions on each GPU to balance memory and computation load across
different pipeline parallelism stages to achieve the highest throughput.}
Similarly, DeepSeek~\cite{deepseekv3} carefully tunes the GPU SMs dedicated to
computation and communication to achieve the optimal balance and trains
DeepSeek-V3 at the peak performance.
Re-distributing the workload to the remaining machines can easily lead to inferior performance, job failures due to unexpected code behavior or out-of-memory errors, or idle GPUs in cases such as removing an entire DP group when distributed optimizer is disabled.

Under such constraints, naively setting up the training cluster to match the job size is undesirable, as a single broken machine can undermine the entire training run.
Instead, operators reserve a pool of healthy standby servers in the cluster for on-the-fly
replacement to reduce downtime.
For example, ByteDance~\cite{robust_bytedance} allocates warm-standby pools based on the 99th percentile of historical GPU failure rates. Alibaba’s HPN~\cite{alihpn} reserves 6\% of its GPUs as backup in each segment. 
Google~\cite{gemini-google} maintains standby TPU cubes within each SuperPod to support rolling maintenance and recovery. Llama-3~\cite{llama3} is trained on 16K GPUs within a 24K-GPU cluster. More discussion on the economic cost of standby machines can be found in \secref{sec:discussion}.

\subsection{Limitation on Existing Solutions}
\label{sec:related_limitation}

The industry has deployed automatic systems~\cite{robust_bytedance, aegis} to reduce interruption downtime. Yet as training scales grow, recovery in the training stack quickly becomes the bottleneck. We evaluate two representative interruption-handling solutions.
Note that, across all experiments, we assume an instant fault localization and isolation. 

\mypar{Solution 1: Restart.}
Restart refers to the standard stop–reschedule–reinitialize mechanism used in production to handle interruptions~\cite{llama3, meta2_realibility, aegis, falcon_ali}. When an interruption occurs, the system first {\em stops} by terminating the training framework and cleaning up active processes and connections. It then {\em reschedules} the cluster by offlining faulty nodes, selecting healthy replacements, and preparing their runtime environments. Finally, the training framework {\em re-initializes} from the most recent checkpoint before training can resume.

This process introduces substantial overhead. We consulted one of the largest cloud providers, and even with an automated recovery stack similar to Llama 3~\cite{llama3}, an 8{,}192-GPU production job still incurs 6.47 minutes of delay per interruption (Table~\ref{tab:restart_time_8192}). Excluding infrastructure overhead, framework initialization alone accounts for 4.45 minutes:
(1) \textbf{Checkpoint loading} (35.1\%), where model and optimizer states are restored from remote storage. This cost grows rapidly with model size~\cite{gemini,falcon_ali,bytecheckpoint}.
(2) \textbf{NCCL instantiation} (24.5\%), where each parallelism dimension (DP, PP, TP) must form its own NCCL group, requiring synchronized metadata exchange and connection setup.
(3) \textbf{Cold warm-up} (40.4\%), including CUDA context creation, GPU memory allocation, JIT compilation, and data-loader initialization.
Because large-scale training is fully synchronized, an interruption affecting only a few nodes~\cite{aegis,meta2_realibility} still forces \emph{all} machines to restart, significantly amplifying downtime.

\mypar{Solution 2: Runtime Reconfiguration.}
Recent works such as Oobleck~\cite{oobleck}, Parcae~\cite{parcae}, and ReCycle~\cite{recycle} aim to avoid global restarts during interruptions by elastically scaling the training cluster up or down without halting the job. These systems adjust training configurations—such as batch sizes or parallelism schemes—to sustain reduced training throughput during failures. When an interruption happens, the system can remove the affected machine (\texttt{-1}) and continue training at reduced throughput, and later adds a replacement machine once one becomes available (\texttt{+1}). 

However, despite avoiding infrastructure overhead, the core restart \emph{critical path} remains. Figure~\ref{fig:boot_breakdown} shows the latency for Oobleck and Parcae when handling interruptions in a 32-GPU scale 6.7B GPT training job. Removing a machine (\texttt{-1}) requires all nodes to redistribute parameters, re-instantiate model components, and reallocate GPU memory to support the new configuration—taking 57 seconds in Oobleck and 21 seconds in Parcae. Adding a machine (\texttt{+1}) is even more expensive: the joiner must initialize NCCL groups and reinitialize software and hardware components, taking over 100 seconds in Oobleck and more than 200 seconds in Parcae. Oobleck is built atop ColossalAI~\cite{colossalai}, while Parcae is built on DeepSpeed~\cite{deepspeed}, leading to different framework warm-up times. Remarkably, this entire cost exceeds even the job initialization overhead in Megatron-LM at the same scale (right bar, around 100 seconds).



\begin{figure*}[tb!]
    \centering
    \begin{minipage}[t]{0.5\textwidth}
        \centering
        \includegraphics[width=\linewidth]{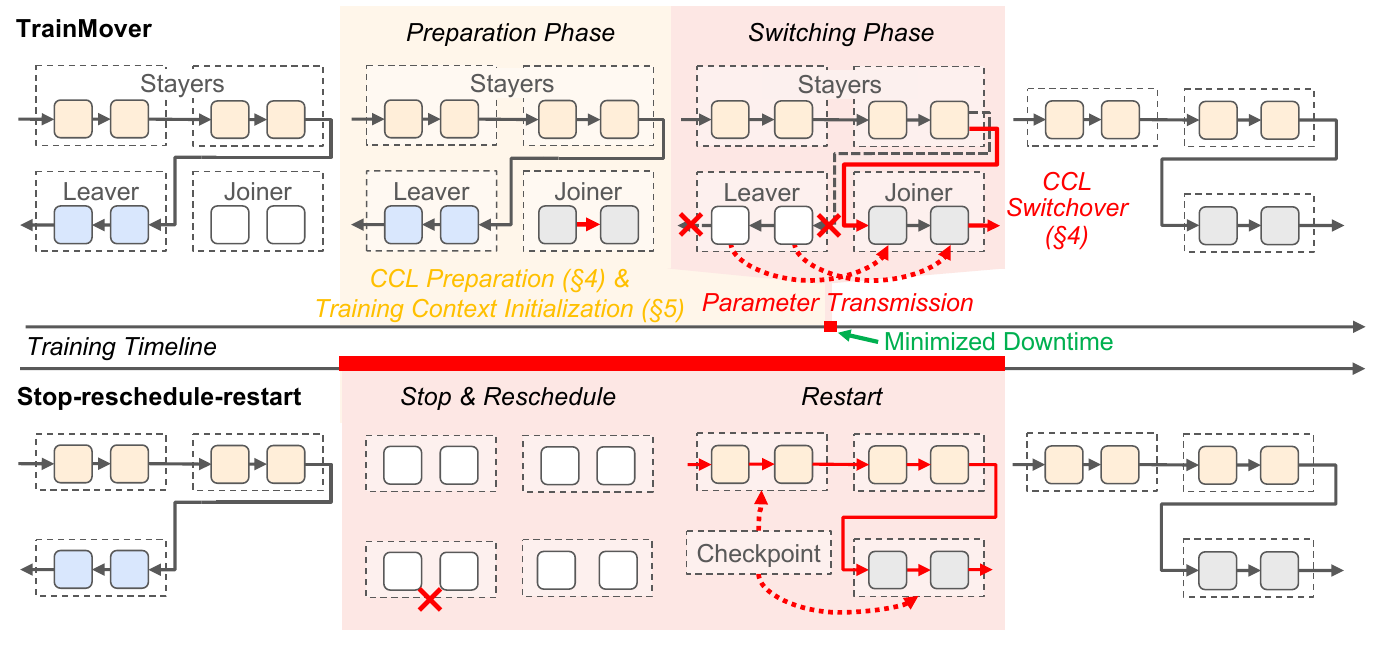}
        \caption{\sysname Overview}
        \label{fig:system_overview}
    \end{minipage}
    \begin{minipage}[t]{0.49\textwidth}
        \centering
        \includegraphics[width=\linewidth]{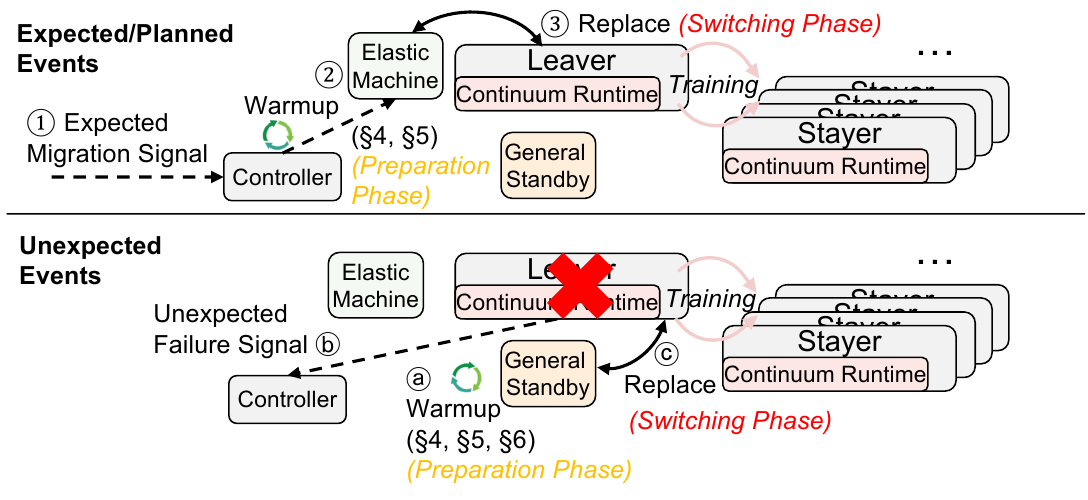}
        \caption{\sysname Workflow}
        \label{fig:system_workflow}
    \end{minipage}
\end{figure*}



In short, although infrastructure overhead can be reduced, \textit{the heavy initialization required to bring a joiner online remains unavoidable.} While this downtime is tolerable at small scales, growing cluster sizes dramatically amplify its impact: higher interruption frequency and the large number of GPUs involved cause downtime to quickly become the dominant bottleneck, substantially degrading ETTR.

This scaling effect is already evident in practical measurements. Although a 4.45-minute restart time in checkpointing appears modest, Figure~\ref{fig:8192_gpu} shows the resulting ETTR when this measurement is combined with the MTTFs reported by Meta~\cite{meta2_realibility} across increasing cluster sizes. As the figure illustrates, effective training throughput collapses at production-scale deployments (64K/128K)~\cite{si2025collectivecommunication100kgpus, robust_bytedance, xai_blog_2024}, resulting in 16.47\% and 31.79\% throughput loss, translating to \$0.95M and \$3.63M of daily monetary loss, respectively.

\section{\sysname Overview}
\label{sec:system_overview}

To break through this scalability barrier, we design \textit{\sysname}, a resilient LLM training runtime for production-scale clusters that sustains high ETTR under frequent interruptions.

\sysname migrates workloads only from interruption-affected \textit{leavers} to healthy backup \textit{joiners}, leaving the remaining \textit{stayers} unchanged. By strategically leveraging readily available machines and overlapping all preparable work ahead of time, \sysname moves most job-setup procedures off the critical path before the joiner participates in training.

This design enables a carefully orchestrated migration procedure that achieves 10× lower downtime for both expected and unexpected interruptions, without any additional GPU memory overhead, thereby avoiding the risk of out-of-memory crashes during migration.

As shown in Figure~\ref{fig:system_overview}, instead of long restarting (bottom), \sysname (top) splits the migration into two phases: in the \textit{preparation
phase}, 
\sysname prepares the joiners' training stack in the background without affecting
the foreground training job; 
once the \textit{joiners} are ready and the foreground job finishes the current iteration, the system enters the \textit{switching phase} and synchronizes each \textit{joiner} to the latest training state before taking over the \textit{leaver}.
Figure~\ref{fig:system_workflow} illustrates \sysname's workflow of handling both expected and unexpected interruptions as described below.
\mypar{Expected Interruptions.}
Expected interruptions—such as scheduled maintenance—provide advance notice. When such an event is planned, the controller issues an expected-migration signal~\textcircled{1}. \sysname then selects elastic machines as \textit{joiners} and immediately places them into a background \textit{preparation phase}~\textcircled{2}, while \textit{stayers} and \textit{leavers} continue training normally. Once preparation completes, \sysname triggers a brief \textit{switching phase}~\textcircled{3}, during which the joiner receives the latest training state one-to-one from its paired leaver and takes over its role. This switching phase contributes to the total training pause downtime.

\mypar{Unexpected Interruptions.}
Unexpected failures occur without prior notice, so a warmed standby machine must remain in the background \textit{preparation phase}~\textcircled{a}. When an unexpected failure is detected~\textcircled{b}, the controller initiates recovery: \sysname immediately promotes the standby machine to a joiner and transitions directly into the \textit{switching phase}~\textcircled{c}, where the joiner retrieves the up-to-date state from a checkpoint, and other machines roll back to the same checkpoint~\cite{gemini, robust_bytedance,check_n_run}\footnote{If such a per-iteration checkpointing system is unavailable, other machines must restore their state from remote checkpoint storage.}. Once the switching phase completes, the \textit{leavers} exit and the \textit{joiners} replace their roles, resuming training seamlessly.




\sysname shifts as much work as possible off the critical path into the \textit{preparation phase} while keeping the \textit{switching phase} as lightweight as possible. However, several inherent properties of large-scale training fundamentally limit how much work can be shifted, creating key challenges for minimizing downtime.


\mypar{(\secref{sec:warmup}) Initialization is implicit and tightly coupled.}
A joiner must complete many initialization steps before starting training, yet modern frameworks trigger these steps implicitly across multiple software layers and often require global coordination.
These heavy behaviors forbid initialization from being invoked independently; performing them in advance requires untangling the tightly coupled procedures and understanding the implicit steps.
\revision{\sysname addresses this with a \textit{communication-free sandbox} (\secref{sec:warmup}): rather than untangling each implicit initialization path, the joiner simply runs a full shadow training iteration in isolation, so all computational initialization fires naturally—without blocking ongoing training or requiring assistance from active machines.}

\mypar{(\secref{sec:nccl}) Communication groups assume static membership.}
Collective communication libraries are designed around fixed group membership. Replacing even a single machine typically requires globally tearing down and rebuilding all communication groups—a slow, synchronized procedure that dominates the interruption-handling critical path.
\revision{\sysname addresses this with a \textit{two-phase delta-based CCL setup} (\secref{sec:nccl}) that allows CCL to decouple and perform all required setup as preparation, and apply only fast, delta connection-level changes when the membership switch actually occurs—with no additional GPU memory overhead.}

\mypar{(\secref{subsec:standby}) Standby resources must cover unexpected failures.}
Unexpected failures can occur on any machine and in any training role, making advance preparation impossible. Warming up standbys for each role is wasteful at scale, while relying solely on elastic machines acquired after a failure introduces large critical path delays. Standby resources must therefore be role-agnostic and able to recover swiftly when failures occur.
\revision{\sysname solves this with a \textit{general standby} design (\secref{subsec:standby}) that exploits the structural symmetry of LLM training ranks: a single standby sequentially warms up for each pipeline role type, making it a universal recovery target without dedicating a separate standby per role.}

\revision{These three techniques compose directly onto the two-phase workflow above. At the start of training, hot standby machines are pre-deployed (\secref{subsec:standby})~\textcircled{a} to guard against unexpected failures; each runs the communication-free sandbox (\secref{sec:warmup}) and Phase~1 CCL setup (\secref{sec:nccl}) upfront, so upon a failure it enters the \textit{switching phase} directly. For expected interruptions, a hot standby is optional—upon receiving the migration signal~\textcircled{2}, an elastic machine can serve as the joiner and undergo the same preparation (\secref{sec:warmup}, \secref{sec:nccl}) overlapping with ongoing training. In both cases, the \textit{switching phase}~\textcircled{3}/\textcircled{c} enters CCL Phase~2 to switch only the delta inter-machine connections (\secref{sec:nccl}) and transfer the state from the leavers to the joiners, keeping downtime to seconds.}
\label{sec:naive_live_migration}

\section{Sandboxed Initialization} \label{sec:warmup}


The joiner should initialize independently and in advance during the \textit{preparation phase} to avoid global blocking; otherwise, all machines must pause until the joiner completes its setup. Yet, training initialization spans both cross-machine and intra-machine dependencies, making it difficult to identify and trigger initialization gracefully and independently (\secref{subsec:lazy_init}).

To address this, we propose a \emph{communication-free sandbox} (\secref{subsec:sandbox} and \secref{subsec:sandbox_optimization}) that allows joiners to execute shadow iterations in isolation to trigger initialization ahead of time. Afterward, the joiner enters the main loop and seamlessly continues training.

\subsection{Prolonged and Complex Training Initialization}
\label{subsec:lazy_init}

Training initialization, once overlooked as a one-time cost, has become a significant source of inefficiency as training scales grow and interruptions become more frequent. Each interruption must redo warm-up, wasting GPU time. In the GPT-10B experiment shown in Table~\ref{tab:nccl_setup}, excluding NCCL instantiation, a single interruption triggers about 150 seconds of warm-up before reaching stable performance.
The first iteration is also much slower—about 6$\times$ longer (around 44 seconds in our GPT-10B experiment)—due to just-in-time (JIT) compilation and initialization dependency chaining. For example, in pipeline-parallel training, each stage depends on the completion of the P2P communication of the previous stage, requiring initialization to proceed sequentially across stages and further amplifying the startup overhead.

Initialization is complex and spans multiple layers of the training stack. While frameworks like JAX provide an eager mode to explicitly trigger initialization, mainstream training frameworks such as Megatron-LM, DeepSpeed, and NeMo rely on hardware-aware optimizations (e.g., memory-layout specialization and fused kernels) that are only activated when real data arrives (e.g., the first iteration). Furthermore, initialization is intertwined with communication, as exemplified by the pipeline-parallelism case discussed above. Addressing these behaviors would require uncovering and manually invoking many hidden initialization paths or thoroughly examining user training code. We need a mechanism that automatically triggers these complex initialization behaviors without requiring user intervention.

\subsection{Shadow Iteration in Communication-free Sandbox}
\label{subsec:sandbox}

To pre-trigger initialization without any other machines' assistance, we create a \textit{communication-free sandbox} that allows joiners to run a shadow iteration independently before joining the main training loop. This sandbox eliminates the need for case-by-case initialization handling: joiners can execute all implicit and framework-specific initialization routines in one shot, without relying on distributed communication or requiring any assumptions about user code or framework behavior.

\begin{figure}[tb]
    \centering
    \includegraphics[width=\linewidth]{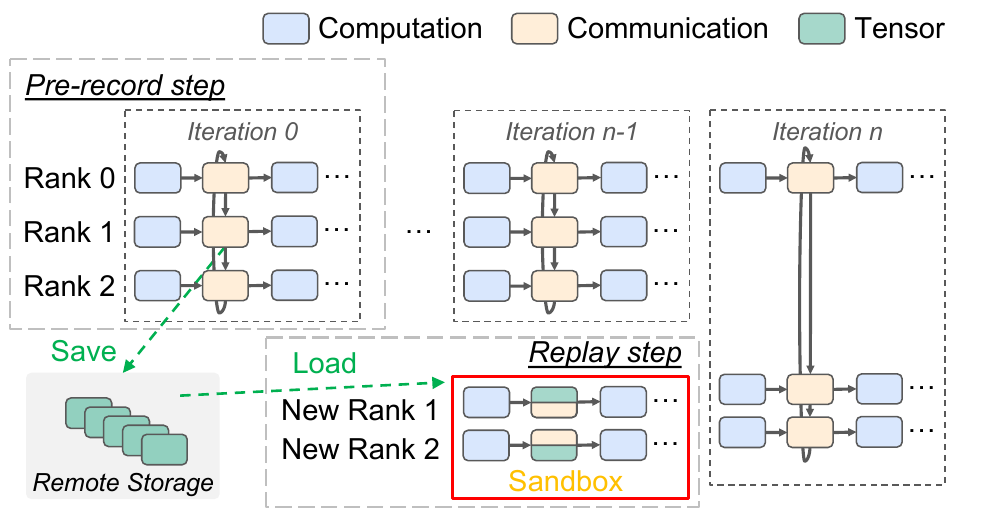}
	\caption{Sandbox initialization workflow}
    \label{fig:sandbox}
\end{figure}

This design introduces two key requirements:
first, the shadow iteration must execute with valid state values and communication results to ensure the program does not crash and that initialization is correctly triggered—using dummy or zero values can easily cause NaNs or assertion errors. Second, communication operations must be handled locally without depending on communication responses from other machines. We address both requirements through valid tensor record–replay, as illustrated in Figure~\ref{fig:sandbox}.

The sandbox warm-up has two steps: \textit{pre-record} and \textit{replay}. In the pre-record step, the designated recording machines capture one or more valid iterations and store the resulting communication outputs in advance. When a migration signal is received or a standby becomes available, the joiners enter the sandbox, where each joiner runs the replay step—reusing the recorded communication results to trigger its initialization.

\mypar{Shield the Warmup --- Pre-record step.}
We build a record hooking layer that sits between PyTorch and the CCL to intercept communication calls issued by the training framework. At the start of a new job as shown in Figure~\ref{fig:sandbox}, designated machines—including Rank 0, Rank 1 and Rank 2—enter the pre-record step to prepare for potential migration. Our hook \revision{monitors and records the output tensors of all collective calls (e.g., all-reduce) to persistent storage once the collective completes.}
They are later replayed in the sandbox to ensure valid training states on the joiner.

This one-time pre-record step runs only during the first training iteration (or a few iterations) at the very beginning. After that, the interception hook is removed, and training proceeds as normal, introducing no additional overhead for the remainder of the training process regardless of the number of interruptions experienced or number of job restarts.

\mypar{Run the Warmup --- Replay step.}
When a migration signal is received or a standby is available (\textit{the preparation phase}), all joiners (new Rank 0 and new Rank 1) enter the \textit{replay step} and are placed into a sandbox, isolating them from the ongoing training in preparation for triggering initialization before joining the main training process. First, each sandboxed joiner pulls the initial state from the checkpoint.
After loading the state, the joiner proceeds to a shadow iteration to trigger initialization.
Our sandbox is designed to be \textit{communication-free}, specifically to avoid any interaction with existing training machines. We deliberately prevent any assistance from active participants—such as responding to collective operations or sending state information—to ensure that the joiner triggers initialization entirely on its own.

We attach another hook to the PyTorch communication layer to intercept communication calls made during the replay step. Any communication call that attempts to reach outside the sandbox is intercepted and instead served with previously recorded tensors as responses. Certain operations—such as \codesm{barrier} and \codesm{send}—are safely bypassed, as invoking these operations does not affect the caller's state.

After the shadow iteration is complete, the joiners (e.g., New Rank 0 and New Rank 1) can replace the leavers (e.g., Rank 0 and Rank 1) at the end of iteration~$n-1$ once they have received the up-to-date states in the \textit{switching phase}, enabling a machine transition with no initialization downtime overhead required after migration.

\mypar{Robustness and Correctness.}
\revision{The sandbox warm-up relies on a degree of determinism in the ML framework that modern LLM training frameworks satisfy. By design—fixed kernel schedules, static memory layouts, and deterministic collective ordering are maintained for reproducibility and peak performance. As a result, almost all initialization—including data-loader setup, CUDA context creation, and memory allocation—is correctly triggered during sandbox warm-up.} Only in rare cases does a component depend on runtime data or require CUDA graph capture whose trigger may be missing. \revision{MoE routing is one such case: token routing is input-dependent and varies across iterations. However, in practice this has minimal impact—MoE frameworks (e.g., Megatron-LM) support preallocating fixed-size expert buffers regardless of the actual routing decision, so the underlying memory layout and communication structure remain static. Any initialization not triggered during sandbox warm-up simply completes during the \textit{switching phase}, introducing only negligible overhead (\textit{ms-level}).}

Also, correctness is always guaranteed because the leaver's latest model and optimizer states will eventually overwrite all local states after sandbox warm-up in the \textit{switching phase}.

\subsection{Low Overhead Record and Replay}
\label{subsec:sandbox_optimization}

Recording and replaying tensors incur a one-time overhead in both storage and tensor loading time. However,
not all communication needs to be recorded or replayed--- each recording is essentially \textit{a preparation for communication edges that might be temporarily missing during sandbox warmup}. We can reduce the recording scope by focusing only on what might be needed. For example, since migrations happen only at the machine level, intra-machine communication (where intensive TP traffic is always involved) can run natively inside the sandbox without record. We further eliminate redundant recordings for duplicated training roles, as discussed in \secref{subsec:standby}. The storage overhead can be eventually reduced to under 300 GB in a GPT 5.12T MoE model setting.

During replay, as illustrated in the lower part of Figure~\ref{fig:sandbox}, only the communication that touches the sandbox boundaries in the training graph relevant to the new joiner needs to be loaded and replayed. When multiple machines are migrated together as a batch (e.g., an entire PP group), the joiners' internal communication can remain real, while only cross-boundary communication is loaded from storage. This boundary-aware record-and-replay design contributes to low I/O overhead.

\section{Dynamic and Lightweight Communication group setup}
\label{sec:nccl}

Migrating one machine to another requires updating the membership of the existing CCL groups. However, CCL groups lack flexibility and are costly—they take time to create, and once created, their membership cannot be modified (\secref{subsec:nccl_high_cost}). Existing approaches~\cite{parcae, oobleck} rely on globally destroying and recreating groups, which introduces unacceptable delays. An alternative strategy is to concurrently pre-allocate new groups in the background to avoid blocking, but this incurs additional GPU memory overhead. To solve this downtime–memory dilemma, we introduce a two-phase asynchronous CCL setup (\secref{subsec:two_stage_control}) that enables dynamic group membership updates with minimal downtime and zero additional GPU memory overhead.

\subsection{Today's Slow and Static CCL Setup}
\label{subsec:nccl_high_cost}

\begin{figure*}[t]
    \centering
    \begin{minipage}[t]{0.35\textwidth}
        \vspace{-20mm}
        \centering
        \footnotesize
        \resizebox{\textwidth}{!}{
        \begin{tabular}{|cc|c|}
        \hline
        \multicolumn{3}{|c|}{\textbf{NCCL Setup Components}} \\ \hline
        \multicolumn{2}{|c|}{Network bootstrap} & 2.48s (4.92\%) \\ \hline
        \multicolumn{2}{|c|}{Topology discovery and computation} & 9.40s (18.63\%) \\ \hline
        \multicolumn{1}{|c|}{\multirow{2}{*}{\shortstack{Connection\\Establishment}}} & Intra-machine & 21.49s (42.59\%) \\ \cline{2-3}
        \multicolumn{1}{|c|}{} & Inter-machine & 17.07s (33.86\%) \\ \hline
        \end{tabular}}
        \captionof{table}{Time breakdown aggregated over all NCCL groups.}
        \label{tab:nccl_setup}
    \end{minipage}
    \hfill
    \begin{minipage}[t]{0.64\textwidth}
        \centering
        \includegraphics[width=\linewidth]{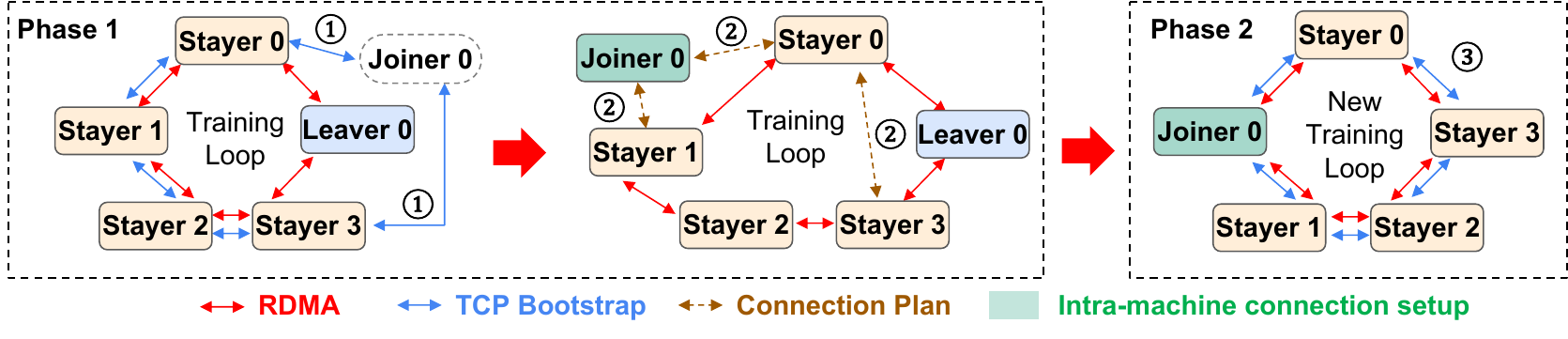}
        \caption{Two-phase CCL Migration workflow}
        \label{fig:migration_workflow}
    \end{minipage}%
\end{figure*}

\textbf{Communication setup is \textit{slow}}. The slowness stems from three main procedures: (1) \textit{Network bootstrap} — participants form a TCP network through multi-round handshakes and consistency checks, which is expensive at scale due to the need for global coordination~\cite{si2025collectivecommunication100kgpus}; (2) \textit{Topology discovery} — each node gathers local device metadata (e.g., NIC bandwidth) and exchanges it via ring all-gather to compute global topology and determine communication neighbors, incurring high synchronization overhead; and (3) \textit{Connection establishment} — nodes allocate GPU buffers and set up inter- and intra-machine connections (e.g., GPU Direct RDMA, NVLink) based on the computed topology. These setup procedures can still be slow despite high parallelism, due to their inherent synchronization behaviors.

Table~\ref{tab:nccl_setup} shows the CCL setup overhead on an 8-machine cluster with 64$\times$40GB A100 GPUs, training a GPT-10B model with TP=4, PP=2, and DP=8. The setup takes around 50 seconds due to the overhead of establishing and managing communication groups—a challenge that grows with scale. For example, training across 1,000 machines results in a connection count that scales with 1,000 × (\# of CCL groups) × (\# of channels inside each group). The number of CCL groups depends on the communication dimensions (e.g., DP, PP, TP); even in this modest 3D setup, each host participates in 7 groups. High-bandwidth links (e.g., 400~Gbps) further increase the number of required channels. Although many setup operations can run in parallel, the scalability of these massive numbers of connections is ultimately limited by global synchronization barriers, straggler effects from heterogeneous network latencies, and driver-level contention under high concurrency (e.g., CUDA and IB verbs serialization), causing setup time to grow with cluster size.

\textbf{Communication setup is \textit{static}}. CCL connections are tightly coupled to hardware (e.g., NVLink, GPU Direct RDMA), requiring memory mappings (via CUDA IPC) and optimized connectivity topology computation—all constructed at initialization. These components are specific to the participating ranks and GPUs, so any group member change requires structural updates and may yield suboptimal performance if not carefully handled.

\subsection{Two-phase Delta-based CCL Setup}
\label{subsec:two_stage_control}

We design a dynamic and lightweight CCL setup that minimizes transition overhead when a joiner migrates into an ongoing training job. Our two-phase design shifts most of the setup to Phase 1, where it is performed in advance and overlapped with training, deferring only minimal connection establishment to Phase 2 to reduce downtime. Additionally, we enable calculating and reestablishing only the delta connections between the old and new topology, avoiding the need to globally reconnect all links.

We demonstrate the migration workflow with four stayers (0–3), one joiner, and one leaver in Figure~\ref{fig:migration_workflow}. Migration begins with a signal and role assignment from the controller. In Phase 1, Stayers 0–3 invoke \texttt{CCL\_prepare\_stayers(new\allowbreak\_topology)} to incrementally modify their existing CCL group in preparation for the new setup with the joiner. This function also enables topology-aware configurations provided by users as input. Meanwhile, the joiner concurrently calls \texttt{CCL\_prepare\allowbreak\_joiners()}. During this process, the original group's communication remains functional \textcircled{1}. Stayers reuse existing TCP bootstrap connections and initiate the bootstrap process with Joiner 0. Since the additional TCP connections and topology state reside in CPU memory, this setup overlaps with training and incurs no GPU memory overhead. Once connected, the joiner and stayers exchange the necessary topology and state information.

After exchanging topology information, a recalculation of connectivity is required, as the original topology may no longer be optimal after migration \textcircled{2} (e.g., Joiner 0). Once the resilience runtime receives the new topology information, all participants locally compute a \textit{delta topology}, identify the minimal set of \codesm{channels} required for the updated configuration, and generate a delta reconfiguration plan that determines how to modify the necessary channels. Since all changes occur at the channel level, the design remains compatible with diverse topologies and communication techniques, such as hierarchical connectivity and SHARP~\cite{nccl_sharp}.

The reconfiguration plan specifies which \codesm{channels} need to be updated. For \codesm{channels} that do not require replacement, they can be directly inherited and reused. Intra-machine channels (e.g., NVLink) are typically fully inherited, as they provide the fastest and highest-bandwidth paths and usually remain unchanged during migration, which operates at the granularity of entire machines. Meanwhile, the joiner (Joiner 0) begins setting up channels that can be initialized locally (e.g., intra-machine connections). If there are multiple joiners that need to connect with each other, their inter-connections are also established accordingly.

Delta \codesm{channel} connections in the reconfiguration plan—typically inter-connections—are intentionally deferred to the second phase. At this point, intra-machine communication is already available through inheritance, while inter-machine connections have not yet been established on the stayer machines—ensuring no additional GPU memory is consumed. Afterward, both the stayer and joiner CCL groups enter the \textit{ready-to-switchout} state. Notably, the stayer and the leaver continue training without interruption during this phase.

Once both the joiners (e.g., Joiner 0) and impacted stayers (e.g., Stayer 0 and Stayer 3) complete Phase 1 and enter the \textit{ready-to-switchout} state, they invoke the second API, \texttt{CCL\_switchover()}, to switch delta inter-machine connections from stayer–leaver to stayer–joiner. For instance, the connection between Stayer 0 and Leaver 0 is replaced by one between Stayer 0 and Joiner 0; likewise, Stayer 3–Leaver 0 is switched to Stayer 3–Joiner 0. The replacement happens at the RDMA queue pair (QP) level, where it re-establishes the inter-machine peers, transitions the CCL group from \textit{ready-to-switchout} to the normal state, and reclaims the resources tied to the old topology \textcircled{3}. This delta operation is the sole contributor to network downtime. The entire procedure incurs zero memory overhead; a detailed workflow can be found in \secref{sec:zero_mem}.

\label{subsec:unexpected_nccl}
\label{subsec:state_sync}

\begin{figure}[tb]
    \centering
    \includegraphics[width=\linewidth]{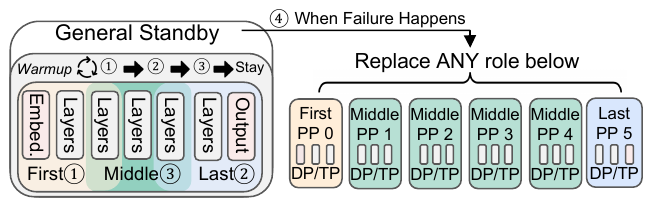}
	\caption{General Standby}
    \label{fig:general_standby}
\end{figure}

\section{General Standby} 
\label{subsec:unexpected_sandbox}
\label{subsec:standby}


Failures can occur unexpectedly at any time. A key challenge in handling such events is that the standby machine does not know in advance which machine it should prepare for, as each machine holds different parallelism roles (e.g., TP and DP IDs). We design a \textit{general standby} mechanism that prepares the standby for \textit{any} rank by leveraging the symmetry of training ranks and the fact that warm-up requires only valid states, independent of role. This design enables both sandboxed initialization (\secref{sec:warmup}) and two-phase CCL warm-up (\secref{sec:nccl}) in unexpected failure cases and significantly reduces the number of standby machines needed.



\subsection{Symmetry on Training Ranks}


LLM architectures exhibit strong symmetry, as maintaining identical model partitions across all machines helps achieve peak synchronous training performance. Parallel methods such as tensor parallelism (TP), data parallelism (DP), expert parallelism (EP), and other structurally symmetric forms of parallelism maintain these identical partitions, which share the same parameter shapes and optimizer-state layouts; for example, running a DP+TP training job in Megatron-LM produces identical memory utilization across all ranks. 
Each rank executes the same set of computational operators and CUDA kernels\footnote{Although some techniques may introduce different execution orders due to scheduling patterns (e.g., VPP), the underlying set of operators remains the same.}, and participates in the same collective communication patterns. This strong, correctness-driven symmetry allows the standby to serve as a universal recovery target.

Pipeline parallelism (PP) introduces an exception in its response to the model’s inherent symmetry: each PP stage contains a subset of layers, while the first and last stages include additional layers such as the \texttt{word\_embedding\_layer} in the \textit{first stage} and the \texttt{output\_layer} in the \textit{last stage}. Despite these variations, the first and last stages remain similar in structure: the \texttt{word\_embedding\_layer} and \texttt{output\_layer} share the same shape (e.g., weight tying). The middle stages consist of repeated transformer blocks and are kept symmetric for optimal performance
. In total, PP introduces at most three distinct role types—the first stage, the middle stages, and the last stage—yet the overall partitioning remains highly symmetric. 

\subsection{General Warm-Up Procedure}




The symmetric design allows the standby machine to warm up and recover any role in the system.
Figure~\ref{fig:general_standby} illustrates the warmup procedure. To prepare it, the standby runs warm-up iterations for initialization~(\secref{sec:warmup}). Without pipeline parallelism (PP), a single iteration is sufficient because all machines share identical roles. When PP is enabled, the standby sequentially and independently runs up to three warm-up iterations \textcircled{1}\textcircled{2}\textcircled{3}, each corresponding to a different role—the \textit{first stage}, \textit{last stage}, and \textit{middle stage}. Running each role during warm-up concretely initializes all role-specific components, such as JIT-compiled fused kernels.

At this stage, the general standby remains in Phase~1 of the CCL setup, where inter-machine connections are not yet established, its overall GPU memory footprint stays hundreds of megabytes below peak usage (detailed measurements in \secref{sec:zero_mem}). \revision{This available headroom easily accommodates the GPU-resident artifacts produced during warm-up across all role types—including JIT-compiled kernels and other compiled GPU artifacts specific to each role's layers—which add at most a few hundred kilobytes per role.}


After warm-up, we retain the middle-stage role because it represents the majority of pipeline instances. If a failure occurs on a middle stage, the standby can replace it directly. For failures in the first or last stage, \sysname updates only the small delta of layers unique to those stages. This delta is negligible because only the memory for the optimizer states and parameters of the additional layers (e.g., embedding or output layers) needs to be allocated. Other artifacts, such as all precompiled kernels, are already initialized. Even at large scale, the standby needs to run only these three role types.


Two-phase CCL preparation~(\secref{sec:nccl}) is also enabled by symmetry, because each machine exposes an identical intra-node and inter-node topology. For example, if a CCL group contains two separate TP groups within a machine, the same configuration must exist on all other machines. This eliminates the need for dynamic topology specialization during recovery. The only exception arises with pipeline parallelism (PP): when PP spans multiple machines, the standby must additionally establish a few connections with its neighboring stages after recovery, according to its assigned role. 



Leveraging this symmetry, we can prepare a general standby for unexpected failures at the beginning of the training job in parallel, with a deployment time no greater than a normal initialization. The general standby achieves a downtime nearly identical to that of the switching phase in an expected event, as evaluated in \secref{eval:benchmark}. The cost effectiveness discussion on general standby machine can be found in \secref{sec:discussion}.

\revision{The number of standbys needed is small in practice (\secref{eval:sys_performance}): the probability of two machines failing simultaneously equals the product of their individual MTTFs, making concurrent failures extremely unlikely~\cite{robust_bytedance,just-in-time-checkpoint,optmeta}. \sysname{} also supports the no-standby case and still recovers via its overlapped recovery path (\secref{sec:implementation}). In practice, 1--2 standbys cover all distinct stage types (first, middle, last) for PP configurations; operators who wish to overprovision can assign dedicated standbys per stage proportion, e.g., with PP degree 8 and a 1:6:1 stage ratio, allocating more standbys to the dominant middle stages.}
\section{Implementation}
\label{sec:implementation}
\revision{\sysname is implemented in Python and C/C++, comprising two main components: the training-node resilience runtime and the controller. \sysname{} comprises about 12K lines of new code: 1.9K LoC of Python for the controller, 7.8K LoC of Python for the Megatron-LM runtime extensions, 1.1K LoC of C++/Python for the PyTorch \codesm{c10d} extensions, and 1.5K LoC of C/C++ for the NCCL extensions. The code will be open-sourced after publication.} 

\mypar{Training Node Resilience Runtime and Controller.} The \sysname runtime builds on Megatron-LM with focused modifications to Megatron-LM, PyTorch, and the CCL layer to support resilient training. In Megatron-LM, it manages interruption signals and preserves CCL ordering during overlapping training and migration. In PyTorch’s \codesm{c10d}, we enable multiple global CCL groups, support the CCL layer’s two-phase initialization API, and add an interception layer that records, replays, or bypasses tensors to facilitate sandbox warm-up. In the CCL layer, we implement the new APIs and demonstrate them in NCCL. The controller coordinates the system by assigning roles, initiating migrations, detecting failures, and distributing topology metadata; each node’s runtime maintains a control channel with the controller to keep metadata synchronized and ensure joiners, leavers, and stayers correctly configure their CCL connections.

\mypar{State Synchronization.}
\revision{Concurrently with CCL switchover, the joiner receives the up-to-date training state before resuming. For expected events, the \textit{leavers} transmit the latest training states—such as model parameters and optimizer states—to their corresponding \textit{joiners}, overwriting the sandbox-warmed state and bringing the joiner fully up to date with the current training progress. For unexpected failures, the \textit{joiners} retrieve the required states based on the redundancy configuration. If a copy is available on other \textit{stayers}—through DP group redundancy or remote CPU memory checkpointing~\cite{gemini}—\sysname directly queries the states from a {\em stayer} machine that maintains a copy. Otherwise, in the absence of redundancy—such as with the distributed optimizer~\cite{megatron-code} or ZeRO optimizer~\cite{zero1, zero-infinity}—the \textit{joiners} fetch the data from remote storage checkpoints~\cite{bytecheckpoint, parcae}. The \textit{leaver}–\textit{joiner} mappings are one-to-one, meaning that state transmission and connection establishment for different \textit{joiners} occur independently and in parallel—the overhead does not grow with cluster size. The transfer is feasible within the $\sim$20-second recovery window because modern GPU memory is bounded (e.g., 80~GB per A100), and the state is transmitted via high-bandwidth RDMA directly from a neighboring stayer's GPU or CPU memory, achieving transfer rates of hundreds of GB/s, well within the recovery budget.}

\section{Evaluation}
\label{sec:eval}

Our experiments show that \sysname scales effectively, with about 20 seconds of downtime at 1024 GPUs and over 55\% less wasted GPU time at 64K GPUs compared to the best baseline (\secref{eval:sys_performance}). We further analyze event downtime (\secref{eval:benchmark}) and additional use cases (\secref{eval:use_cases}), where \sysname sustains 97\% training efficiency during periodic 10-minute load rebalancing in 1024-GPU scale. A detailed breakdown of the designs and the \textit{zero} memory overhead is shown in~\secref{eval:breakdown}.

\subsection{Experiment Setup}
\label{eval:setup}

\begin{figure}[t]
    \centering
    \begin{minipage}{0.44\columnwidth}
        \centering
        \scriptsize
        \setlength{\tabcolsep}{2pt}
        \resizebox{\linewidth}{!}{
        \begin{tabular}{|c|c|c|c|c|c|}
        \hline
        \textbf{\# GPU} & \textbf{32} & \textbf{128} & \textbf{256} & \textbf{512} & \textbf{1024} \\ \hline
        \textbf{Type} & Dense & MoE & MoE & MoE & MoE \\ \hline
        \textbf{Model} & 175B & 960B & 1.28T & 2.56T & 5.12T \\ \hline
        \textbf{TP} & 4 & 8 & 8 & 8 & 8 \\ \hline
        \textbf{PP} & 8 & 8 & 8 & 8 & 8 \\ \hline
        \textbf{EP} & / & 2 & 4 & 8 & 16 \\ \hline
        \end{tabular}}
        \captionof{table}{Large-scale model configurations used in experiments.}
        \label{tab:large_scale_model}
    \end{minipage}
    \hfill
    \begin{minipage}{0.55\columnwidth}
        \centering
        \ifdefined\ArxivVersion
            \includegraphics[width=\linewidth]{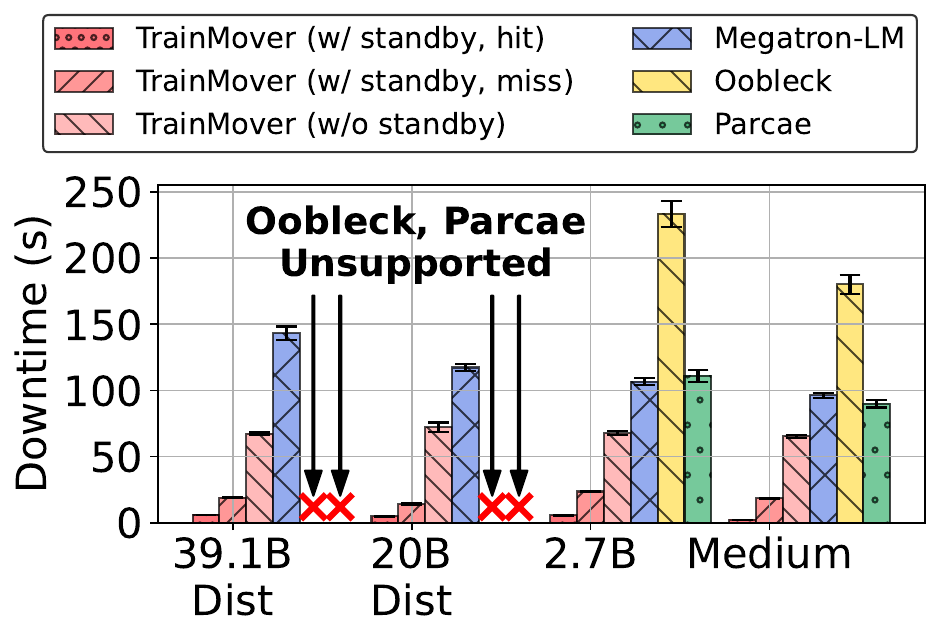}
        \else
            \includegraphics[width=\linewidth]{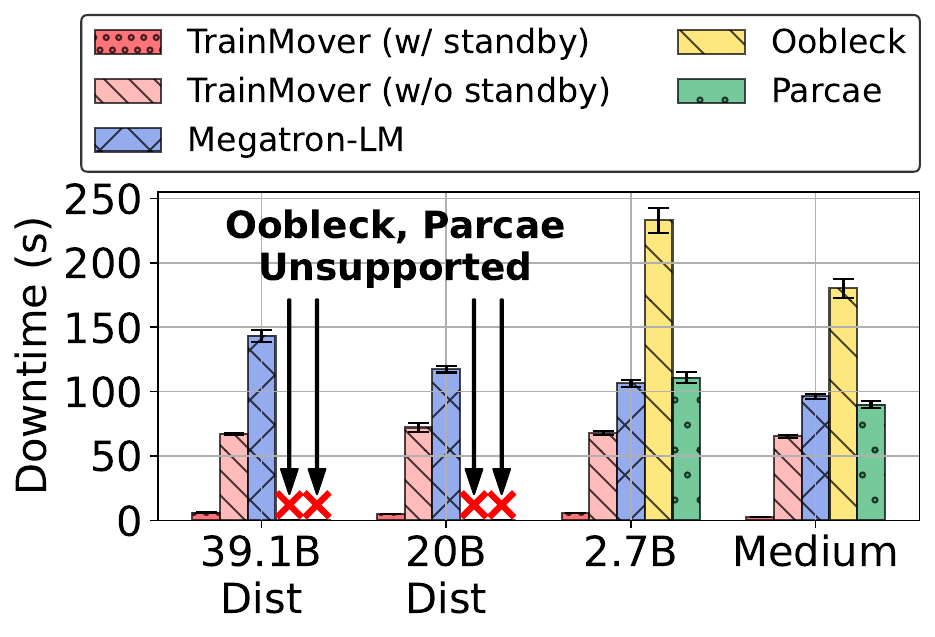}
        \fi
        \caption{Unexpected failure downtime}
        \label{fig:unexpected_failure}
    \end{minipage}
\end{figure}

\mypar{Testbed and Model Settings.}
We conduct our experiments on a 1024-GPU testbed and evaluate a range of models—including GPT 5.12T MoE, GPT-175B, GPT-39.1B~\cite{391b}, GPT-20B~\cite{gemini}, Medium~\cite{gpt_models}, GPT-2.7B~\cite{gpt_models} and others—to cover diverse model sizes across different scales. Large-scale model configurations are summarized in Table~\ref{tab:large_scale_model}. \revision{Detailed downtime analysis is additionally conducted at the 32-GPU scale to enable fair comparison with Oobleck and Parcae, which cannot scale beyond 32 GPUs (e.g., Oobleck's template construction time becomes prohibitively long at larger scales). For these experiments, models are tested with various TP, DP, and PP configurations including (TP1, PP8, DP3), (TP4, PP8, DP3), and (TP8, PP8, DP3). Default profiles are: GPT-Medium and GPT-2.7B use TP1, PP8, DP3 with a global batch size of 96 and microbatch size of 2; GPT-20B uses TP1, PP8, DP3 with the distributed optimizer, global batch size of 36, and microbatch size of 1; GPT-39.1B uses TP4, PP2, DP3 with the distributed optimizer, global batch size of 36, and microbatch size of 1. The Wikitext dataset~\cite{wiki_dataset} is used, following prior work~\cite{oobleck,parcae}.}





\mypar{Baseline.} Our primary baselines fall into two categories: the stop-and-restart approach, represented by Megatron-LM, and reconfiguration-based approaches, represented by Oobleck and Parcae. For expected events, since Oobleck and Parcae do not natively support live migration, we perform it by removing a node (-1) and adding a new one (+1). Parcae does not support tensor parallelism, and neither Oobleck nor Parcae supports distributed optimizers~\cite{zero1}, as both rely on redundancy within data parallelism. We include various 32-GPU experiments because Oobleck cannot scale due to its prolonged template-generation overhead. 

\mypar{Metrics.}
We primarily use three metrics in our evaluation. The first is \textit{downtime}, which measures the duration of interruption handling for both expected events and unexpected failures, reflecting the additional system time required to complete training per interruption. The second is \textit{GPU hours wasted per week}, an end-to-end metric that quantifies the total GPU time lost during event handling\revision{---essentially overhead expressed as an absolute resource cost}, which can be directly translated into training capital cost. The third one is \textit{ETTR}. 

\begin{figure*}[t]
\centering
\begin{minipage}{0.47\linewidth}
    \centering
    \begin{minipage}{\linewidth}
        \centering
        \includegraphics[width=\linewidth]{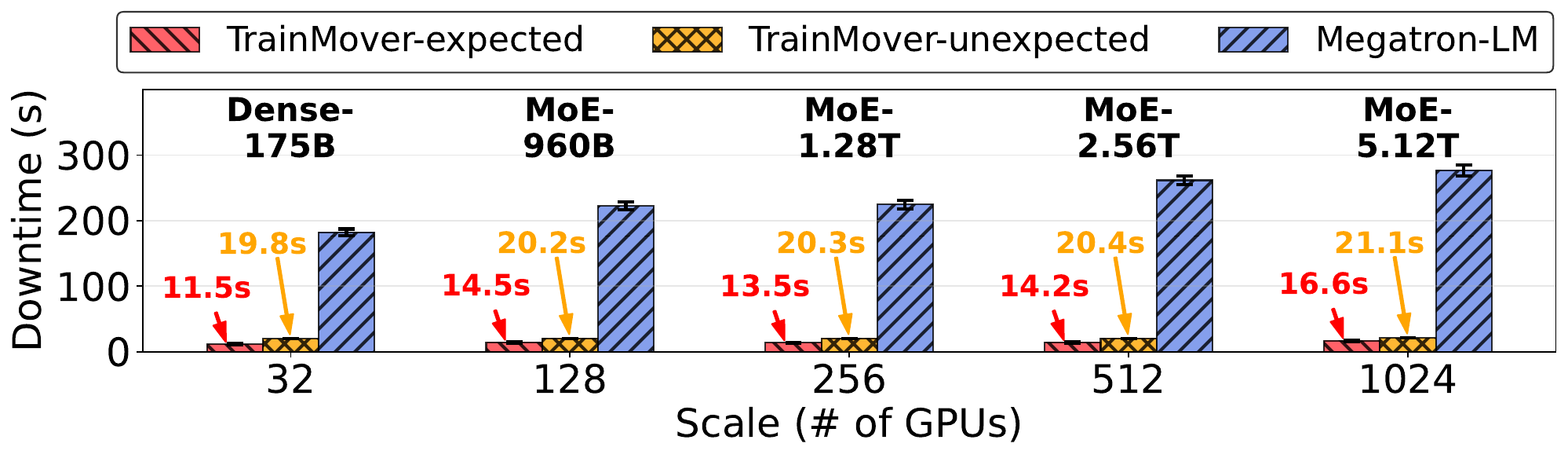}
        \caption{Downtime comparison across different GPU scales.}
        \label{fig:large_scale}
    \end{minipage}

    \begin{minipage}{\linewidth}
        \centering
        \includegraphics[width=0.7\linewidth]{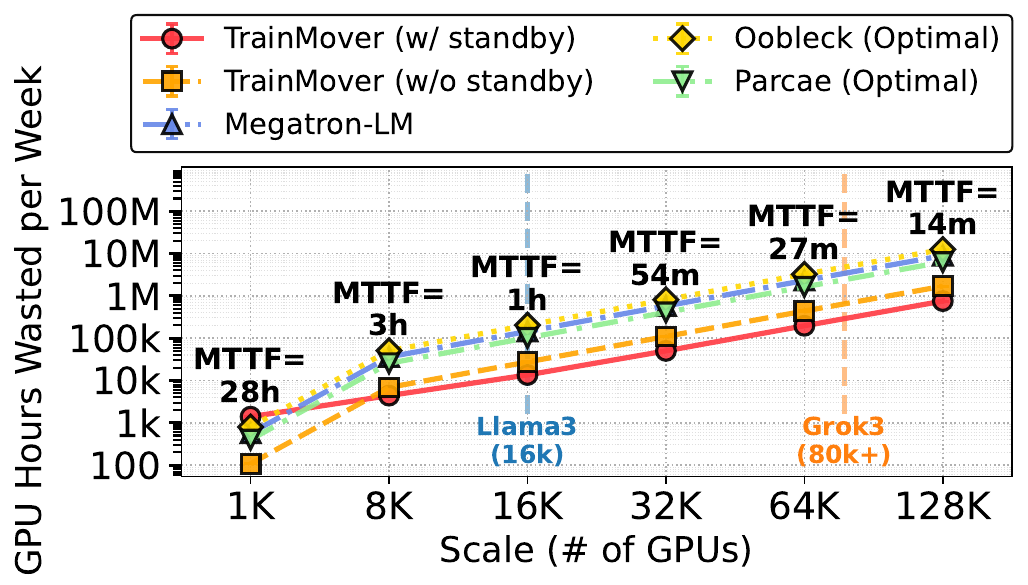}
        \caption{Projected GPU hour wasted in large scale deployment}
        \label{fig:large_scale_waste}
    \end{minipage}
\end{minipage}
\hfill
\begin{minipage}{0.52\linewidth}
    \centering
    \ifdefined\ArxivVersion
        \includegraphics[width=\linewidth]{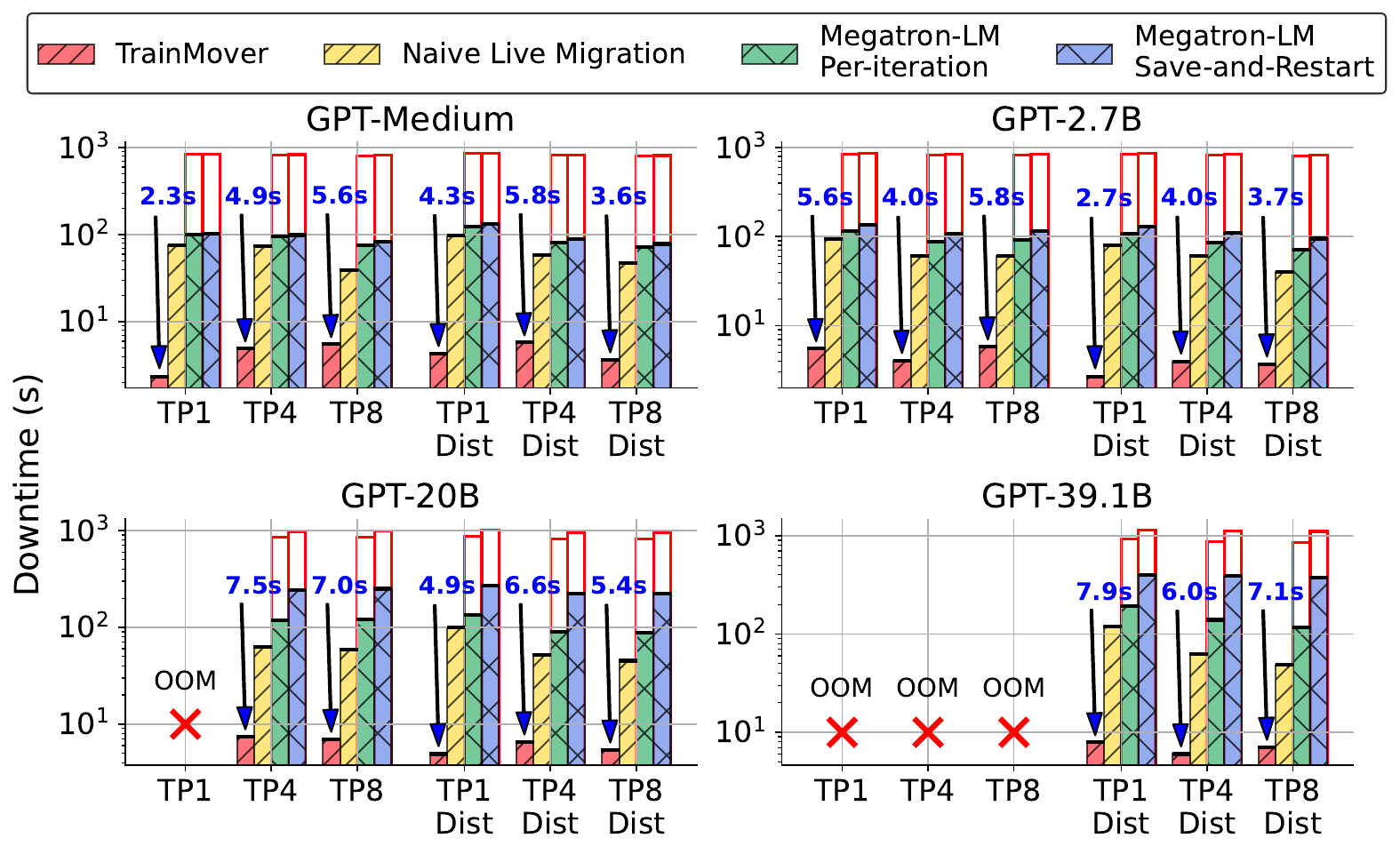}
    \else
        \includegraphics[width=\linewidth]{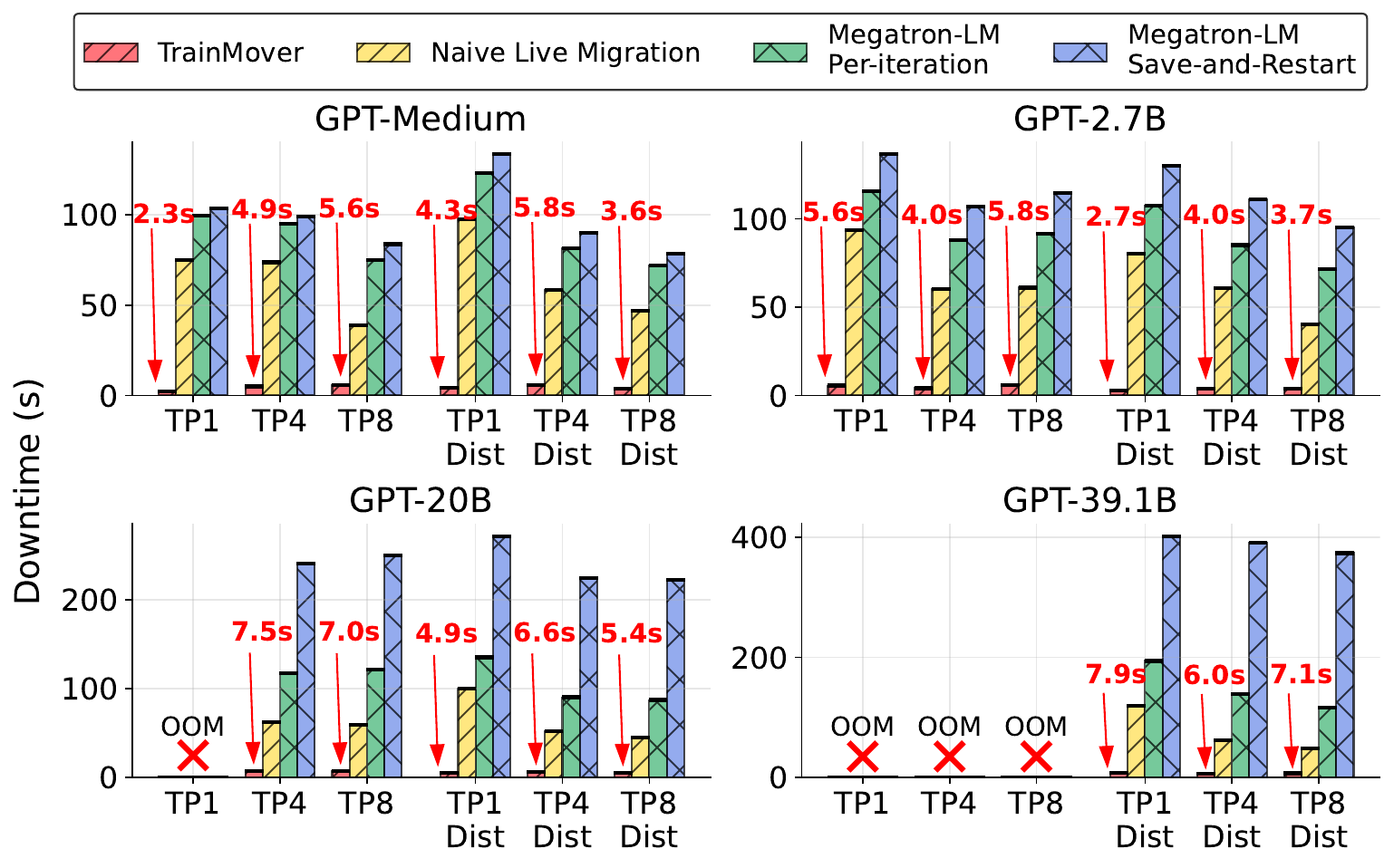}
    \fi
    \caption{Migration downtime with different models and parallel settings}
    \label{fig:overall_performance}
\end{minipage}
\end{figure*}

\subsection{End-to-end Experiments}

\label{eval:sys_performance}

\mypar{\sysname Performance at Scale. }
We evaluate downtime for both expected and unexpected interruptions on a large-scale testbed with up to 1024 GPUs, demonstrating \sysname's scalability and low downtime in Figure~\ref{fig:large_scale}. In \sysname, migration downtime during expected events consistently remains below 20 seconds, while downtime for unexpected failures is also below 30 seconds even at the 1024-GPU scale. The downtime increases by no more than 10 seconds when scaling from 32 to 1024 GPUs, as \sysname's design is fundamentally insensitive to job size.


The scalability benefit comes from our delta-based design. For expected events, during the switching phase, \sysname only updates the connections between the leaver and joiner, while model states are transferred in parallel between corresponding leaver–joiner pairs. All other machines and connections remain unaffected. The small increase in downtime primarily results from the larger number of RDMA connection re-establishments required at higher scales. For unexpected failures, only \sysname's general standby machine needs to retrieve its up-to-date state from a neighbor~\cite{gemini,robust_bytedance}. If such a system is not present, the failed machine must recover its state from the remote checkpoint storage. Other procedures remain nearly identical to those for handling expected events, as the general standby machine performs CCL initialization and sandbox initialization in advance. 




In contrast, Megatron-LM restarts the entire job during migration, requiring NCCL group re-instantiation and other initialization steps across all machines, taking up to nearly 300 seconds at the 1024-GPU scale. We exclude the job rescheduling and cleanup overhead in Megatron-LM, as these depend heavily on the underlying cluster management system, and this procedure typically takes minutes-level latency (Table~\ref{tab:scale_comparison}).

\mypar{Projected GPU-Hour Savings at Production Scale.}
In practice, large-scale training jobs often run on tens of thousands of GPUs. We project GPU-hour waste per week by extrapolating from our 1K-GPU measurements to production scales of up to 128K GPUs~\cite{robust_bytedance,xai_blog_2024} (Figure~\ref{fig:large_scale_waste}), using MTTF values from Meta~\cite{meta2_realibility} and a 1:8.9 expected-to-unexpected failure ratio~\cite{llama3}. Each system's downtime is held constant from its measured value: \revision{\sysname{} uses 1024-GPU measurements, while Oobleck and Parcae use 32-GPU measurements (\secref{eval:benchmark}) since they cannot scale beyond 32 GPUs. Using 32-GPU measurements gives Oobleck and Parcae an optimistic advantage, as downtime typically grows with scale; for \sysname{}, holding the 1024-GPU downtime constant is justified by its delta-based design, which only updates leaver-joiner connections and is scale-insensitive.} We also add a 2-minute infrastructure rescheduling cost to all systems from the 8192-GPU measurements (Table~\ref{tab:scale_comparison}), which is conservative as this overhead also increases with scale.


Two versions of \sysname{} are included for comparison. The first, \sysname{} (w/ standby), pre-allocates a standby machine, which is counted as a reserved resource and considered wasted if unused. We use one standby machine as backup because simultaneous multi-machine failures are rare~\cite{robust_bytedance,just-in-time-checkpoint,optmeta}. The second, \sysname{} (w/o standby), uses all GPUs without a dedicated standby. Without a standby machine, the handling downtime for unexpected failures increases, while expected events remain unaffected because their handling does not require standby resources. \revision{This configuration represents the case where all GPUs are dedicated to training with no reserved standby, and only \sysname{}'s recovery path optimizations are applied upon failure. It also characterizes the performance when the pool of reserved standby machines is exhausted.}

At the 1K scale in Figure~\ref{fig:large_scale_waste}, where failures are infrequent, the wasted GPU hours across all systems are similar. In fact, \sysname{} (w/ standby) incurs slightly higher waste than Oobleck and Parcae, as the standby GPU remains idle during training. However, as scale increases to production levels—for example, at 8K GPUs where interruptions become more frequent (MTTF = 3h)—\sysname{} (w/ standby) achieves the lowest GPU-hour waste, reducing waste by 35.91\% over \sysname{} (w/o standby) and by 82.77\% over Parcae. Beyond 8K GPUs, \sysname{} (w/ standby) continues to deliver the highest overall efficiency, as the marginal benefit of adding more training GPUs diminishes while failures become more frequent and restarts take longer. At the 128K scale, \sysname{} (w/ standby) further reduces weekly GPU-hour waste by 55\% compared to \sysname w/o standby and 88\% to Parcae.




\subsection{Downtime Analysis}
\label{eval:benchmark}


\label{eval:fault_tolerant}
\mypar{Fast Recovery from Unexpected Failures.} We evaluate \sysname's downtime under unexpected failures in two settings: with a general standby machine and without one, as shown in Figure~\ref{fig:unexpected_failure}. Since Oobleck cannot scale to 1024 GPUs because of the prolonged template-generation time, we run other systems at the same 32-GPU scale for a fair comparison \revision{with the 32-GPU model settings (\secref{eval:setup}).}
Looking at the Medium and 2.7B models without the distributed optimizer enabled, \sysname—despite having no standby—still outperforms all baselines. This is because \sysname accelerates the initialization procedure by overlapping CCL instantiation, warm-up, framework initialization, and state transmission even when a general standby is not present. In contrast, Megatron-LM executes all components sequentially after rebooting, resulting in up to 3.48× longer downtime than \sysname. Parcae and Oobleck, although enabling direct state transfer to reduce checkpoint loading time, still incur high restart overheads: initializing framework components, reinstantiating NCCL groups, and performing warm-up. These steps lead to higher downtime than checkpoint approaches.

For the GPT-39.1B and GPT-20B models, where the distributed optimizer (DO) must be enabled to fit the models into GPU memory, Oobleck and Parcae are not supported because redundancy is eliminated. \sysname (no standby) demonstrates 2.1× and 1.6× shorter downtime compared to Megatron-LM for the GPT-39.1B and GPT-20B models, respectively. This overlapping advantage becomes significant as model sizes grow.
In the case where a standby machine is available, \sysname achieves performance nearly identical to live migration, with less than 10 seconds of downtime across all models.

\mypar{Fast Migration on Expected Events.} Migration occurs when an expected event takes place. We present the migration downtime comparison in Figure~\ref{fig:overall_performance}.
In addition to the \textit{Megatron-LM per-iteration} checkpoint baseline, which assumes that checkpoint saving is cost-free and can always be overlapped, we include two additional baselines for a comprehensive comparison of expected-event migrations.
\textit{Megatron-LM Save-and-Restart} is a more practical baseline. Before shutdown, training stops and waits for the checkpoint to be saved. 
\textit{Naive live migration} directly transfers states from the leaver to the joiner without using the initialization warm-up design~(\secref{sec:warmup}) and the two-phase CCL setup~(\secref{sec:nccl}), and without requiring all machines to pull states from remote storage.





\textit{Megatron-LM Save-and-Restart} performs the worst among all baselines due to significant checkpoint-saving overhead, which increases with model size—reaching approximately 400 seconds for the GPT-39.1B model. \textit{Naive live migration} outperforms Megatron-LM Per-iteration by enabling direct transfer of model states from migration leaver machines to migration joiner machines. This optimization achieves up to a 1.745× speedup on the GPT-39.1B model with TP4 and DO (distributed optimizer) enabled.

\sysname delivers the best performance by overlapping CCL setup and warm-up processes with the non-critical path. This design constantly achieves second level downtime and provides at least a 15× speedup compared to the \textit{Megatron-LM Save-and-Restart} baseline across all models. Both \sysname and the naive system maintain zero additional memory overhead during migration.

\begin{figure}[t]
    \centering
    \begin{subfigure}[t]{0.49\linewidth}
        \centering
        \ifdefined\ArxivVersion
            \includegraphics[width=\linewidth]{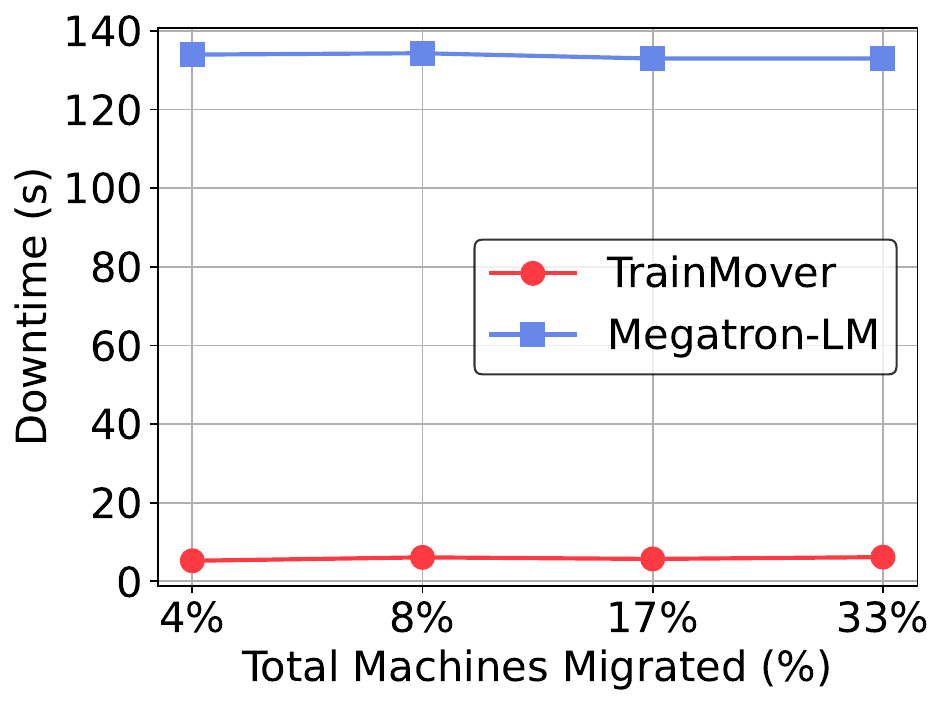}
        \else
            \includegraphics[width=\linewidth]{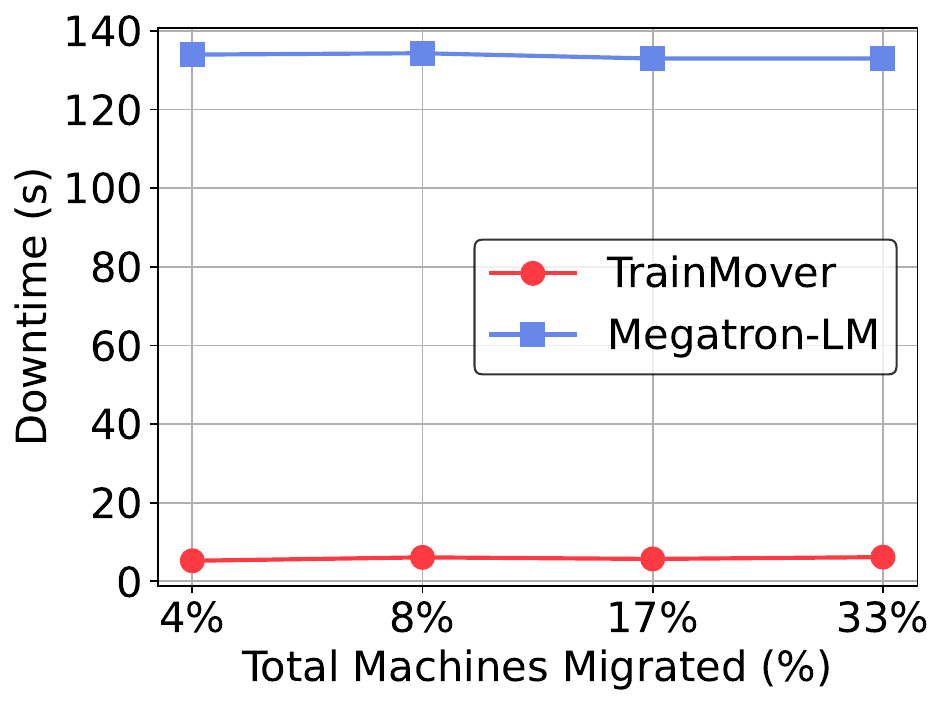}
        \fi
        \caption{GPT-20B}
        \label{fig:migrate_multiple_machine_20b}
    \end{subfigure}
    \hfill
    \begin{subfigure}[t]{0.49\linewidth}
        \centering
        \ifdefined\ArxivVersion
            \includegraphics[width=\linewidth]{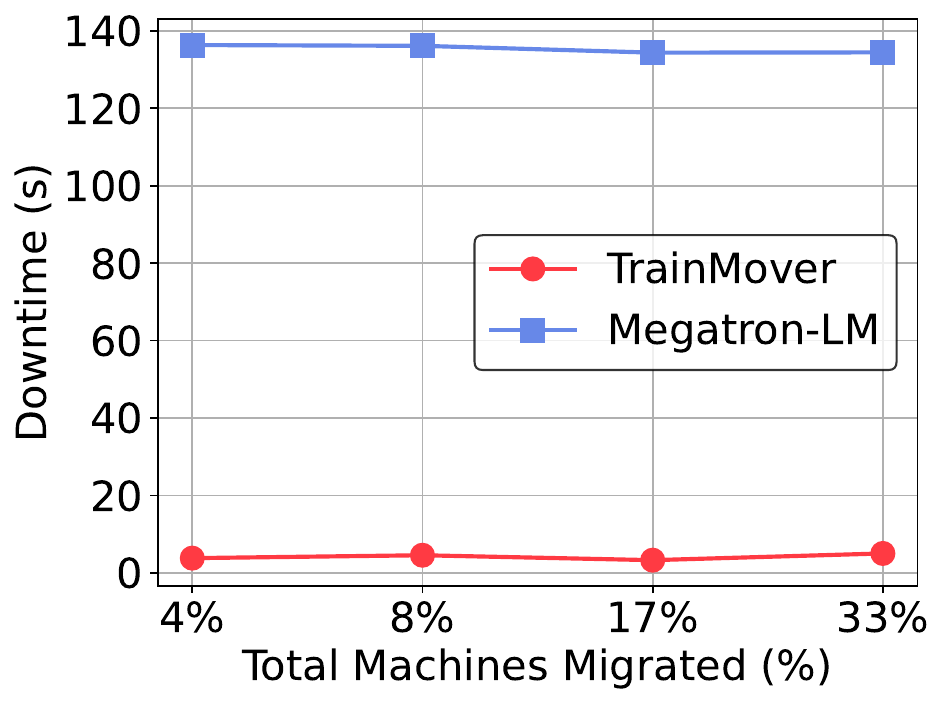}
        \else
            \includegraphics[width=\linewidth]{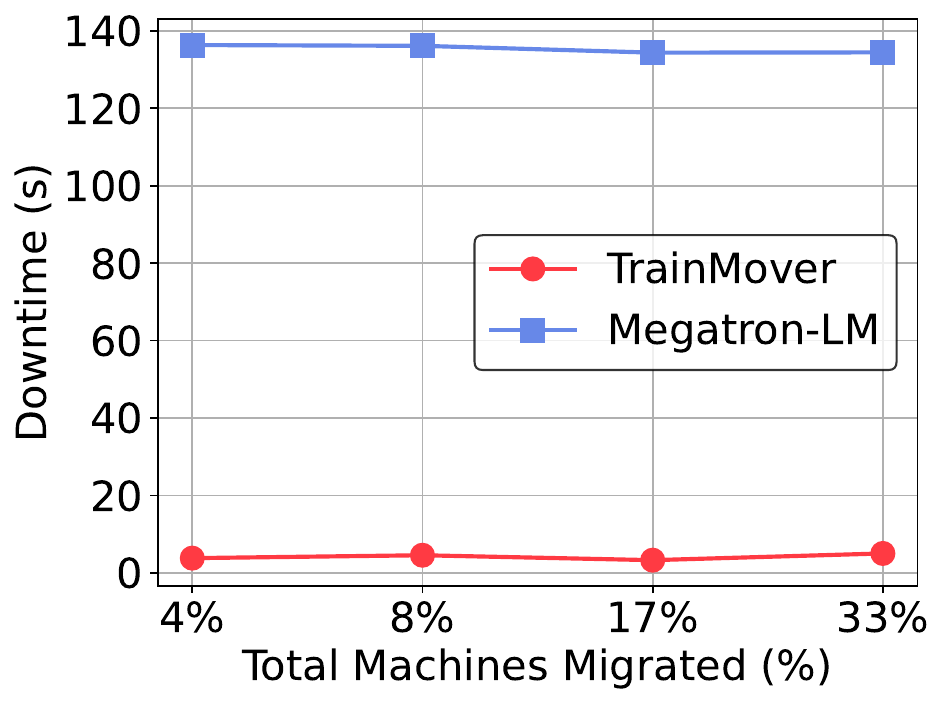}
        \fi
        \caption{GPT-39.1B}
        \label{fig:migrate_multiple_machine_39b}
    \end{subfigure}
    \caption{Downtime varying migration scale}
    \label{fig:migrate_multiple_machine}
\end{figure}

\mypar{Migrating Multiple Machines at Once. }
\label{app:migrate_multiple_at_once}
Migrating multiple machines is essential for rapid large-scale operations like rebalancing and maintenance~\cite{meta_maintainence_blog}. Figures~\ref{fig:migrate_multiple_machine_20b} and~\ref{fig:migrate_multiple_machine_39b} show the performance of GPT-20B and GPT-39.1B during migrations of 4\% to 33\% of the total GPUs.

Both models maintain stable downtime overheads: up to 6.15 seconds across both GPT-20B and GPT-39.1B. This consistency stems from \sysname's design, where each migration leaver and joiner operates in parallel, performing one-to-one data transfers. This approach ensures constant overhead, preserving efficiency regardless of migration scale.

In contrast, checkpointing systems like Megatron-LM are highly inefficient. Even migrating just 4\% of machines forces all machines to reboot and retrieve checkpoints from remote storage, causing 138 seconds overhead.


\subsection{Use Cases}
\label{eval:use_cases}
\label{eval:stragger}
We also present other use cases of \sysname. Note that none of these use cases require hot standby machines.

\begin{figure}[t]

\begin{minipage}[t]{\linewidth}
    \centering
    \includegraphics[width=\linewidth]{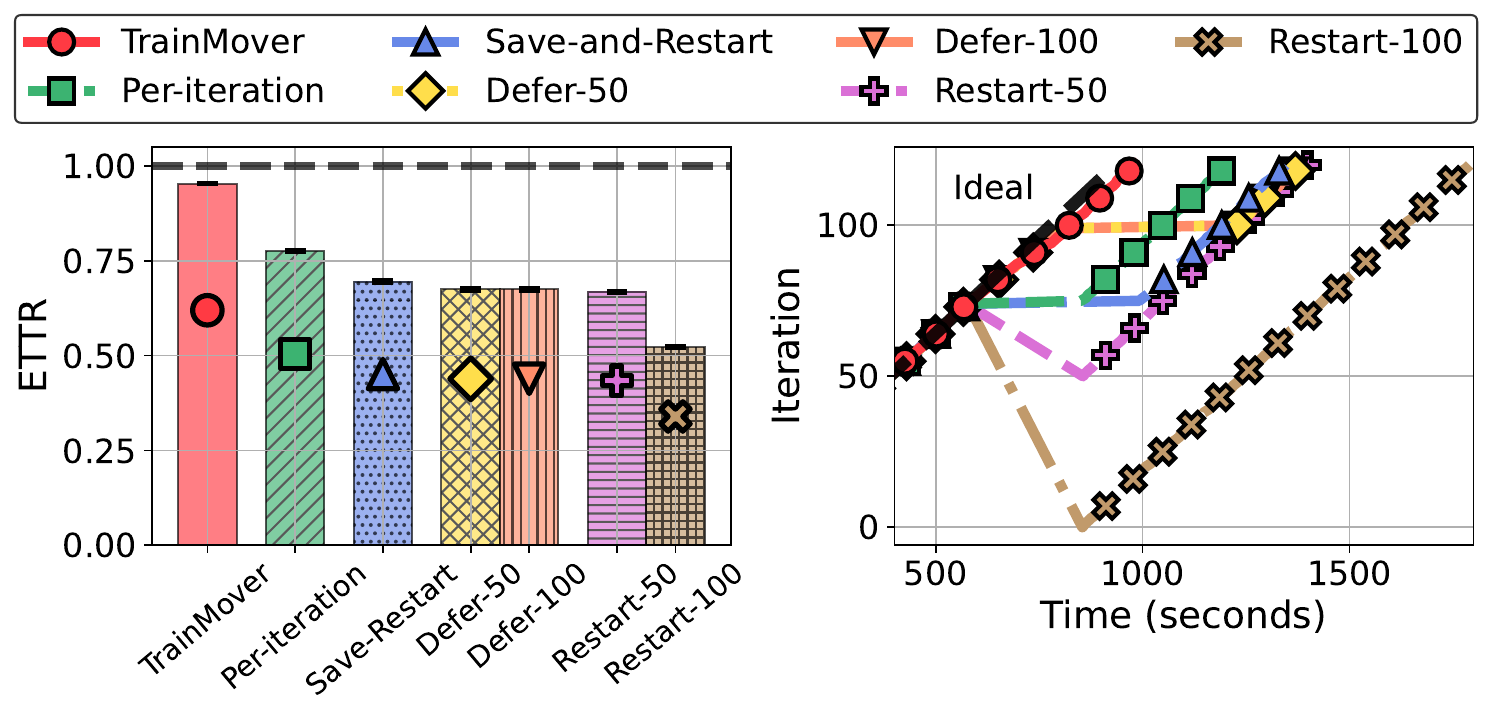}
    \caption{GPT 5.12T MoE model with 20\% slowdown starting at the 75th iteration}
    \label{fig:straggler_time}
\end{minipage}

\begin{minipage}[b]{0.46\linewidth}
    \centering
    \includegraphics[width=\linewidth]{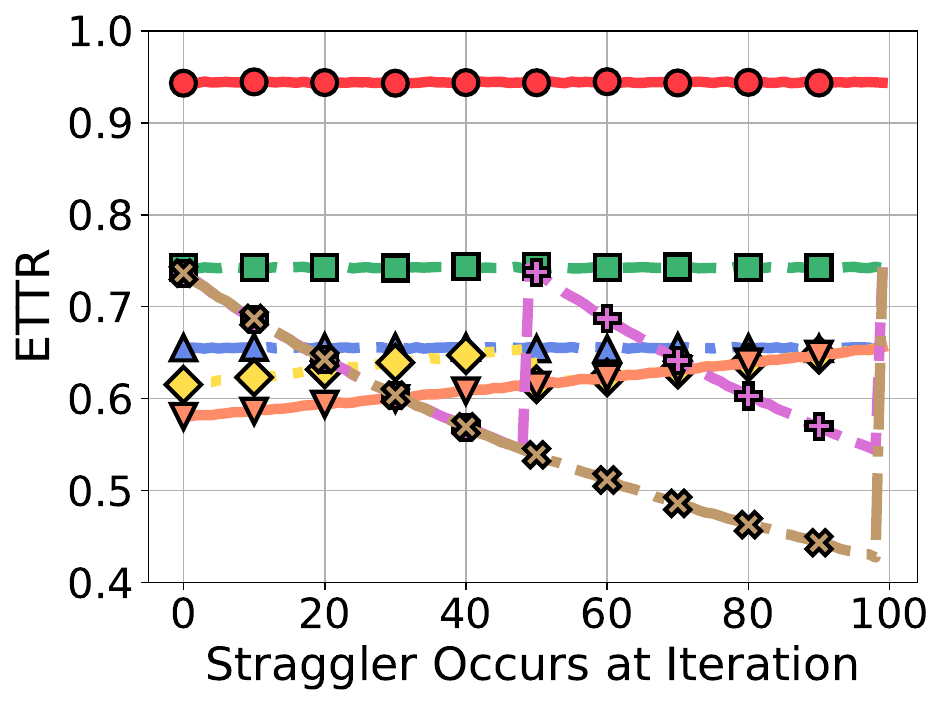}
    \caption{Straggler occurs at different iteration}
    \label{fig:straggler_iteration}
\end{minipage}
\hfill
\begin{minipage}[b]{0.53\linewidth}
    \centering
    \includegraphics[width=\linewidth]{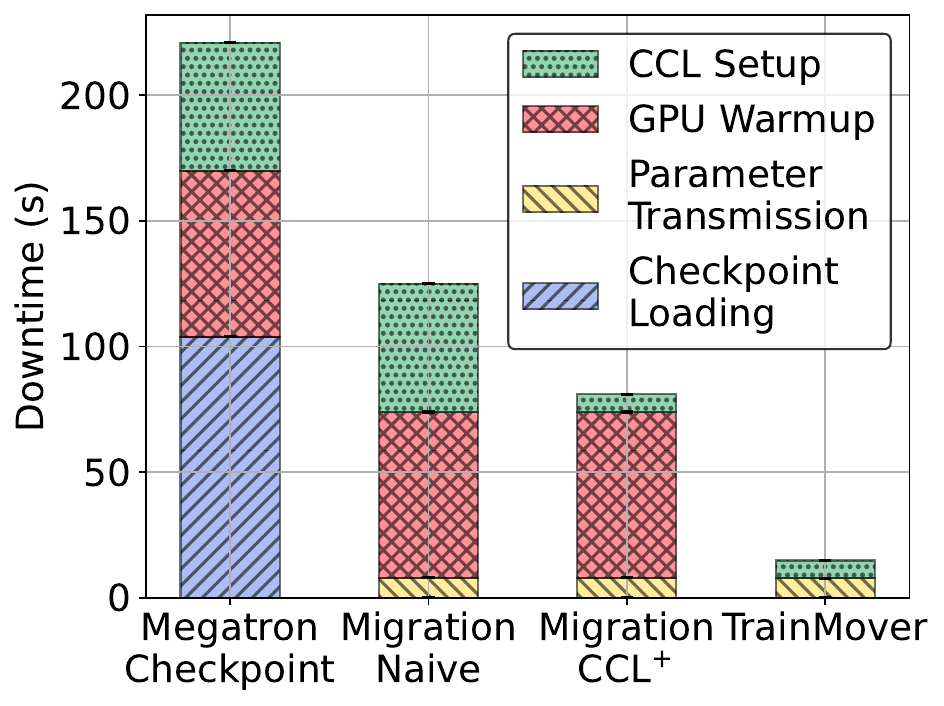}
    \caption{Design Breakdown}
    \label{fig:design_breakdown}
\end{minipage}

\end{figure}

\mypar{Handling Stragglers.} We evaluate the use case of handling stragglers, as shown in Figure~\ref{fig:straggler_time}. In this experiment, we inject a straggler by slowing down one GPU by 20\%~\cite{falcon_ali} at the 75th iteration of a 100-iteration, 1024-GPU job. Figure~\ref{fig:straggler_time} reports the training efficiency and visualizes the training process. While this setup represents a specific case where the straggler appears at iteration 75, Figure~\ref{fig:straggler_iteration} extends the analysis to all possible iterations (1st–100th), showing the full spectrum of when stragglers occur.

Alongside the previously evaluated \textit{Per-iteration} and \textit{Save-and-Restart} approaches, we include two additional strategies for broader comparison: deferred checkpointing and restart with progress loss.
\textit{Save-and-Restart} saves a checkpoint immediately and restarts training when a straggler occurs.
\textit{Defer-50/100} continues training and defers restarting until the next scheduled checkpoint (e.g., every 50 or 100 iterations). 
\textit{Restart-50/100} restarts training immediately, losing progress based on the most recent checkpoint at iteration 50 or 100.

As shown in Figure~\ref{fig:straggler_time}, \sysname loses only 4.7\% of training efficiency—the lowest among all approaches. During the slowdown period, \sysname continues training while performing migration in parallel, minimizing downtime when switching from the straggler machine to a healthy one.

Defer-type baselines suffer reduced performance until the next checkpoint, restarting only afterward and thus causing delays. Restart-type baselines avoid checkpoint-saving overhead but lose progress proportional to the checkpoint interval, leading to up to 43\% lower efficiency than \sysname.

Figure~\ref{fig:straggler_iteration} further illustrates the full performance spectrum across all possible straggler injection points. \sysname consistently outperforms all baselines, maintaining high efficiency without relying on specific checkpointing frequencies. Remarkably, it even surpasses the ideal per-iteration checkpointing system, improving training efficiency by 22.8\% in all cases. Restart-type methods peak immediately after a checkpoint when progress loss is minimal, but their performance deteriorates as the distance from the last checkpoint increases.

\begin{figure}[t]
\centering
\scriptsize

\setlength{\tabcolsep}{4pt}
\renewcommand{\arraystretch}{0.9}

\textbf{Scaling up to 1024 GPUs.}
\vspace{3pt}

\begin{tabular}{|l|c|c|c|c|}
\hline
\textbf{System $\backslash$ Scale} & 128 & 256 & 512 & 1024 \\
\Xhline{3\arrayrulewidth}
\sysname & 0.98 & 0.98 & 0.98 & 0.97 \\
\hline
Megatron-LM & 0.50 & 0.49 & 0.44 & 0.42 \\
\hline
\end{tabular}

\vspace{3pt}


\textbf{Breakdown at 32 GPUs.}
\vspace{3pt}

\setlength{\tabcolsep}{1pt}
\renewcommand{\arraystretch}{0.85}

\resizebox{\columnwidth}{!}{
\begin{tabular}{|l|l|c|c|c|c||l|c|c|c|c|} 
\hline 
& \multicolumn{5}{c||}{Enable Distributed Optimizer} & \multicolumn{5}{c|}{Disable Distributed Optimizer} \\ 
\hline\hline 
& & Med. & 2.7B & 20B & 39.1B & & Med. & 2.7B & 20B & 39.1B \\ 
\hline 
\multirow{3}{*}{\sysname} 
& TP1 & 0.99 & 0.99 & 0.99 & 0.99 & TP1 & 0.99 & 0.99 & OOM & OOM \\ 
& TP4 & 0.99 & 0.99 & 0.99 & 0.99 & TP4 & 0.99 & 0.99 & 0.99 & OOM \\ 
& TP8 & 0.99 & 0.99 & 0.99 & 0.99 & TP8 & 0.99 & 0.99 & 0.99 & OOM \\ 
\hline 
\multirow{3}{*}{Megatron-LM} 
& TP1 & 0.79 & 0.82 & 0.77 & 0.68 & TP1 & 0.83 & 0.81 & OOM & OOM \\ 
& TP4 & 0.86 & 0.86 & 0.85 & 0.77 & TP4 & 0.84 & 0.85 & 0.92 & OOM \\ 
& TP8 & 0.88 & 0.88 & 0.85 & 0.81 & TP8 & 0.87 & 0.85 & 0.96 & OOM \\ 
\hline 
\multirow{3}{*}{Oobleck\cite{oobleck}} 
& \multicolumn{5}{c||}{} & TP1 & 0.70 & 0.61 & OOM & OOM \\ 
& \multicolumn{5}{c||}{\textbf{Unsupported.}} & TP4 & 0.62 & 0.49 & 0.03 & OOM \\ 
& \multicolumn{5}{c||}{} & TP8 & 0.57 & 0.49 & 0.00 & OOM \\ 
\hline 
\multirow{3}{*}{Parcae\cite{parcae}} 
& \multicolumn{5}{c||}{} & TP1 & 0.85 & 0.82 & OOM & OOM \\ 
& \multicolumn{5}{c||}{\centering\textbf{Unsupported.}} & \multicolumn{5}{c|}{\multirow{2}{*}{\centering\textbf{Unsupported.}}} \\ 
& \multicolumn{5}{c||}{} & \multicolumn{5}{c|}{} \\
\hline 
\end{tabular}
}

\vspace{4pt}

\caption{ETTR across large cluster scales (128–1024 GPUs, top) 
and detailed performance breakdown at scale 32 (bottom).}
\label{tab:combined-performance}
\label{tab:performance}
\end{figure}


\mypar{Rebalancing / Maintenance. }
\label{eval:rebalance}
\sysname handles machine changes swiftly, making it suitable for rebalancing or maintenance that require frequent GPU or network updates. 
The upper part of Table~\ref{tab:performance} compares the training efficiency of \sysname and Megatron-LM per-iteration checkpointing across scales from 128 to 1024 GPUs. \sysname consistently maintains an ETTR of above 0.97 across all scales, showing that rebalancing remains effective even as cluster size grows. In contrast, Megatron-LM’s ETTR stays low under high rebalancing frequency, even dropping to 0.42 at 1024 GPUs due to the increasing cost of repeated full-job restarts.

We also include comparisons with Parcae and Oobleck at the 32-GPU scale covering various model sizes and training parameters in the lower part of the table.
During machine changes, online reconfiguration systems like Oobleck sometimes perform worse than the checkpointing system Megatron-LM. This is because reconfiguration involves starting new machines, tearing down and reestablishing NCCL groups, and performing warmup procedures, which saves little time compared to checkpoint-based systems and performs even worse if the system is not well-engineered. For instance, with the Oobleck 20B TP8 model, reconfiguration delays result in negligible progress with rebalancing every 10 minutes. Additionally, Oobleck and Parcae lack support for distributed optimizers, and Parcae also does not support tensor parallelism, limiting their scalability for larger models. 

\sysname delivers the best performance among all approaches, maintaining a training efficiency of 0.99 across all models with and without distributed optimizer. In comparison, the efficiency drops to 0.68 when using Megatron-LM per-iteration checkpointing. This table highlights that even in high-frequency (10-minute) rebalancing scenarios, \sysname loses no more than 0.1\% of performance, making it a robust and efficient solution.

\subsection{\sysname Key Designs}
\label{eval:breakdown}
\label{sec:design_breakdown}
\label{eval:design_breakdown}




\mypar{Design Breakdown.}    
We break down our designs and evaluate their incremental impact on system performance in Figure~\ref{fig:design_breakdown} with the GPT-5.12T MoE model. In this setup, Megatron-LM's total downtime is approximately 221 seconds, with checkpoint loading, GPU warmup, and CCL setup times accounting for 47\%, 30\%, and 23\% of the total, respectively.

\textit{Migration Naive}, represents a naive migration system that excludes the two-layer CCL designs and lazy initialization designs. Unlike Megatron-LM, which loads checkpoints from remote storage, it retrieves parameters directly from the leaver to the joiner. This approach eliminates checkpoint loading time, reducing the total downtime by approximately 1.8×. The benefit becomes more pronounced as model sizes increase.

\textit{Migration CCL$^{+}$} builds on the naive migration baseline by introducing the CCL two-layer designs. This design reduces the critical path, leaving only the second stage of CCL initialization in the critical path. Consequently, CCL time decreases from 51 seconds to 7 seconds. \revision{This means Phase~1 absorbs $\sim$44 seconds off the critical path, leaving only Phase~2's 7 seconds as downtime---an 86\% reduction in CCL-related recovery time.} 

The full system, \textit{\sysname}, further integrates the self-detached warmup design, effectively eliminating GPU warmup time. This enables training to resume immediately after migration, reducing downtime to 16 seconds.

\begin{figure}[!tb]
    \centering

    \begin{minipage}{0.48\linewidth}
        \centering
        \ifdefined\ArxivVersion
            \includegraphics[width=\linewidth]{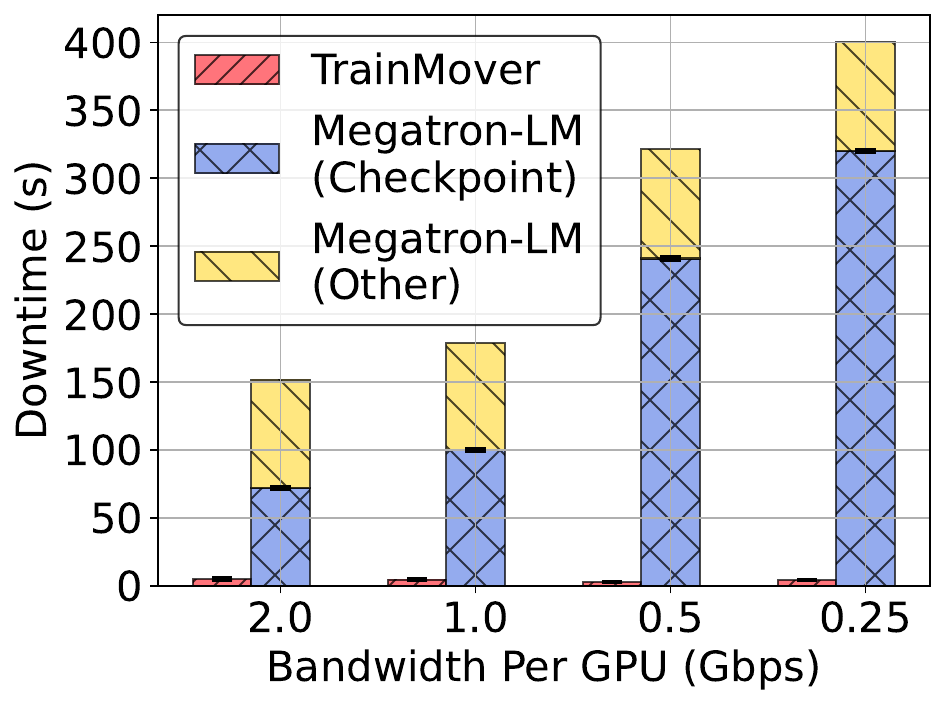}
        \else
            \includegraphics[width=\linewidth]{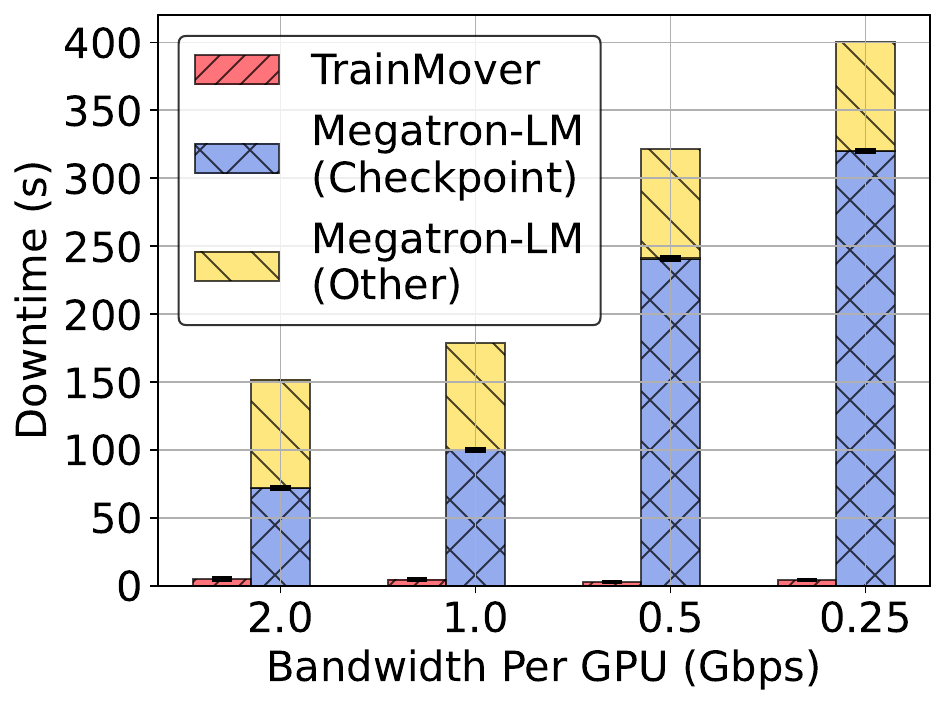}
        \fi
        \caption{Downtime with different bandwidth (GPT-20B)}
        \label{fig:bandwidth_per_gpu_20b}
    \end{minipage}
    \hfill
    \begin{minipage}{0.48\linewidth}
        \centering
        \ifdefined\ArxivVersion
            \includegraphics[width=\linewidth]{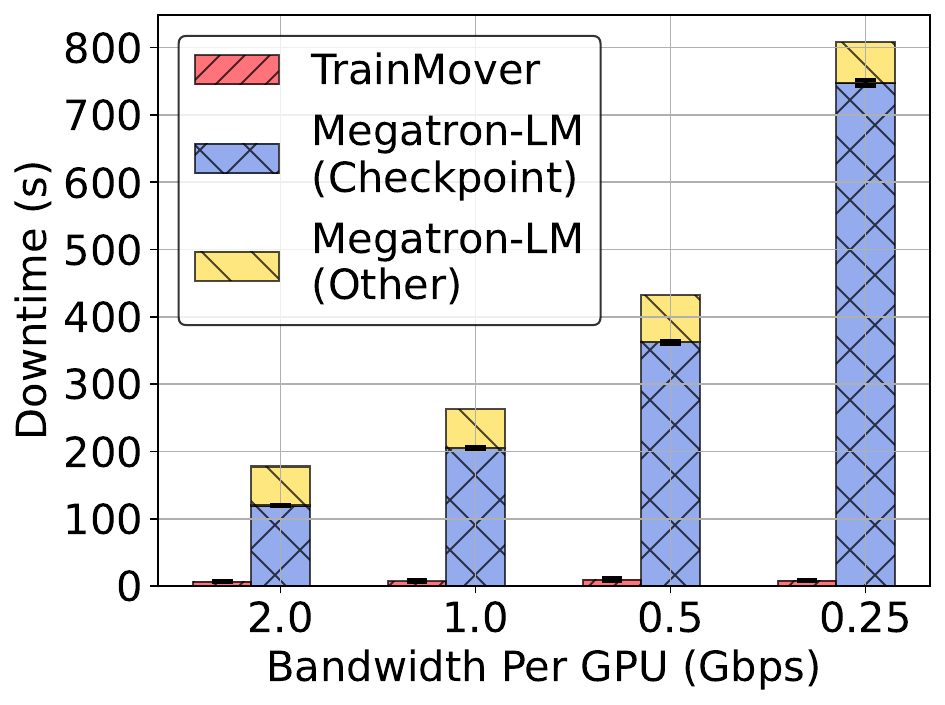}
        \else
            \includegraphics[width=\linewidth]{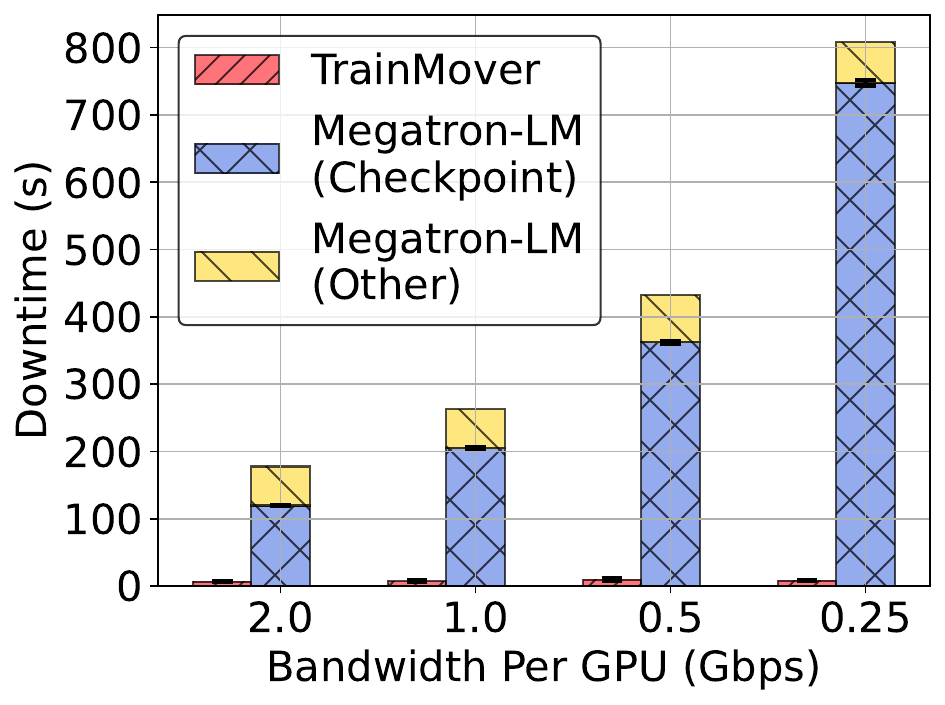}
        \fi
        \caption{Downtime with different bandwidth per GPU (GPT-39.1B)}
        \label{fig:bandwidth_per_gpu_39b}
    \end{minipage}

\end{figure}

\mypar{\sysname eliminates Network Bottleneck. }
\label{sec:elimiate_network}
\sysname significantly alleviates network bandwidth bottlenecks during machine changes. LLaMA reports~\cite{llama3} their storage system bandwidth ranging from 2 TB/s to 7 TB/s while serving approximately 7,500 machines, translating to an effective bandwidth of 0.267 GB/s to 0.933 GB/s per machine. In this experiment, we evaluate checkpointing overhead under bandwidth ranging from 0.25 GB/s to 2 GB/s per GPU, encompassing the bandwidth range reported by LLaMA.

In both GPT-20B (Figure~\ref{fig:bandwidth_per_gpu_20b}) and GPT-39.1B (Figure~\ref{fig:bandwidth_per_gpu_39b}) setups, \sysname maintains stable overheads of 6-9 seconds, due to parallel one-to-one state transfers between migration leaver and joiner. By contrast, Megatron-LM, which requires all GPUs to pull checkpoints from remote storage, incurs increasing overhead as bandwidth decreases and model size grows. Specifically, checkpoint loading overheads reach 320 seconds for GPT-20B and 750 seconds for GPT-39.1B at 0.25 GB/s per GPU.

\begin{figure}[t]
    \centering
    \includegraphics[width=0.9\linewidth]{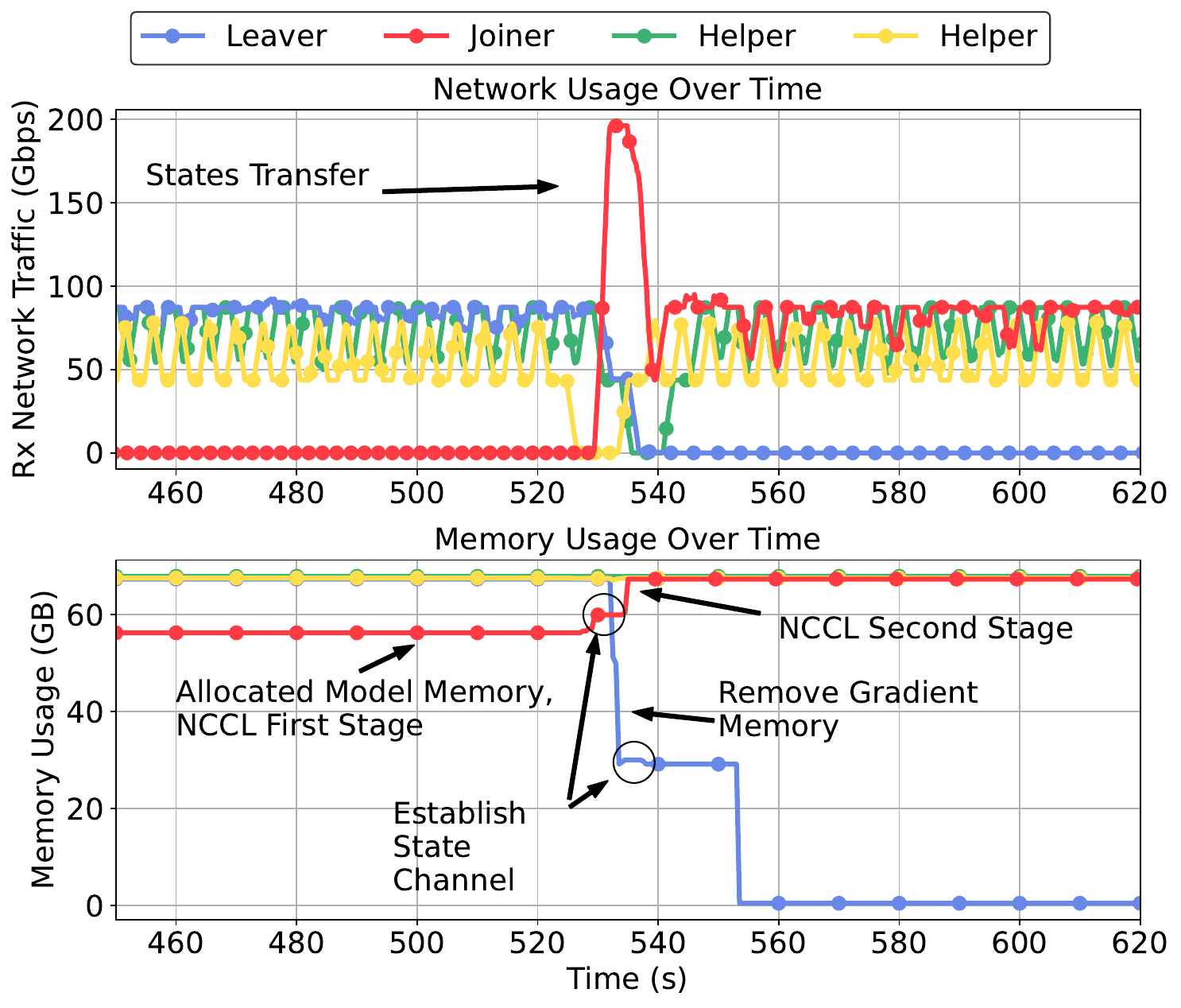}
    \caption{Network traffic and memory usage timeline during Migration}
    \label{fig:workflow_breakdown}
    \vspace{3mm}
\end{figure}

\mypar{Zero Memory Overhead Workflow. }
\label{sec:zero_mem}
Figure~\ref{fig:workflow_breakdown} illustrates the network (top) and memory (bottom) usage during a migration in \sysname. A key step in the process is state transfer. As shown in the network breakdown, the leaver sends traffic to the joiner around the 530th second, resulting in high Rx bandwidth utilization on the joiner. However, model transfer is not without costs. Each leaver-joiner pair requires a new NCCL group for state delivery, which consumes GPU memory. To achieve zero memory overhead, \sysname implements specific optimizations during state transmission:

The migration leaver does not need to continue training after the migration is complete, it can free up and repurpose the pre-allocated GPU gradient buffer for use as the NCCL transmission channel at around the 530th second.
On the migration joiner, during the first stage of NCCL initialization, inter-machine connections are not fully established, creating temporary memory availability at the 530th second. This allows the state transfer channel to operate without exceeding memory limits. Once the transfer is complete, the channel is immediately destroyed, freeing memory before the second stage of NCCL initialization. This process ensures zero memory overhead during migration.

A detailed analysis of NCCL design choices is provided in Appendix~\ref{eval:ncclbreakdown}.




\section{Related Work and Discussion}
\label{sec:relatedwork}
\label{sec:discussion}
\mypar{Economics of Standby.}
At massive scales, simply adding more GPUs yields marginal returns, whereas leveraging a standby machine can dramatically reduce downtime and provide far more cost-effective gains. For instance, at the 32K-GPU scale in Figure~\ref{fig:8192_gpu}, using a single standby machine to shorten recovery from 4.45 minutes to 20 seconds delivers a benefit equivalent to adding roughly 2,400 GPUs (300 machines!) under ideal linear scaling. In other words, the acceleration provided by one or few standby machines' fast recovery is comparable to provisioning thousands of additional GPUs—making standby resources vastly more economical.

\mypar{CCL Reconfiguration and Generality.}
MCCS~\cite{mccs} enables runtime changes to CCL topologies but lacks dynamic membership. HotSpa~\cite{hotspa} hides reconfiguration overhead by reusing CCL groups, at the cost of added GPU-memory overhead. NCCLX~\cite{si2025collectivecommunication100kgpus} speeds up CCL initialization by reducing bootstrap costs and adaptively configuring channels. In contrast, \sysname supports dynamic CCL membership without GPU memory overhead and moves all non-critical initialization off the critical setup path.
This two-stage setup is general and applies to systems that frequently reconfigure communication patterns, including reconfiguration frameworks and inference autoschedulers~\cite{recycle,hotspa,parcae,bamboo,oobleck,alpaserve,adapcc,mccs,photonic_rachee, si2025collectivecommunication100kgpus}.

\mypar{Scalability.}
\sysname scales to very large clusters. Transitional downtime comes from only two steps: inter-connection establishment and state transfer. State transfer is one-to-one—either from the leaver to the joiner during expected failures or from the leaver pulling from a checkpoint during unexpected failures—so its cost is effectively constant and the transfer size is bounded by the GPU size. Inter-connection establishment affects only the leaver’s neighbors, which update their channels to the joiner. Neither step requires global coordination, enabling \sysname to scale efficiently even in large production clusters.

\mypar{GPU-granularity Migration.} 
\sysname's implementation supports GPU-granularity migration.
The training framework can still benefit from minimal downtime and zero memory overhead.
However, we need to forgo the intra-machine optimizations introduced in \secref{subsec:sandbox_optimization} and pay extra storage overhead.

\mypar{Heterogeneity and Warm-up Coverage.}
\revision{
Heterogeneous or multimodal models may break symmetry assumptions or introduce non-static training paths. In practice, even if certain layer-specific or data-dependent execution paths are not fully covered, their impact is limited because the vast majority of initialization steps (e.g., data-loader setup, CUDA context creation, memory allocation, and runtime setup) remain shared across execution paths and are already covered by the warm-up procedure. These shared components dominate initialization time and are largely independent of specific layer behaviors. 
Extending \sysname to fully support such dynamism remains future work.
}
 
\mypar{Snapshot-based Warm Start.}
\revision{
Recent advances such as CUDA checkpoint/restore (e.g., cuda-checkpoint~\cite{cuda_checkpoint} combined with CRIU~\cite{criu}) provide an emerging mechanism for capturing and restoring GPU runtime state, offering a potential solution for warm-starting initialization—analogous to cold-start mitigation in serverless computing~\cite{catalyzer, faasnap}. These techniques enable preserving device memory and CUDA contexts, and are effective when the target role is known.
\sysname targets a different goal: supporting a \emph{general standby} that can recover \emph{any} role without prior knowledge. In large-scale training, different ranks (e.g., pipeline stages) may follow different execution paths. Supporting this with snapshots would require maintaining multiple role-specific snapshots or adapting them at runtime, increasing system complexity. In contrast, \sysname uses record--replay to enable role-agnostic preparation without pre-defined artifacts.
}

\vspace{-1mm}
\section{Conclusion}
\label{sec:conclusion}
We presented TrainMover, a resilient runtime that handles both expected and unexpected interruptions with minimal downtime and zero memory overhead. TrainMover leverages two-phase communication setup, communication-free sandboxed warmup, and general standby to enable rapid machine replacement. Our evaluation shows that TrainMover achieves 20-second downtime at the 1K-GPU scale and wastes 55\% fewer GPU-hours at 64K GPUs, enabling interruption-tolerant large-scale training.

\section*{Acknowledgment}

We would like to thank Ion Stoica for insightful discussions on the general standby design.

\bibliographystyle{plain_limit}
\bibliography{reference}

@inproceedings{catalyzer,
author = {Du, Dong and Yu, Tianyi and Xia, Yubin and Zang, Binyu and Yan, Guanglu and Qin, Chenggang and Wu, Qixuan and Chen, Haibo},
title = {Catalyzer: Sub-millisecond Startup for Serverless Computing with Initialization-less Booting},
year = {2020},
isbn = {9781450371025},
publisher = {Association for Computing Machinery},
address = {New York, NY, USA},
url = {https://doi.org/10.1145/3373376.3378512},
doi = {10.1145/3373376.3378512},
booktitle = {Proceedings of the Twenty-Fifth International Conference on Architectural Support for Programming Languages and Operating Systems},
pages = {467–481},
numpages = {15},
location = {Lausanne, Switzerland},
series = {ASPLOS '20}
}

@inproceedings{faasnap,
author = {Ao, Lixiang and Porter, George and Voelker, Geoffrey M.},
title = {FaaSnap: FaaS made fast using snapshot-based VMs},
year = {2022},
isbn = {9781450391627},
publisher = {Association for Computing Machinery},
address = {New York, NY, USA},
url = {https://doi.org/10.1145/3492321.3524270},
doi = {10.1145/3492321.3524270},
booktitle = {Proceedings of the Seventeenth European Conference on Computer Systems},
pages = {730–746},
numpages = {17},
location = {Rennes, France},
series = {EuroSys '22}
}

@misc{nemo,
      title={NeMo: a toolkit for building AI applications using Neural Modules}, 
      author={Oleksii Kuchaiev and Jason Li and Huyen Nguyen and Oleksii Hrinchuk and Ryan Leary and Boris Ginsburg and Samuel Kriman and Stanislav Beliaev and Vitaly Lavrukhin and Jack Cook and Patrice Castonguay and Mariya Popova and Jocelyn Huang and Jonathan M. Cohen},
      year={2019},
      eprint={1909.09577},
      archivePrefix={arXiv},
      primaryClass={cs.LG},
      url={https://arxiv.org/abs/1909.09577}, 
}

@misc{meta_maintainence_blog,
    title = {{Maintaining large-scale AI capacity at Meta.}},
    howpublished = {\url{https://engineering.fb.com/2024/06/12/production-engineering/maintaining-large-scale-ai-capacity-meta/}}
    ,year = {2024}
}

@misc{xai_blog_2024,
  author       = {{xAI}},
  title        = {xAI's Colossus supercomputer cluster},
  howpublished = {\url{https://x.ai/colossus/}},
  year         = {2024},
  note         = {Accessed 2024}
}

@article{aegis,
  title={Evolution of Aegis: Fault Diagnosis for AI Model Training Cloud Service in Production (Experience Track)},
  author={Dong, Jianbo and Qian, Kun and Zhang, Pengcheng and Zheng, Zhilong and Chen, Liang and Feng, Fei and Zhu, Yikai and Lu, Gang and Ren, Zhihui and Li, Xue and others}
}

@misc{falcon_ali,
      title={FALCON: Pinpointing and Mitigating Stragglers for Large-Scale Hybrid-Parallel Training}, 
      author={Tianyuan Wu and Wei Wang and Yinghao Yu and Siran Yang and Wenchao Wu and Qinkai Duan and Guodong Yang and Jiamang Wang and Lin Qu and Liping Zhang},
      year={2024},
      eprint={2410.12588},
      archivePrefix={arXiv},
      primaryClass={cs.DC},
      url={https://arxiv.org/abs/2410.12588}, 
}

@inproceedings{zero1,
author = {Rajbhandari, Samyam and Rasley, Jeff and Ruwase, Olatunji and He, Yuxiong},
title = {ZeRO: memory optimizations toward training trillion parameter models},
year = {2020},
isbn = {9781728199986},
publisher = {IEEE Press},
booktitle = {Proceedings of the International Conference for High Performance Computing, Networking, Storage and Analysis},
articleno = {20},
numpages = {16},
location = {Atlanta, Georgia},
series = {SC '20}
}

@misc{391b,
      title={Automated Tensor Model Parallelism with Overlapped Communication for Efficient Foundation Model Training}, 
      author={Shengwei Li and Zhiquan Lai and Yanqi Hao and Weijie Liu and Keshi Ge and Xiaoge Deng and Dongsheng Li and Kai Lu},
      year={2023},
      eprint={2305.16121},
      archivePrefix={arXiv},
      primaryClass={cs.DC},
      url={https://arxiv.org/abs/2305.16121}, 
}

@misc{gpt_models,
      title={Language Models are Few-Shot Learners}, 
      author={Tom B. Brown and Benjamin Mann and Nick Ryder and Melanie Subbiah and Jared Kaplan and Prafulla Dhariwal and Arvind Neelakantan and Pranav Shyam and Girish Sastry and Amanda Askell and Sandhini Agarwal and Ariel Herbert-Voss and Gretchen Krueger and Tom Henighan and Rewon Child and Aditya Ramesh and Daniel M. Ziegler and Jeffrey Wu and Clemens Winter and Christopher Hesse and Mark Chen and Eric Sigler and Mateusz Litwin and Scott Gray and Benjamin Chess and Jack Clark and Christopher Berner and Sam McCandlish and Alec Radford and Ilya Sutskever and Dario Amodei},
      year={2020},
      eprint={2005.14165},
      archivePrefix={arXiv},
      primaryClass={cs.CL},
      url={https://arxiv.org/abs/2005.14165}, 
}

@misc{wiki_dataset,
      title={Pointer Sentinel Mixture Models}, 
      author={Stephen Merity and Caiming Xiong and James Bradbury and Richard Socher},
      year={2016},
      eprint={1609.07843},
      archivePrefix={arXiv},
      primaryClass={cs.CL},
      url={https://arxiv.org/abs/1609.07843}, 
}

@inproceedings{alihpn,
author = {Qian, Kun and Xi, Yongqing and Cao, Jiamin and Gao, Jiaqi and Xu, Yichi and Guan, Yu and Fu, Binzhang and Shi, Xuemei and Zhu, Fangbo and Miao, Rui and Wang, Chao and Wang, Peng and Zhang, Pengcheng and Zeng, Xianlong and Ruan, Eddie and Yao, Zhiping and Zhai, Ennan and Cai, Dennis},
title = {Alibaba HPN: A Data Center Network for Large Language Model Training},
year = {2024},
isbn = {9798400706141},
publisher = {Association for Computing Machinery},
address = {New York, NY, USA},
url = {https://doi-org.ezp-prod1.hul.harvard.edu/10.1145/3651890.3672265},
doi = {10.1145/3651890.3672265},
booktitle = {Proceedings of the ACM SIGCOMM 2024 Conference},
pages = {691–706},
numpages = {16},
location = {Sydney, NSW, Australia},
series = {ACM SIGCOMM '24}
}

@misc{meta2_realibility,
      title={Revisiting Reliability in Large-Scale Machine Learning Research Clusters}, 
      author={Apostolos Kokolis and Michael Kuchnik and John Hoffman and Adithya Kumar and Parth Malani and Faye Ma and Zachary DeVito and Shubho Sengupta and Kalyan Saladi and Carole-Jean Wu},
      year={2024},
      eprint={2410.21680},
      archivePrefix={arXiv},
      primaryClass={cs.DC},
      url={https://arxiv.org/abs/2410.21680}, 
}

@inproceedings{gemini,
author = {Wang, Zhuang and Jia, Zhen and Zheng, Shuai and Zhang, Zhen and Fu, Xinwei and Ng, T. S. Eugene and Wang, Yida},
title = {GEMINI: Fast Failure Recovery in Distributed Training with In-Memory Checkpoints},
year = {2023},
isbn = {9798400702297},
publisher = {Association for Computing Machinery},
address = {New York, NY, USA},
url = {https://doi.org/10.1145/3600006.3613145},
doi = {10.1145/3600006.3613145},
booktitle = {Proceedings of the 29th Symposium on Operating Systems Principles},
pages = {364–381},
numpages = {18},
location = {Koblenz, Germany},
series = {SOSP '23}
}

@inproceedings{just-in-time-checkpoint,
author = {Gupta, Tanmaey and Krishnan, Sanjeev and Kumar, Rituraj and Vijeev, Abhishek and Gulavani, Bhargav and Kwatra, Nipun and Ramjee, Ramachandran and Sivathanu, Muthian},
title = {Just-In-Time Checkpointing: Low Cost Error Recovery from Deep Learning Training Failures},
year = {2024},
isbn = {9798400704376},
publisher = {Association for Computing Machinery},
address = {New York, NY, USA},
url = {https://doi-org.ezp-prod1.hul.harvard.edu/10.1145/3627703.3650085},
doi = {10.1145/3627703.3650085},
booktitle = {Proceedings of the Nineteenth European Conference on Computer Systems},
pages = {1110–1125},
numpages = {16},
location = {Athens, Greece},
series = {EuroSys '24}
}

@inproceedings {bamboo,
author = {John Thorpe and Pengzhan Zhao and Jonathan Eyolfson and Yifan Qiao and Zhihao Jia and Minjia Zhang and Ravi Netravali and Guoqing Harry Xu},
title = {Bamboo: Making Preemptible Instances Resilient for Affordable Training of Large {DNNs}},
booktitle = {20th USENIX Symposium on Networked Systems Design and Implementation (NSDI 23)},
year = {2023},
isbn = {978-1-939133-33-5},
address = {Boston, MA},
pages = {497--513},
url = {https://www.usenix.org/conference/nsdi23/presentation/thorpe},
publisher = {USENIX Association},
month = apr
}

@inproceedings{deepspeed,
author = {Rasley, Jeff and Rajbhandari, Samyam and Ruwase, Olatunji and He, Yuxiong},
title = {DeepSpeed: System Optimizations Enable Training Deep Learning Models with Over 100 Billion Parameters},
year = {2020},
isbn = {9781450379984},
publisher = {Association for Computing Machinery},
address = {New York, NY, USA},
url = {https://doi-org.ezp-prod1.hul.harvard.edu/10.1145/3394486.3406703},
doi = {10.1145/3394486.3406703},
booktitle = {Proceedings of the 26th ACM SIGKDD International Conference on Knowledge Discovery \& Data Mining},
pages = {3505–3506},
numpages = {2},
location = {Virtual Event, CA, USA},
series = {KDD '20}
}

@inproceedings{colossalai,
author = {Li, Shenggui and Liu, Hongxin and Bian, Zhengda and Fang, Jiarui and Huang, Haichen and Liu, Yuliang and Wang, Boxiang and You, Yang},
title = {Colossal-AI: A Unified Deep Learning System For Large-Scale Parallel Training},
year = {2023},
isbn = {9798400708435},
publisher = {Association for Computing Machinery},
address = {New York, NY, USA},
url = {https://doi.org/10.1145/3605573.3605613},
doi = {10.1145/3605573.3605613},
booktitle = {Proceedings of the 52nd International Conference on Parallel Processing},
pages = {766–775},
numpages = {10},
location = {Salt Lake City, UT, USA},
series = {ICPP '23}
}

@inproceedings {parcae,
author = {Jiangfei Duan and Ziang Song and Xupeng Miao and Xiaoli Xi and Dahua Lin and Harry Xu and Minjia Zhang and Zhihao Jia},
title = {Parcae: Proactive, {Liveput-Optimized} {DNN} Training on Preemptible Instances},
booktitle = {21st USENIX Symposium on Networked Systems Design and Implementation (NSDI 24)},
year = {2024},
isbn = {978-1-939133-39-7},
address = {Santa Clara, CA},
pages = {1121--1139},
url = {https://www.usenix.org/conference/nsdi24/presentation/duan},
publisher = {USENIX Association},
month = apr
}

@inproceedings{recycle,
author = {Gandhi, Swapnil and Zhao, Mark and Skiadopoulos, Athinagoras and Kozyrakis, Christos},
title = {ReCycle: Resilient Training of Large DNNs using Pipeline Adaptation},
year = {2024},
isbn = {9798400712517},
publisher = {Association for Computing Machinery},
address = {New York, NY, USA},
url = {https://doi-org.ezp-prod1.hul.harvard.edu/10.1145/3694715.3695960},
doi = {10.1145/3694715.3695960},
booktitle = {Proceedings of the ACM SIGOPS 30th Symposium on Operating Systems Principles},
pages = {211–228},
numpages = {18},
location = {Austin, TX, USA},
series = {SOSP '24}
}

@inproceedings{hotspa,
author = {Ge, Hao and Fu, Fangcheng and Li, Haoyang and Wang, Xuanyu and Lin, Sheng and Wang, Yujie and Nie, Xiaonan and Zhang, Hailin and Miao, Xupeng and Cui, Bin},
title = {Enabling Parallelism Hot Switching for Efficient Training of Large Language Models},
year = {2024},
isbn = {9798400712517},
publisher = {Association for Computing Machinery},
address = {New York, NY, USA},
url = {https://doi-org.ezp-prod1.hul.harvard.edu/10.1145/3694715.3695969},
doi = {10.1145/3694715.3695969},
booktitle = {Proceedings of the ACM SIGOPS 30th Symposium on Operating Systems Principles},
pages = {178–194},
numpages = {17},
location = {Austin, TX, USA},
series = {SOSP '24}
}

@inproceedings{oobleck,
author = {Jang, Insu and Yang, Zhenning and Zhang, Zhen and Jin, Xin and Chowdhury, Mosharaf},
title = {Oobleck: Resilient Distributed Training of Large Models Using Pipeline Templates},
year = {2023},
isbn = {9798400702297},
publisher = {Association for Computing Machinery},
address = {New York, NY, USA},
url = {https://doi-org.ezp-prod1.hul.harvard.edu/10.1145/3600006.3613152},
doi = {10.1145/3600006.3613152},
booktitle = {Proceedings of the 29th Symposium on Operating Systems Principles},
pages = {382–395},
numpages = {14},
location = {Koblenz, Germany},
series = {SOSP '23}
}

@misc{google-blog,
    title = {{Balance of power: A full-stack approach to power and thermal fluctuations in ML infrastructure}},
    howpublished = {\url{https://cloud.google.com/blog/topics/systems/mitigating-power-and-thermal-fluctuations-in-ml-infrastructure}}
    ,year = {2024}
}

@misc{ec2-princing,
    title = {{Amazon EC2 Capacity Blocks for ML pricing}},
    howpublished = {\url{https://aws.amazon.com/ec2/capacityblocks/pricing/}}
    ,year = {2025}
}

@misc{megatron-code,
    title = {{Megatron-LM Github Repository}},
    howpublished = {\url{https://github.com/NVIDIA/Megatron-LM}}
    ,year = {2024}
}

@article{gpt1,
  title={Improving language understanding by generative pre-training},
  author={Radford, Alec and Narasimhan, Karthik and Salimans, Tim and Sutskever, Ilya and others},
  year={2018},
  publisher={OpenAI}
}

@article{gpt2,
  title={Language Models are Unsupervised Multitask Learners},
  author={Radford, Alec and Wu, Jeff and Child, Rewon and Luan, David and Amodei, Dario and Sutskever, Ilya},
  year={2019}
}

@inproceedings{gpt3,
author = {Brown, Tom B. and Mann, Benjamin and Ryder, Nick and Subbiah, Melanie and Kaplan, Jared and Dhariwal, Prafulla and Neelakantan, Arvind and Shyam, Pranav and Sastry, Girish and Askell, Amanda and Agarwal, Sandhini and Herbert-Voss, Ariel and Krueger, Gretchen and Henighan, Tom and Child, Rewon and Ramesh, Aditya and Ziegler, Daniel M. and Wu, Jeffrey and Winter, Clemens and Hesse, Christopher and Chen, Mark and Sigler, Eric and Litwin, Mateusz and Gray, Scott and Chess, Benjamin and Clark, Jack and Berner, Christopher and McCandlish, Sam and Radford, Alec and Sutskever, Ilya and Amodei, Dario},
title = {Language Models Are Few-Shot Learners},
year = {2020},
isbn = {9781713829546},
publisher = {Curran Associates Inc.},
address = {Red Hook, NY, USA},
booktitle = {Proceedings of the 34th International Conference on Neural Information Processing Systems},
articleno = {159},
numpages = {25},
location = {Vancouver, BC, Canada},
series = {NIPS'20}
}

@article{Llama,
  title={Llama: Open and efficient foundation language models},
  author={Touvron, Hugo and Lavril, Thibaut and Izacard, Gautier and Martinet, Xavier and Lachaux, Marie-Anne and Lacroix, Timoth{\'e}e and Rozi{\`e}re, Baptiste and Goyal, Naman and Hambro, Eric and Azhar, Faisal and others},
  journal={arXiv preprint arXiv:2302.13971},
  year={2023}
}

@article{Llama2,
  title={Llama 2: Open foundation and fine-tuned chat models},
  author={Touvron, Hugo and Martin, Louis and Stone, Kevin and Albert, Peter and Almahairi, Amjad and Babaei, Yasmine and Bashlykov, Nikolay and Batra, Soumya and Bhargava, Prajjwal and Bhosale, Shruti and others},
  journal={arXiv preprint arXiv:2307.09288},
  year={2023}
}

@misc{llama3,
      title={The Llama 3 Herd of Models}, 
      author={Aaron Grattafiori and Abhimanyu Dubey and Abhinav Jauhri and Abhinav Pandey and Abhishek Kadian and Ahmad Al-Dahle and Aiesha Letman and Akhil Mathur and Alan Schelten and Alex Vaughan and Amy Yang and Angela Fan and Anirudh Goyal and Anthony Hartshorn and Aobo Yang and Archi Mitra and Archie Sravankumar and Artem Korenev and Arthur Hinsvark and Arun Rao and Aston Zhang and Aurelien Rodriguez and Austen Gregerson and Ava Spataru and Baptiste Roziere and Bethany Biron and Binh Tang and Bobbie Chern and Charlotte Caucheteux and Chaya Nayak and Chloe Bi and Chris Marra and Chris McConnell and Christian Keller and Christophe Touret and Chunyang Wu and Corinne Wong and Cristian Canton Ferrer and Cyrus Nikolaidis and Damien Allonsius and Daniel Song and Danielle Pintz and Danny Livshits and Danny Wyatt and David Esiobu and Dhruv Choudhary and Dhruv Mahajan and Diego Garcia-Olano and Diego Perino and Dieuwke Hupkes and Egor Lakomkin and Ehab AlBadawy and Elina Lobanova and Emily Dinan and Eric Michael Smith and Filip Radenovic and Francisco Guzmán and Frank Zhang and Gabriel Synnaeve and Gabrielle Lee and Georgia Lewis Anderson and Govind Thattai and Graeme Nail and Gregoire Mialon and Guan Pang and Guillem Cucurell and Hailey Nguyen and Hannah Korevaar and Hu Xu and Hugo Touvron and Iliyan Zarov and Imanol Arrieta Ibarra and Isabel Kloumann and Ishan Misra and Ivan Evtimov and Jack Zhang and Jade Copet and Jaewon Lee and Jan Geffert and Jana Vranes and Jason Park and Jay Mahadeokar and Jeet Shah and Jelmer van der Linde and Jennifer Billock and Jenny Hong and Jenya Lee and Jeremy Fu and Jianfeng Chi and Jianyu Huang and Jiawen Liu and Jie Wang and Jiecao Yu and Joanna Bitton and Joe Spisak and Jongsoo Park and Joseph Rocca and Joshua Johnstun and Joshua Saxe and Junteng Jia and Kalyan Vasuden Alwala and Karthik Prasad and Kartikeya Upasani and Kate Plawiak and Ke Li and Kenneth Heafield and Kevin Stone and Khalid El-Arini and Krithika Iyer and Kshitiz Malik and Kuenley Chiu and Kunal Bhalla and Kushal Lakhotia and Lauren Rantala-Yeary and Laurens van der Maaten and Lawrence Chen and Liang Tan and Liz Jenkins and Louis Martin and Lovish Madaan and Lubo Malo and Lukas Blecher and Lukas Landzaat and Luke de Oliveira and Madeline Muzzi and Mahesh Pasupuleti and Mannat Singh and Manohar Paluri and Marcin Kardas and Maria Tsimpoukelli and Mathew Oldham and Mathieu Rita and Maya Pavlova and Melanie Kambadur and Mike Lewis and Min Si and Mitesh Kumar Singh and Mona Hassan and Naman Goyal and Narjes Torabi and Nikolay Bashlykov and Nikolay Bogoychev and Niladri Chatterji and Ning Zhang and Olivier Duchenne and Onur Çelebi and Patrick Alrassy and Pengchuan Zhang and Pengwei Li and Petar Vasic and Peter Weng and Prajjwal Bhargava and Pratik Dubal and Praveen Krishnan and Punit Singh Koura and Puxin Xu and Qing He and Qingxiao Dong and Ragavan Srinivasan and Raj Ganapathy and Ramon Calderer and Ricardo Silveira Cabral and Robert Stojnic and Roberta Raileanu and Rohan Maheswari and Rohit Girdhar and Rohit Patel and Romain Sauvestre and Ronnie Polidoro and Roshan Sumbaly and Ross Taylor and Ruan Silva and Rui Hou and Rui Wang and Saghar Hosseini and Sahana Chennabasappa and Sanjay Singh and Sean Bell and Seohyun Sonia Kim and Sergey Edunov and Shaoliang Nie and Sharan Narang and Sharath Raparthy and Sheng Shen and Shengye Wan and Shruti Bhosale and Shun Zhang and Simon Vandenhende and Soumya Batra and Spencer Whitman and Sten Sootla and Stephane Collot and Suchin Gururangan and Sydney Borodinsky and Tamar Herman and Tara Fowler and Tarek Sheasha and Thomas Georgiou and Thomas Scialom and Tobias Speckbacher and Todor Mihaylov and Tong Xiao and Ujjwal Karn and Vedanuj Goswami and Vibhor Gupta and Vignesh Ramanathan and Viktor Kerkez and Vincent Gonguet and Virginie Do and Vish Vogeti and Vítor Albiero and Vladan Petrovic and Weiwei Chu and Wenhan Xiong and Wenyin Fu and Whitney Meers and Xavier Martinet and Xiaodong Wang and Xiaofang Wang and Xiaoqing Ellen Tan and Xide Xia and Xinfeng Xie and Xuchao Jia and Xuewei Wang and Yaelle Goldschlag and Yashesh Gaur and Yasmine Babaei and Yi Wen and Yiwen Song and Yuchen Zhang and Yue Li and Yuning Mao and Zacharie Delpierre Coudert and Zheng Yan and Zhengxing Chen and Zoe Papakipos and Aaditya Singh and Aayushi Srivastava and Abha Jain and Adam Kelsey and Adam Shajnfeld and Adithya Gangidi and Adolfo Victoria and Ahuva Goldstand and Ajay Menon and Ajay Sharma and Alex Boesenberg and Alexei Baevski and Allie Feinstein and Amanda Kallet and Amit Sangani and Amos Teo and Anam Yunus and Andrei Lupu and Andres Alvarado and Andrew Caples and Andrew Gu and Andrew Ho and Andrew Poulton and Andrew Ryan and Ankit Ramchandani and Annie Dong and Annie Franco and Anuj Goyal and Aparajita Saraf and Arkabandhu Chowdhury and Ashley Gabriel and Ashwin Bharambe and Assaf Eisenman and Azadeh Yazdan and Beau James and Ben Maurer and Benjamin Leonhardi and Bernie Huang and Beth Loyd and Beto De Paola and Bhargavi Paranjape and Bing Liu and Bo Wu and Boyu Ni and Braden Hancock and Bram Wasti and Brandon Spence and Brani Stojkovic and Brian Gamido and Britt Montalvo and Carl Parker and Carly Burton and Catalina Mejia and Ce Liu and Changhan Wang and Changkyu Kim and Chao Zhou and Chester Hu and Ching-Hsiang Chu and Chris Cai and Chris Tindal and Christoph Feichtenhofer and Cynthia Gao and Damon Civin and Dana Beaty and Daniel Kreymer and Daniel Li and David Adkins and David Xu and Davide Testuggine and Delia David and Devi Parikh and Diana Liskovich and Didem Foss and Dingkang Wang and Duc Le and Dustin Holland and Edward Dowling and Eissa Jamil and Elaine Montgomery and Eleonora Presani and Emily Hahn and Emily Wood and Eric-Tuan Le and Erik Brinkman and Esteban Arcaute and Evan Dunbar and Evan Smothers and Fei Sun and Felix Kreuk and Feng Tian and Filippos Kokkinos and Firat Ozgenel and Francesco Caggioni and Frank Kanayet and Frank Seide and Gabriela Medina Florez and Gabriella Schwarz and Gada Badeer and Georgia Swee and Gil Halpern and Grant Herman and Grigory Sizov and Guangyi and Zhang and Guna Lakshminarayanan and Hakan Inan and Hamid Shojanazeri and Han Zou and Hannah Wang and Hanwen Zha and Haroun Habeeb and Harrison Rudolph and Helen Suk and Henry Aspegren and Hunter Goldman and Hongyuan Zhan and Ibrahim Damlaj and Igor Molybog and Igor Tufanov and Ilias Leontiadis and Irina-Elena Veliche and Itai Gat and Jake Weissman and James Geboski and James Kohli and Janice Lam and Japhet Asher and Jean-Baptiste Gaya and Jeff Marcus and Jeff Tang and Jennifer Chan and Jenny Zhen and Jeremy Reizenstein and Jeremy Teboul and Jessica Zhong and Jian Jin and Jingyi Yang and Joe Cummings and Jon Carvill and Jon Shepard and Jonathan McPhie and Jonathan Torres and Josh Ginsburg and Junjie Wang and Kai Wu and Kam Hou U and Karan Saxena and Kartikay Khandelwal and Katayoun Zand and Kathy Matosich and Kaushik Veeraraghavan and Kelly Michelena and Keqian Li and Kiran Jagadeesh and Kun Huang and Kunal Chawla and Kyle Huang and Lailin Chen and Lakshya Garg and Lavender A and Leandro Silva and Lee Bell and Lei Zhang and Liangpeng Guo and Licheng Yu and Liron Moshkovich and Luca Wehrstedt and Madian Khabsa and Manav Avalani and Manish Bhatt and Martynas Mankus and Matan Hasson and Matthew Lennie and Matthias Reso and Maxim Groshev and Maxim Naumov and Maya Lathi and Meghan Keneally and Miao Liu and Michael L. Seltzer and Michal Valko and Michelle Restrepo and Mihir Patel and Mik Vyatskov and Mikayel Samvelyan and Mike Clark and Mike Macey and Mike Wang and Miquel Jubert Hermoso and Mo Metanat and Mohammad Rastegari and Munish Bansal and Nandhini Santhanam and Natascha Parks and Natasha White and Navyata Bawa and Nayan Singhal and Nick Egebo and Nicolas Usunier and Nikhil Mehta and Nikolay Pavlovich Laptev and Ning Dong and Norman Cheng and Oleg Chernoguz and Olivia Hart and Omkar Salpekar and Ozlem Kalinli and Parkin Kent and Parth Parekh and Paul Saab and Pavan Balaji and Pedro Rittner and Philip Bontrager and Pierre Roux and Piotr Dollar and Polina Zvyagina and Prashant Ratanchandani and Pritish Yuvraj and Qian Liang and Rachad Alao and Rachel Rodriguez and Rafi Ayub and Raghotham Murthy and Raghu Nayani and Rahul Mitra and Rangaprabhu Parthasarathy and Raymond Li and Rebekkah Hogan and Robin Battey and Rocky Wang and Russ Howes and Ruty Rinott and Sachin Mehta and Sachin Siby and Sai Jayesh Bondu and Samyak Datta and Sara Chugh and Sara Hunt and Sargun Dhillon and Sasha Sidorov and Satadru Pan and Saurabh Mahajan and Saurabh Verma and Seiji Yamamoto and Sharadh Ramaswamy and Shaun Lindsay and Shaun Lindsay and Sheng Feng and Shenghao Lin and Shengxin Cindy Zha and Shishir Patil and Shiva Shankar and Shuqiang Zhang and Shuqiang Zhang and Sinong Wang and Sneha Agarwal and Soji Sajuyigbe and Soumith Chintala and Stephanie Max and Stephen Chen and Steve Kehoe and Steve Satterfield and Sudarshan Govindaprasad and Sumit Gupta and Summer Deng and Sungmin Cho and Sunny Virk and Suraj Subramanian and Sy Choudhury and Sydney Goldman and Tal Remez and Tamar Glaser and Tamara Best and Thilo Koehler and Thomas Robinson and Tianhe Li and Tianjun Zhang and Tim Matthews and Timothy Chou and Tzook Shaked and Varun Vontimitta and Victoria Ajayi and Victoria Montanez and Vijai Mohan and Vinay Satish Kumar and Vishal Mangla and Vlad Ionescu and Vlad Poenaru and Vlad Tiberiu Mihailescu and Vladimir Ivanov and Wei Li and Wenchen Wang and Wenwen Jiang and Wes Bouaziz and Will Constable and Xiaocheng Tang and Xiaojian Wu and Xiaolan Wang and Xilun Wu and Xinbo Gao and Yaniv Kleinman and Yanjun Chen and Ye Hu and Ye Jia and Ye Qi and Yenda Li and Yilin Zhang and Ying Zhang and Yossi Adi and Youngjin Nam and Yu and Wang and Yu Zhao and Yuchen Hao and Yundi Qian and Yunlu Li and Yuzi He and Zach Rait and Zachary DeVito and Zef Rosnbrick and Zhaoduo Wen and Zhenyu Yang and Zhiwei Zhao and Zhiyu Ma},
      year={2024},
      eprint={2407.21783},
      archivePrefix={arXiv},
      primaryClass={cs.AI},
      url={https://arxiv.org/abs/2407.21783}, 
}

@article{dlrm,
  title={Deep learning recommendation model for personalization and recommendation systems},
  author={Naumov, Maxim and Mudigere, Dheevatsa and Shi, Hao-Jun Michael and Huang, Jianyu and Sundaraman, Narayanan and Park, Jongsoo and Wang, Xiaodong and Gupta, Udit and Wu, Carole-Jean and Azzolini, Alisson G and others},
  journal={arXiv preprint arXiv:1906.00091},
  year={2019}
}

@inproceedings{megatron-lm-in-scale,
author = {Narayanan, Deepak and Shoeybi, Mohammad and Casper, Jared and LeGresley, Patrick and Patwary, Mostofa and Korthikanti, Vijay and Vainbrand, Dmitri and Kashinkunti, Prethvi and Bernauer, Julie and Catanzaro, Bryan and Phanishayee, Amar and Zaharia, Matei},
title = {Efficient Large-Scale Language Model Training on GPU Clusters Using Megatron-LM},
year = {2021},
isbn = {9781450384421},
publisher = {Association for Computing Machinery},
address = {New York, NY, USA},
url = {https://doi-org.ezp-prod1.hul.harvard.edu/10.1145/3458817.3476209},
doi = {10.1145/3458817.3476209},
booktitle = {Proceedings of the International Conference for High Performance Computing, Networking, Storage and Analysis},
articleno = {58},
numpages = {15},
location = {St. Louis, Missouri},
series = {SC '21}
}

@misc{palm,
  doi = {10.48550/ARXIV.2204.02311},
  
  url = {https://arxiv.org/abs/2204.02311},
  
  author = {Chowdhery, Aakanksha and Narang, Sharan and Devlin, Jacob and Bosma, Maarten and Mishra, Gaurav and Roberts, Adam and Barham, Paul and Chung, Hyung Won and Sutton, Charles and Gehrmann, Sebastian and Schuh, Parker and Shi, Kensen and Tsvyashchenko, Sasha and Maynez, Joshua and Rao, Abhishek and Barnes, Parker and Tay, Yi and Shazeer, Noam and Prabhakaran, Vinodkumar and Reif, Emily and Du, Nan and Hutchinson, Ben and Pope, Reiner and Bradbury, James and Austin, Jacob and Isard, Michael and Gur-Ari, Guy and Yin, Pengcheng and Duke, Toju and Levskaya, Anselm and Ghemawat, Sanjay and Dev, Sunipa and Michalewski, Henryk and Garcia, Xavier and Misra, Vedant and Robinson, Kevin and Fedus, Liam and Zhou, Denny and Ippolito, Daphne and Luan, David and Lim, Hyeontaek and Zoph, Barret and Spiridonov, Alexander and Sepassi, Ryan and Dohan, David and Agrawal, Shivani and Omernick, Mark and Dai, Andrew M. and Pillai, Thanumalayan Sankaranarayana and Pellat, Marie and Lewkowycz, Aitor and Moreira, Erica and Child, Rewon and Polozov, Oleksandr and Lee, Katherine and Zhou, Zongwei and Wang, Xuezhi and Saeta, Brennan and Diaz, Mark and Firat, Orhan and Catasta, Michele and Wei, Jason and Meier-Hellstern, Kathy and Eck, Douglas and Dean, Jeff and Petrov, Slav and Fiedel, Noah},
  
  title = {PaLM: Scaling Language Modeling with Pathways},
  
  publisher = {arXiv},
  
  year = {2022},
  
  copyright = {Creative Commons Attribution 4.0 International}
}

@article{megatron-lm,
  title={Megatron-lm: Training multi-billion parameter language models using model parallelism},
  author={Shoeybi, Mohammad and Patwary, Mostofa and Puri, Raul and LeGresley, Patrick and Casper, Jared and Catanzaro, Bryan},
  journal={arXiv preprint arXiv:1909.08053},
  year={2019}
}

@inproceedings{bytecheckpoint,
author = {Wan, Borui and Han, Mingji and Sheng, Yiyao and Peng, Yanghua and Lin, Haibin and Zhang, Mofan and Lai, Zhichao and Yu, Menghan and Zhang, Junda and Song, Zuquan and Liu, Xin and Wu, Chuan},
title = {ByteCheckpoint: a unified checkpointing system for large foundation model development},
year = {2025},
isbn = {978-1-939133-46-5},
publisher = {USENIX Association},
address = {USA},
booktitle = {Proceedings of the 22nd USENIX Symposium on Networked Systems Design and Implementation},
articleno = {30},
numpages = {20},
location = {Philadelphia, PA, USA},
series = {NSDI '25}
}

@inproceedings {check_n_run,
author = {Assaf Eisenman and Kiran Kumar Matam and Steven Ingram and Dheevatsa Mudigere and Raghuraman Krishnamoorthi and Krishnakumar Nair and Misha Smelyanskiy and Murali Annavaram},
title = {{Check-N-Run}: a Checkpointing System for Training Deep Learning Recommendation Models},
booktitle = {19th USENIX Symposium on Networked Systems Design and Implementation (NSDI 22)},
year = {2022},
isbn = {978-1-939133-27-4},
address = {Renton, WA},
pages = {929--943},
url = {https://www.usenix.org/conference/nsdi22/presentation/eisenman},
publisher = {USENIX Association},
month = apr
}

@inproceedings {bytedance10000nsdi24,
author = {Ziheng Jiang and Haibin Lin and Yinmin Zhong and Qi Huang and Yangrui Chen and Zhi Zhang and Yanghua Peng and Xiang Li and Cong Xie and Shibiao Nong and Yulu Jia and Sun He and Hongmin Chen and Zhihao Bai and Qi Hou and Shipeng Yan and Ding Zhou and Yiyao Sheng and Zhuo Jiang and Haohan Xu and Haoran Wei and Zhang Zhang and Pengfei Nie and Leqi Zou and Sida Zhao and Liang Xiang and Zherui Liu and Zhe Li and Xiaoying Jia and Jianxi Ye and Xin Jin and Xin Liu},
title = {{MegaScale}: Scaling Large Language Model Training to More Than 10,000 {GPUs}},
booktitle = {21st USENIX Symposium on Networked Systems Design and Implementation (NSDI 24)},
year = {2024},
isbn = {978-1-939133-39-7},
address = {Santa Clara, CA},
pages = {745--760},
url = {https://www.usenix.org/conference/nsdi24/presentation/jiang-ziheng},
publisher = {USENIX Association},
month = apr
}

@inproceedings{zero-infinity,
author = {Rajbhandari, Samyam and Ruwase, Olatunji and Rasley, Jeff and Smith, Shaden and He, Yuxiong},
title = {ZeRO-infinity: breaking the GPU memory wall for extreme scale deep learning},
year = {2021},
isbn = {9781450384421},
publisher = {Association for Computing Machinery},
address = {New York, NY, USA},
url = {https://doi-org.ezp-prod1.hul.harvard.edu/10.1145/3458817.3476205},
doi = {10.1145/3458817.3476205},
booktitle = {Proceedings of the International Conference for High Performance Computing, Networking, Storage and Analysis},
articleno = {59},
numpages = {14},
location = {St. Louis, Missouri},
series = {SC '21}
}

@inproceedings {cant_be_late,
author = {Zhanghao Wu and Wei-Lin Chiang and Ziming Mao and Zongheng Yang and Eric Friedman and Scott Shenker and Ion Stoica},
title = {Can{\textquoteright}t Be Late: Optimizing Spot Instance Savings under Deadlines},
booktitle = {21st USENIX Symposium on Networked Systems Design and Implementation (NSDI 24)},
year = {2024},
isbn = {978-1-939133-39-7},
address = {Santa Clara, CA},
pages = {185--203},
url = {https://www.usenix.org/conference/nsdi24/presentation/wu-zhanghao},
publisher = {USENIX Association},
month = apr
}

@inproceedings{borg,
author = {Verma, Abhishek and Pedrosa, Luis and Korupolu, Madhukar and Oppenheimer, David and Tune, Eric and Wilkes, John},
title = {Large-scale cluster management at Google with Borg},
year = {2015},
isbn = {9781450332385},
publisher = {Association for Computing Machinery},
address = {New York, NY, USA},
url = {https://doi-org.ezp-prod1.hul.harvard.edu/10.1145/2741948.2741964},
doi = {10.1145/2741948.2741964},
booktitle = {Proceedings of the Tenth European Conference on Computer Systems},
articleno = {18},
numpages = {17},
location = {Bordeaux, France},
series = {EuroSys '15}
}

@inproceedings{mccs,
author = {Wu, Yongji and Xu, Yechen and Chen, Jingrong and Wang, Zhaodong and Zhang, Ying and Lentz, Matthew and Zhuo, Danyang},
title = {MCCS: A Service-based Approach to Collective Communication for Multi-Tenant Cloud},
year = {2024},
isbn = {9798400706141},
publisher = {Association for Computing Machinery},
address = {New York, NY, USA},
url = {https://doi.org/10.1145/3651890.3672252},
doi = {10.1145/3651890.3672252},
booktitle = {Proceedings of the ACM SIGCOMM 2024 Conference},
pages = {679–690},
numpages = {12},
location = {Sydney, NSW, Australia},
series = {ACM SIGCOMM '24}
}

@inproceedings{photonic_rachee,
author = {Kumar, Abhishek Vijaya and Devraj, Arjun and Bunandar, Darius and Singh, Rachee},
title = {A case for server-scale photonic connectivity},
year = {2024},
isbn = {9798400712722},
publisher = {Association for Computing Machinery},
address = {New York, NY, USA},
url = {https://doi.org/10.1145/3696348.3696856},
doi = {10.1145/3696348.3696856},
booktitle = {Proceedings of the 23rd ACM Workshop on Hot Topics in Networks},
pages = {290–299},
numpages = {10},
location = {Irvine, CA, USA},
series = {HotNets '24}
}

@INPROCEEDINGS{adapcc,
  author={Zhao, Xiaoyang and Zhang, Zhe and Wu, Chuan},
  booktitle={2024 IEEE 44th International Conference on Distributed Computing Systems (ICDCS)}, 
  title={AdapCC: Making Collective Communication in Distributed Machine Learning Adaptive}, 
  year={2024},
  volume={},
  number={},
  pages={25-35},
  doi={10.1109/ICDCS60910.2024.00012}}

@inproceedings{astral,
author = {Meng, Qingkai and Zheng, Hao and Zhang, Zhenhui and Lao, ChonLam and Huang, Chengyuan and Li, Baojia and Zhu, Ziyuan and Lu, Hao and Dang, Weizhen and Lin, Zitong and Zhang, Weifeng and Liu, Lingfeng and Gong, Yuanyuan and He, Chunzhi and Hu, Xiaoyuan and Xia, Yinben and Li, Xiang and He, Zekun and Wang, Yachen and Zou, Xianneng and Yang, Kun and Antichi, Gianni and Chen, Guihai and Tian, Chen},
title = {Astral: A Datacenter Infrastructure for Large Language Model Training at Scale},
year = {2025},
isbn = {9798400715242},
publisher = {Association for Computing Machinery},
address = {New York, NY, USA},
url = {https://doi.org/10.1145/3718958.3750521},
doi = {10.1145/3718958.3750521},
booktitle = {Proceedings of the ACM SIGCOMM 2025 Conference},
pages = {609–625},
numpages = {17},
location = {S\~{a}o Francisco Convent, Coimbra, Portugal},
series = {SIGCOMM '25}
}

@inproceedings{bytedance_straggler,
author = {Lin, Jinkun and Jiang, Ziheng and Song, Zuquan and Zhao, Sida and Yu, Menghan and Wang, Zhanghan and Wang, Chenyuan and Shi, Zuocheng and Shi, Xiang and Jia, Wei and Liu, Zherui and Wang, Shuguang and Lin, Haibin and Liu, Xin and Panda, Aurojit and Li, Jinyang},
title = {Understanding stragglers in large model training using what-if analysis},
year = {2025},
isbn = {978-1-939133-47-2},
publisher = {USENIX Association},
address = {USA},
booktitle = {Proceedings of the 19th USENIX Conference on Operating Systems Design and Implementation},
articleno = {27},
numpages = {16},
location = {Boston, MA, USA},
series = {OSDI '25}
}

@misc{gemini-google,
      title={Gemini: A Family of Highly Capable Multimodal Models}, 
      author={Gemini Team and Rohan Anil and Sebastian Borgeaud and Jean-Baptiste Alayrac and Jiahui Yu and Radu Soricut and Johan Schalkwyk and Andrew M. Dai and Anja Hauth and Katie Millican and David Silver and Melvin Johnson and Ioannis Antonoglou and Julian Schrittwieser and Amelia Glaese and Jilin Chen and Emily Pitler and Timothy Lillicrap and Angeliki Lazaridou and Orhan Firat and James Molloy and Michael Isard and Paul R. Barham and Tom Hennigan and Benjamin Lee and Fabio Viola and Malcolm Reynolds and Yuanzhong Xu and Ryan Doherty and Eli Collins and Clemens Meyer and Eliza Rutherford and Erica Moreira and Kareem Ayoub and Megha Goel and Jack Krawczyk and Cosmo Du and Ed Chi and Heng-Tze Cheng and Eric Ni and Purvi Shah and Patrick Kane and Betty Chan and Manaal Faruqui and Aliaksei Severyn and Hanzhao Lin and YaGuang Li and Yong Cheng and Abe Ittycheriah and Mahdis Mahdieh and Mia Chen and Pei Sun and Dustin Tran and Sumit Bagri and Balaji Lakshminarayanan and Jeremiah Liu and Andras Orban and Fabian Güra and Hao Zhou and Xinying Song and Aurelien Boffy and Harish Ganapathy and Steven Zheng and HyunJeong Choe and Ágoston Weisz and Tao Zhu and Yifeng Lu and Siddharth Gopal and Jarrod Kahn and Maciej Kula and Jeff Pitman and Rushin Shah and Emanuel Taropa and Majd Al Merey and Martin Baeuml and Zhifeng Chen and Laurent El Shafey and Yujing Zhang and Olcan Sercinoglu and George Tucker and Enrique Piqueras and Maxim Krikun and Iain Barr and Nikolay Savinov and Ivo Danihelka and Becca Roelofs and Anaïs White and Anders Andreassen and Tamara von Glehn and Lakshman Yagati and Mehran Kazemi and Lucas Gonzalez and Misha Khalman and Jakub Sygnowski and Alexandre Frechette and Charlotte Smith and Laura Culp and Lev Proleev and Yi Luan and Xi Chen and James Lottes and Nathan Schucher and Federico Lebron and Alban Rrustemi and Natalie Clay and Phil Crone and Tomas Kocisky and Jeffrey Zhao and Bartek Perz and Dian Yu and Heidi Howard and Adam Bloniarz and Jack W. Rae and Han Lu and Laurent Sifre and Marcello Maggioni and Fred Alcober and Dan Garrette and Megan Barnes and Shantanu Thakoor and Jacob Austin and Gabriel Barth-Maron and William Wong and Rishabh Joshi and Rahma Chaabouni and Deeni Fatiha and Arun Ahuja and Gaurav Singh Tomar and Evan Senter and Martin Chadwick and Ilya Kornakov and Nithya Attaluri and Iñaki Iturrate and Ruibo Liu and Yunxuan Li and Sarah Cogan and Jeremy Chen and Chao Jia and Chenjie Gu and Qiao Zhang and Jordan Grimstad and Ale Jakse Hartman and Xavier Garcia and Thanumalayan Sankaranarayana Pillai and Jacob Devlin and Michael Laskin and Diego de Las Casas and Dasha Valter and Connie Tao and Lorenzo Blanco and Adrià Puigdomènech Badia and David Reitter and Mianna Chen and Jenny Brennan and Clara Rivera and Sergey Brin and Shariq Iqbal and Gabriela Surita and Jane Labanowski and Abhi Rao and Stephanie Winkler and Emilio Parisotto and Yiming Gu and Kate Olszewska and Ravi Addanki and Antoine Miech and Annie Louis and Denis Teplyashin and Geoff Brown and Elliot Catt and Jan Balaguer and Jackie Xiang and Pidong Wang and Zoe Ashwood and Anton Briukhov and Albert Webson and Sanjay Ganapathy and Smit Sanghavi and Ajay Kannan and Ming-Wei Chang and Axel Stjerngren and Josip Djolonga and Yuting Sun and Ankur Bapna and Matthew Aitchison and Pedram Pejman and Henryk Michalewski and Tianhe Yu and Cindy Wang and Juliette Love and Junwhan Ahn and Dawn Bloxwich and Kehang Han and Peter Humphreys and Thibault Sellam and James Bradbury and Varun Godbole and Sina Samangooei and Bogdan Damoc and Alex Kaskasoli and Sébastien M. R. Arnold and Vijay Vasudevan and Shubham Agrawal and Jason Riesa and Dmitry Lepikhin and Richard Tanburn and Srivatsan Srinivasan and Hyeontaek Lim and Sarah Hodkinson and Pranav Shyam and Johan Ferret and Steven Hand and Ankush Garg and Tom Le Paine and Jian Li and Yujia Li and Minh Giang and Alexander Neitz and Zaheer Abbas and Sarah York and Machel Reid and Elizabeth Cole and Aakanksha Chowdhery and Dipanjan Das and Dominika Rogozińska and Vitaliy Nikolaev and Pablo Sprechmann and Zachary Nado and Lukas Zilka and Flavien Prost and Luheng He and Marianne Monteiro and Gaurav Mishra and Chris Welty and Josh Newlan and Dawei Jia and Miltiadis Allamanis and Clara Huiyi Hu and Raoul de Liedekerke and Justin Gilmer and Carl Saroufim and Shruti Rijhwani and Shaobo Hou and Disha Shrivastava and Anirudh Baddepudi and Alex Goldin and Adnan Ozturel and Albin Cassirer and Yunhan Xu and Daniel Sohn and Devendra Sachan and Reinald Kim Amplayo and Craig Swanson and Dessie Petrova and Shashi Narayan and Arthur Guez and Siddhartha Brahma and Jessica Landon and Miteyan Patel and Ruizhe Zhao and Kevin Villela and Luyu Wang and Wenhao Jia and Matthew Rahtz and Mai Giménez and Legg Yeung and James Keeling and Petko Georgiev and Diana Mincu and Boxi Wu and Salem Haykal and Rachel Saputro and Kiran Vodrahalli and James Qin and Zeynep Cankara and Abhanshu Sharma and Nick Fernando and Will Hawkins and Behnam Neyshabur and Solomon Kim and Adrian Hutter and Priyanka Agrawal and Alex Castro-Ros and George van den Driessche and Tao Wang and Fan Yang and Shuo-yiin Chang and Paul Komarek and Ross McIlroy and Mario Lučić and Guodong Zhang and Wael Farhan and Michael Sharman and Paul Natsev and Paul Michel and Yamini Bansal and Siyuan Qiao and Kris Cao and Siamak Shakeri and Christina Butterfield and Justin Chung and Paul Kishan Rubenstein and Shivani Agrawal and Arthur Mensch and Kedar Soparkar and Karel Lenc and Timothy Chung and Aedan Pope and Loren Maggiore and Jackie Kay and Priya Jhakra and Shibo Wang and Joshua Maynez and Mary Phuong and Taylor Tobin and Andrea Tacchetti and Maja Trebacz and Kevin Robinson and Yash Katariya and Sebastian Riedel and Paige Bailey and Kefan Xiao and Nimesh Ghelani and Lora Aroyo and Ambrose Slone and Neil Houlsby and Xuehan Xiong and Zhen Yang and Elena Gribovskaya and Jonas Adler and Mateo Wirth and Lisa Lee and Music Li and Thais Kagohara and Jay Pavagadhi and Sophie Bridgers and Anna Bortsova and Sanjay Ghemawat and Zafarali Ahmed and Tianqi Liu and Richard Powell and Vijay Bolina and Mariko Iinuma and Polina Zablotskaia and James Besley and Da-Woon Chung and Timothy Dozat and Ramona Comanescu and Xiance Si and Jeremy Greer and Guolong Su and Martin Polacek and Raphaël Lopez Kaufman and Simon Tokumine and Hexiang Hu and Elena Buchatskaya and Yingjie Miao and Mohamed Elhawaty and Aditya Siddhant and Nenad Tomasev and Jinwei Xing and Christina Greer and Helen Miller and Shereen Ashraf and Aurko Roy and Zizhao Zhang and Ada Ma and Angelos Filos and Milos Besta and Rory Blevins and Ted Klimenko and Chih-Kuan Yeh and Soravit Changpinyo and Jiaqi Mu and Oscar Chang and Mantas Pajarskas and Carrie Muir and Vered Cohen and Charline Le Lan and Krishna Haridasan and Amit Marathe and Steven Hansen and Sholto Douglas and Rajkumar Samuel and Mingqiu Wang and Sophia Austin and Chang Lan and Jiepu Jiang and Justin Chiu and Jaime Alonso Lorenzo and Lars Lowe Sjösund and Sébastien Cevey and Zach Gleicher and Thi Avrahami and Anudhyan Boral and Hansa Srinivasan and Vittorio Selo and Rhys May and Konstantinos Aisopos and Léonard Hussenot and Livio Baldini Soares and Kate Baumli and Michael B. Chang and Adrià Recasens and Ben Caine and Alexander Pritzel and Filip Pavetic and Fabio Pardo and Anita Gergely and Justin Frye and Vinay Ramasesh and Dan Horgan and Kartikeya Badola and Nora Kassner and Subhrajit Roy and Ethan Dyer and Víctor Campos Campos and Alex Tomala and Yunhao Tang and Dalia El Badawy and Elspeth White and Basil Mustafa and Oran Lang and Abhishek Jindal and Sharad Vikram and Zhitao Gong and Sergi Caelles and Ross Hemsley and Gregory Thornton and Fangxiaoyu Feng and Wojciech Stokowiec and Ce Zheng and Phoebe Thacker and Çağlar Ünlü and Zhishuai Zhang and Mohammad Saleh and James Svensson and Max Bileschi and Piyush Patil and Ankesh Anand and Roman Ring and Katerina Tsihlas and Arpi Vezer and Marco Selvi and Toby Shevlane and Mikel Rodriguez and Tom Kwiatkowski and Samira Daruki and Keran Rong and Allan Dafoe and Nicholas FitzGerald and Keren Gu-Lemberg and Mina Khan and Lisa Anne Hendricks and Marie Pellat and Vladimir Feinberg and James Cobon-Kerr and Tara Sainath and Maribeth Rauh and Sayed Hadi Hashemi and Richard Ives and Yana Hasson and Eric Noland and Yuan Cao and Nathan Byrd and Le Hou and Qingze Wang and Thibault Sottiaux and Michela Paganini and Jean-Baptiste Lespiau and Alexandre Moufarek and Samer Hassan and Kaushik Shivakumar and Joost van Amersfoort and Amol Mandhane and Pratik Joshi and Anirudh Goyal and Matthew Tung and Andrew Brock and Hannah Sheahan and Vedant Misra and Cheng Li and Nemanja Rakićević and Mostafa Dehghani and Fangyu Liu and Sid Mittal and Junhyuk Oh and Seb Noury and Eren Sezener and Fantine Huot and Matthew Lamm and Nicola De Cao and Charlie Chen and Sidharth Mudgal and Romina Stella and Kevin Brooks and Gautam Vasudevan and Chenxi Liu and Mainak Chain and Nivedita Melinkeri and Aaron Cohen and Venus Wang and Kristie Seymore and Sergey Zubkov and Rahul Goel and Summer Yue and Sai Krishnakumaran and Brian Albert and Nate Hurley and Motoki Sano and Anhad Mohananey and Jonah Joughin and Egor Filonov and Tomasz Kępa and Yomna Eldawy and Jiawern Lim and Rahul Rishi and Shirin Badiezadegan and Taylor Bos and Jerry Chang and Sanil Jain and Sri Gayatri Sundara Padmanabhan and Subha Puttagunta and Kalpesh Krishna and Leslie Baker and Norbert Kalb and Vamsi Bedapudi and Adam Kurzrok and Shuntong Lei and Anthony Yu and Oren Litvin and Xiang Zhou and Zhichun Wu and Sam Sobell and Andrea Siciliano and Alan Papir and Robby Neale and Jonas Bragagnolo and Tej Toor and Tina Chen and Valentin Anklin and Feiran Wang and Richie Feng and Milad Gholami and Kevin Ling and Lijuan Liu and Jules Walter and Hamid Moghaddam and Arun Kishore and Jakub Adamek and Tyler Mercado and Jonathan Mallinson and Siddhinita Wandekar and Stephen Cagle and Eran Ofek and Guillermo Garrido and Clemens Lombriser and Maksim Mukha and Botu Sun and Hafeezul Rahman Mohammad and Josip Matak and Yadi Qian and Vikas Peswani and Pawel Janus and Quan Yuan and Leif Schelin and Oana David and Ankur Garg and Yifan He and Oleksii Duzhyi and Anton Älgmyr and Timothée Lottaz and Qi Li and Vikas Yadav and Luyao Xu and Alex Chinien and Rakesh Shivanna and Aleksandr Chuklin and Josie Li and Carrie Spadine and Travis Wolfe and Kareem Mohamed and Subhabrata Das and Zihang Dai and Kyle He and Daniel von Dincklage and Shyam Upadhyay and Akanksha Maurya and Luyan Chi and Sebastian Krause and Khalid Salama and Pam G Rabinovitch and Pavan Kumar Reddy M and Aarush Selvan and Mikhail Dektiarev and Golnaz Ghiasi and Erdem Guven and Himanshu Gupta and Boyi Liu and Deepak Sharma and Idan Heimlich Shtacher and Shachi Paul and Oscar Akerlund and François-Xavier Aubet and Terry Huang and Chen Zhu and Eric Zhu and Elico Teixeira and Matthew Fritze and Francesco Bertolini and Liana-Eleonora Marinescu and Martin Bölle and Dominik Paulus and Khyatti Gupta and Tejasi Latkar and Max Chang and Jason Sanders and Roopa Wilson and Xuewei Wu and Yi-Xuan Tan and Lam Nguyen Thiet and Tulsee Doshi and Sid Lall and Swaroop Mishra and Wanming Chen and Thang Luong and Seth Benjamin and Jasmine Lee and Ewa Andrejczuk and Dominik Rabiej and Vipul Ranjan and Krzysztof Styrc and Pengcheng Yin and Jon Simon and Malcolm Rose Harriott and Mudit Bansal and Alexei Robsky and Geoff Bacon and David Greene and Daniil Mirylenka and Chen Zhou and Obaid Sarvana and Abhimanyu Goyal and Samuel Andermatt and Patrick Siegler and Ben Horn and Assaf Israel and Francesco Pongetti and Chih-Wei "Louis" Chen and Marco Selvatici and Pedro Silva and Kathie Wang and Jackson Tolins and Kelvin Guu and Roey Yogev and Xiaochen Cai and Alessandro Agostini and Maulik Shah and Hung Nguyen and Noah Ó Donnaile and Sébastien Pereira and Linda Friso and Adam Stambler and Adam Kurzrok and Chenkai Kuang and Yan Romanikhin and Mark Geller and ZJ Yan and Kane Jang and Cheng-Chun Lee and Wojciech Fica and Eric Malmi and Qijun Tan and Dan Banica and Daniel Balle and Ryan Pham and Yanping Huang and Diana Avram and Hongzhi Shi and Jasjot Singh and Chris Hidey and Niharika Ahuja and Pranab Saxena and Dan Dooley and Srividya Pranavi Potharaju and Eileen O'Neill and Anand Gokulchandran and Ryan Foley and Kai Zhao and Mike Dusenberry and Yuan Liu and Pulkit Mehta and Ragha Kotikalapudi and Chalence Safranek-Shrader and Andrew Goodman and Joshua Kessinger and Eran Globen and Prateek Kolhar and Chris Gorgolewski and Ali Ibrahim and Yang Song and Ali Eichenbaum and Thomas Brovelli and Sahitya Potluri and Preethi Lahoti and Cip Baetu and Ali Ghorbani and Charles Chen and Andy Crawford and Shalini Pal and Mukund Sridhar and Petru Gurita and Asier Mujika and Igor Petrovski and Pierre-Louis Cedoz and Chenmei Li and Shiyuan Chen and Niccolò Dal Santo and Siddharth Goyal and Jitesh Punjabi and Karthik Kappaganthu and Chester Kwak and Pallavi LV and Sarmishta Velury and Himadri Choudhury and Jamie Hall and Premal Shah and Ricardo Figueira and Matt Thomas and Minjie Lu and Ting Zhou and Chintu Kumar and Thomas Jurdi and Sharat Chikkerur and Yenai Ma and Adams Yu and Soo Kwak and Victor Ähdel and Sujeevan Rajayogam and Travis Choma and Fei Liu and Aditya Barua and Colin Ji and Ji Ho Park and Vincent Hellendoorn and Alex Bailey and Taylan Bilal and Huanjie Zhou and Mehrdad Khatir and Charles Sutton and Wojciech Rzadkowski and Fiona Macintosh and Roopali Vij and Konstantin Shagin and Paul Medina and Chen Liang and Jinjing Zhou and Pararth Shah and Yingying Bi and Attila Dankovics and Shipra Banga and Sabine Lehmann and Marissa Bredesen and Zifan Lin and John Eric Hoffmann and Jonathan Lai and Raynald Chung and Kai Yang and Nihal Balani and Arthur Bražinskas and Andrei Sozanschi and Matthew Hayes and Héctor Fernández Alcalde and Peter Makarov and Will Chen and Antonio Stella and Liselotte Snijders and Michael Mandl and Ante Kärrman and Paweł Nowak and Xinyi Wu and Alex Dyck and Krishnan Vaidyanathan and Raghavender R and Jessica Mallet and Mitch Rudominer and Eric Johnston and Sushil Mittal and Akhil Udathu and Janara Christensen and Vishal Verma and Zach Irving and Andreas Santucci and Gamaleldin Elsayed and Elnaz Davoodi and Marin Georgiev and Ian Tenney and Nan Hua and Geoffrey Cideron and Edouard Leurent and Mahmoud Alnahlawi and Ionut Georgescu and Nan Wei and Ivy Zheng and Dylan Scandinaro and Heinrich Jiang and Jasper Snoek and Mukund Sundararajan and Xuezhi Wang and Zack Ontiveros and Itay Karo and Jeremy Cole and Vinu Rajashekhar and Lara Tumeh and Eyal Ben-David and Rishub Jain and Jonathan Uesato and Romina Datta and Oskar Bunyan and Shimu Wu and John Zhang and Piotr Stanczyk and Ye Zhang and David Steiner and Subhajit Naskar and Michael Azzam and Matthew Johnson and Adam Paszke and Chung-Cheng Chiu and Jaume Sanchez Elias and Afroz Mohiuddin and Faizan Muhammad and Jin Miao and Andrew Lee and Nino Vieillard and Jane Park and Jiageng Zhang and Jeff Stanway and Drew Garmon and Abhijit Karmarkar and Zhe Dong and Jong Lee and Aviral Kumar and Luowei Zhou and Jonathan Evens and William Isaac and Geoffrey Irving and Edward Loper and Michael Fink and Isha Arkatkar and Nanxin Chen and Izhak Shafran and Ivan Petrychenko and Zhe Chen and Johnson Jia and Anselm Levskaya and Zhenkai Zhu and Peter Grabowski and Yu Mao and Alberto Magni and Kaisheng Yao and Javier Snaider and Norman Casagrande and Evan Palmer and Paul Suganthan and Alfonso Castaño and Irene Giannoumis and Wooyeol Kim and Mikołaj Rybiński and Ashwin Sreevatsa and Jennifer Prendki and David Soergel and Adrian Goedeckemeyer and Willi Gierke and Mohsen Jafari and Meenu Gaba and Jeremy Wiesner and Diana Gage Wright and Yawen Wei and Harsha Vashisht and Yana Kulizhskaya and Jay Hoover and Maigo Le and Lu Li and Chimezie Iwuanyanwu and Lu Liu and Kevin Ramirez and Andrey Khorlin and Albert Cui and Tian LIN and Marcus Wu and Ricardo Aguilar and Keith Pallo and Abhishek Chakladar and Ginger Perng and Elena Allica Abellan and Mingyang Zhang and Ishita Dasgupta and Nate Kushman and Ivo Penchev and Alena Repina and Xihui Wu and Tom van der Weide and Priya Ponnapalli and Caroline Kaplan and Jiri Simsa and Shuangfeng Li and Olivier Dousse and Fan Yang and Jeff Piper and Nathan Ie and Rama Pasumarthi and Nathan Lintz and Anitha Vijayakumar and Daniel Andor and Pedro Valenzuela and Minnie Lui and Cosmin Paduraru and Daiyi Peng and Katherine Lee and Shuyuan Zhang and Somer Greene and Duc Dung Nguyen and Paula Kurylowicz and Cassidy Hardin and Lucas Dixon and Lili Janzer and Kiam Choo and Ziqiang Feng and Biao Zhang and Achintya Singhal and Dayou Du and Dan McKinnon and Natasha Antropova and Tolga Bolukbasi and Orgad Keller and David Reid and Daniel Finchelstein and Maria Abi Raad and Remi Crocker and Peter Hawkins and Robert Dadashi and Colin Gaffney and Ken Franko and Anna Bulanova and Rémi Leblond and Shirley Chung and Harry Askham and Luis C. Cobo and Kelvin Xu and Felix Fischer and Jun Xu and Christina Sorokin and Chris Alberti and Chu-Cheng Lin and Colin Evans and Alek Dimitriev and Hannah Forbes and Dylan Banarse and Zora Tung and Mark Omernick and Colton Bishop and Rachel Sterneck and Rohan Jain and Jiawei Xia and Ehsan Amid and Francesco Piccinno and Xingyu Wang and Praseem Banzal and Daniel J. Mankowitz and Alex Polozov and Victoria Krakovna and Sasha Brown and MohammadHossein Bateni and Dennis Duan and Vlad Firoiu and Meghana Thotakuri and Tom Natan and Matthieu Geist and Ser tan Girgin and Hui Li and Jiayu Ye and Ofir Roval and Reiko Tojo and Michael Kwong and James Lee-Thorp and Christopher Yew and Danila Sinopalnikov and Sabela Ramos and John Mellor and Abhishek Sharma and Kathy Wu and David Miller and Nicolas Sonnerat and Denis Vnukov and Rory Greig and Jennifer Beattie and Emily Caveness and Libin Bai and Julian Eisenschlos and Alex Korchemniy and Tomy Tsai and Mimi Jasarevic and Weize Kong and Phuong Dao and Zeyu Zheng and Frederick Liu and Fan Yang and Rui Zhu and Tian Huey Teh and Jason Sanmiya and Evgeny Gladchenko and Nejc Trdin and Daniel Toyama and Evan Rosen and Sasan Tavakkol and Linting Xue and Chen Elkind and Oliver Woodman and John Carpenter and George Papamakarios and Rupert Kemp and Sushant Kafle and Tanya Grunina and Rishika Sinha and Alice Talbert and Diane Wu and Denese Owusu-Afriyie and Cosmo Du and Chloe Thornton and Jordi Pont-Tuset and Pradyumna Narayana and Jing Li and Saaber Fatehi and John Wieting and Omar Ajmeri and Benigno Uria and Yeongil Ko and Laura Knight and Amélie Héliou and Ning Niu and Shane Gu and Chenxi Pang and Yeqing Li and Nir Levine and Ariel Stolovich and Rebeca Santamaria-Fernandez and Sonam Goenka and Wenny Yustalim and Robin Strudel and Ali Elqursh and Charlie Deck and Hyo Lee and Zonglin Li and Kyle Levin and Raphael Hoffmann and Dan Holtmann-Rice and Olivier Bachem and Sho Arora and Christy Koh and Soheil Hassas Yeganeh and Siim Põder and Mukarram Tariq and Yanhua Sun and Lucian Ionita and Mojtaba Seyedhosseini and Pouya Tafti and Zhiyu Liu and Anmol Gulati and Jasmine Liu and Xinyu Ye and Bart Chrzaszcz and Lily Wang and Nikhil Sethi and Tianrun Li and Ben Brown and Shreya Singh and Wei Fan and Aaron Parisi and Joe Stanton and Vinod Koverkathu and Christopher A. Choquette-Choo and Yunjie Li and TJ Lu and Abe Ittycheriah and Prakash Shroff and Mani Varadarajan and Sanaz Bahargam and Rob Willoughby and David Gaddy and Guillaume Desjardins and Marco Cornero and Brona Robenek and Bhavishya Mittal and Ben Albrecht and Ashish Shenoy and Fedor Moiseev and Henrik Jacobsson and Alireza Ghaffarkhah and Morgane Rivière and Alanna Walton and Clément Crepy and Alicia Parrish and Zongwei Zhou and Clement Farabet and Carey Radebaugh and Praveen Srinivasan and Claudia van der Salm and Andreas Fidjeland and Salvatore Scellato and Eri Latorre-Chimoto and Hanna Klimczak-Plucińska and David Bridson and Dario de Cesare and Tom Hudson and Piermaria Mendolicchio and Lexi Walker and Alex Morris and Matthew Mauger and Alexey Guseynov and Alison Reid and Seth Odoom and Lucia Loher and Victor Cotruta and Madhavi Yenugula and Dominik Grewe and Anastasia Petrushkina and Tom Duerig and Antonio Sanchez and Steve Yadlowsky and Amy Shen and Amir Globerson and Lynette Webb and Sahil Dua and Dong Li and Surya Bhupatiraju and Dan Hurt and Haroon Qureshi and Ananth Agarwal and Tomer Shani and Matan Eyal and Anuj Khare and Shreyas Rammohan Belle and Lei Wang and Chetan Tekur and Mihir Sanjay Kale and Jinliang Wei and Ruoxin Sang and Brennan Saeta and Tyler Liechty and Yi Sun and Yao Zhao and Stephan Lee and Pandu Nayak and Doug Fritz and Manish Reddy Vuyyuru and John Aslanides and Nidhi Vyas and Martin Wicke and Xiao Ma and Evgenii Eltyshev and Nina Martin and Hardie Cate and James Manyika and Keyvan Amiri and Yelin Kim and Xi Xiong and Kai Kang and Florian Luisier and Nilesh Tripuraneni and David Madras and Mandy Guo and Austin Waters and Oliver Wang and Joshua Ainslie and Jason Baldridge and Han Zhang and Garima Pruthi and Jakob Bauer and Feng Yang and Riham Mansour and Jason Gelman and Yang Xu and George Polovets and Ji Liu and Honglong Cai and Warren Chen and XiangHai Sheng and Emily Xue and Sherjil Ozair and Christof Angermueller and Xiaowei Li and Anoop Sinha and Weiren Wang and Julia Wiesinger and Emmanouil Koukoumidis and Yuan Tian and Anand Iyer and Madhu Gurumurthy and Mark Goldenson and Parashar Shah and MK Blake and Hongkun Yu and Anthony Urbanowicz and Jennimaria Palomaki and Chrisantha Fernando and Ken Durden and Harsh Mehta and Nikola Momchev and Elahe Rahimtoroghi and Maria Georgaki and Amit Raul and Sebastian Ruder and Morgan Redshaw and Jinhyuk Lee and Denny Zhou and Komal Jalan and Dinghua Li and Blake Hechtman and Parker Schuh and Milad Nasr and Kieran Milan and Vladimir Mikulik and Juliana Franco and Tim Green and Nam Nguyen and Joe Kelley and Aroma Mahendru and Andrea Hu and Joshua Howland and Ben Vargas and Jeffrey Hui and Kshitij Bansal and Vikram Rao and Rakesh Ghiya and Emma Wang and Ke Ye and Jean Michel Sarr and Melanie Moranski Preston and Madeleine Elish and Steve Li and Aakash Kaku and Jigar Gupta and Ice Pasupat and Da-Cheng Juan and Milan Someswar and Tejvi M. and Xinyun Chen and Aida Amini and Alex Fabrikant and Eric Chu and Xuanyi Dong and Amruta Muthal and Senaka Buthpitiya and Sarthak Jauhari and Nan Hua and Urvashi Khandelwal and Ayal Hitron and Jie Ren and Larissa Rinaldi and Shahar Drath and Avigail Dabush and Nan-Jiang Jiang and Harshal Godhia and Uli Sachs and Anthony Chen and Yicheng Fan and Hagai Taitelbaum and Hila Noga and Zhuyun Dai and James Wang and Chen Liang and Jenny Hamer and Chun-Sung Ferng and Chenel Elkind and Aviel Atias and Paulina Lee and Vít Listík and Mathias Carlen and Jan van de Kerkhof and Marcin Pikus and Krunoslav Zaher and Paul Müller and Sasha Zykova and Richard Stefanec and Vitaly Gatsko and Christoph Hirnschall and Ashwin Sethi and Xingyu Federico Xu and Chetan Ahuja and Beth Tsai and Anca Stefanoiu and Bo Feng and Keshav Dhandhania and Manish Katyal and Akshay Gupta and Atharva Parulekar and Divya Pitta and Jing Zhao and Vivaan Bhatia and Yashodha Bhavnani and Omar Alhadlaq and Xiaolin Li and Peter Danenberg and Dennis Tu and Alex Pine and Vera Filippova and Abhipso Ghosh and Ben Limonchik and Bhargava Urala and Chaitanya Krishna Lanka and Derik Clive and Yi Sun and Edward Li and Hao Wu and Kevin Hongtongsak and Ianna Li and Kalind Thakkar and Kuanysh Omarov and Kushal Majmundar and Michael Alverson and Michael Kucharski and Mohak Patel and Mudit Jain and Maksim Zabelin and Paolo Pelagatti and Rohan Kohli and Saurabh Kumar and Joseph Kim and Swetha Sankar and Vineet Shah and Lakshmi Ramachandruni and Xiangkai Zeng and Ben Bariach and Laura Weidinger and Tu Vu and Alek Andreev and Antoine He and Kevin Hui and Sheleem Kashem and Amar Subramanya and Sissie Hsiao and Demis Hassabis and Koray Kavukcuoglu and Adam Sadovsky and Quoc Le and Trevor Strohman and Yonghui Wu and Slav Petrov and Jeffrey Dean and Oriol Vinyals},
      year={2025},
      eprint={2312.11805},
      archivePrefix={arXiv},
      primaryClass={cs.CL},
      url={https://arxiv.org/abs/2312.11805}, 
}

@inproceedings{robust_bytedance,
author = {Wan, Borui and Liu, Gaohong and Song, Zuquan and Wang, Jun and Zhang, Yun and Sheng, Guangming and Wang, Shuguang and Wei, Houmin and Wang, Chenyuan and Lou, Weiqiang and Yang, Xi and Zhang, Mofan and Jiang, Kaihua and Ren, Cheng and Zhi, Xiaoyun and Yu, Menghan and Nan, Zhe and Zheng, Zhuolin and Zhong, Baoquan and Wang, Qinlong and Yu, Huan and Chi, Jinxin and Zhang, Wang and Li, Yuhan and Du, Zixian and Zhao, Sida and Zhang, Yongqiang and Tang, Jingzhe and Liu, Zherui and Wu, Chuan and Peng, Yanghua and Lin, Haibin and Xiao, Wencong and Liu, Xin and Xiang, Liang},
title = {Robust LLM Training Infrastructure at ByteDance},
year = {2025},
isbn = {9798400718700},
publisher = {Association for Computing Machinery},
address = {New York, NY, USA},
url = {https://doi.org/10.1145/3731569.3764838},
doi = {10.1145/3731569.3764838},
booktitle = {Proceedings of the ACM SIGOPS 31st Symposium on Operating Systems Principles},
pages = {186–203},
numpages = {18},
location = {Lotte Hotel World, Seoul, Republic of Korea},
series = {SOSP '25}
}

@misc{si2025collectivecommunication100kgpus,
      title={Collective Communication for 100k+ GPUs}, 
      author={Min Si and Pavan Balaji and Yongzhou Chen and Ching-Hsiang Chu and Adi Gangidi and Saif Hasan and Subodh Iyengar and Dan Johnson and Bingzhe Liu and Regina Ren and Ashmitha Jeevaraj Shetty and Greg Steinbrecher and Yulun Wang and Bruce Wu and Xinfeng Xie and Jingyi Yang and Mingran Yang and Kenny Yu and Minlan Yu and Cen Zhao and Wes Bland and Denis Boyda and Suman Gumudavelli and Prashanth Kannan and Cristian Lumezanu and Rui Miao and Zhe Qu and Venkat Ramesh and Maxim Samoylov and Jan Seidel and Srikanth Sundaresan and Feng Tian and Qiye Tan and Shuqiang Zhang and Yimeng Zhao and Shengbao Zheng and Art Zhu and Hongyi Zeng},
      year={2025},
      eprint={2510.20171},
      archivePrefix={arXiv},
      primaryClass={cs.DC},
      url={https://arxiv.org/abs/2510.20171}, 
}

@inproceedings {alpaserve,
author = {Zhuohan Li and Lianmin Zheng and Yinmin Zhong and Vincent Liu and Ying Sheng and Xin Jin and Yanping Huang and Zhifeng Chen and Hao Zhang and Joseph E. Gonzalez and Ion Stoica},
title = {{AlpaServe}: Statistical Multiplexing with Model Parallelism for Deep Learning Serving},
booktitle = {17th USENIX Symposium on Operating Systems Design and Implementation (OSDI 23)},
year = {2023},
isbn = {978-1-939133-34-2},
address = {Boston, MA},
pages = {663--679},
url = {https://www.usenix.org/conference/osdi23/presentation/li-zhouhan},
publisher = {USENIX Association},
month = jul
}

@misc{optmeta,
      title={OPT: Open Pre-trained Transformer Language Models}, 
      author={Susan Zhang and Stephen Roller and Naman Goyal and Mikel Artetxe and Moya Chen and Shuohui Chen and Christopher Dewan and Mona Diab and Xian Li and Xi Victoria Lin and Todor Mihaylov and Myle Ott and Sam Shleifer and Kurt Shuster and Daniel Simig and Punit Singh Koura and Anjali Sridhar and Tianlu Wang and Luke Zettlemoyer},
      year={2022},
      eprint={2205.01068},
      archivePrefix={arXiv},
      primaryClass={cs.CL},
      url={https://arxiv.org/abs/2205.01068}, 
}

@misc{deepseekv3,
      title={DeepSeek-V3 Technical Report}, 
      author={DeepSeek-AI and Aixin Liu and Bei Feng and Bing Xue and Bingxuan Wang and Bochao Wu and Chengda Lu and Chenggang Zhao and Chengqi Deng and Chenyu Zhang and Chong Ruan and Damai Dai and Daya Guo and Dejian Yang and Deli Chen and Dongjie Ji and Erhang Li and Fangyun Lin and Fucong Dai and Fuli Luo and Guangbo Hao and Guanting Chen and Guowei Li and H. Zhang and Han Bao and Hanwei Xu and Haocheng Wang and Haowei Zhang and Honghui Ding and Huajian Xin and Huazuo Gao and Hui Li and Hui Qu and J. L. Cai and Jian Liang and Jianzhong Guo and Jiaqi Ni and Jiashi Li and Jiawei Wang and Jin Chen and Jingchang Chen and Jingyang Yuan and Junjie Qiu and Junlong Li and Junxiao Song and Kai Dong and Kai Hu and Kaige Gao and Kang Guan and Kexin Huang and Kuai Yu and Lean Wang and Lecong Zhang and Lei Xu and Leyi Xia and Liang Zhao and Litong Wang and Liyue Zhang and Meng Li and Miaojun Wang and Mingchuan Zhang and Minghua Zhang and Minghui Tang and Mingming Li and Ning Tian and Panpan Huang and Peiyi Wang and Peng Zhang and Qiancheng Wang and Qihao Zhu and Qinyu Chen and Qiushi Du and R. J. Chen and R. L. Jin and Ruiqi Ge and Ruisong Zhang and Ruizhe Pan and Runji Wang and Runxin Xu and Ruoyu Zhang and Ruyi Chen and S. S. Li and Shanghao Lu and Shangyan Zhou and Shanhuang Chen and Shaoqing Wu and Shengfeng Ye and Shengfeng Ye and Shirong Ma and Shiyu Wang and Shuang Zhou and Shuiping Yu and Shunfeng Zhou and Shuting Pan and T. Wang and Tao Yun and Tian Pei and Tianyu Sun and W. L. Xiao and Wangding Zeng and Wanjia Zhao and Wei An and Wen Liu and Wenfeng Liang and Wenjun Gao and Wenqin Yu and Wentao Zhang and X. Q. Li and Xiangyue Jin and Xianzu Wang and Xiao Bi and Xiaodong Liu and Xiaohan Wang and Xiaojin Shen and Xiaokang Chen and Xiaokang Zhang and Xiaosha Chen and Xiaotao Nie and Xiaowen Sun and Xiaoxiang Wang and Xin Cheng and Xin Liu and Xin Xie and Xingchao Liu and Xingkai Yu and Xinnan Song and Xinxia Shan and Xinyi Zhou and Xinyu Yang and Xinyuan Li and Xuecheng Su and Xuheng Lin and Y. K. Li and Y. Q. Wang and Y. X. Wei and Y. X. Zhu and Yang Zhang and Yanhong Xu and Yanhong Xu and Yanping Huang and Yao Li and Yao Zhao and Yaofeng Sun and Yaohui Li and Yaohui Wang and Yi Yu and Yi Zheng and Yichao Zhang and Yifan Shi and Yiliang Xiong and Ying He and Ying Tang and Yishi Piao and Yisong Wang and Yixuan Tan and Yiyang Ma and Yiyuan Liu and Yongqiang Guo and Yu Wu and Yuan Ou and Yuchen Zhu and Yuduan Wang and Yue Gong and Yuheng Zou and Yujia He and Yukun Zha and Yunfan Xiong and Yunxian Ma and Yuting Yan and Yuxiang Luo and Yuxiang You and Yuxuan Liu and Yuyang Zhou and Z. F. Wu and Z. Z. Ren and Zehui Ren and Zhangli Sha and Zhe Fu and Zhean Xu and Zhen Huang and Zhen Zhang and Zhenda Xie and Zhengyan Zhang and Zhewen Hao and Zhibin Gou and Zhicheng Ma and Zhigang Yan and Zhihong Shao and Zhipeng Xu and Zhiyu Wu and Zhongyu Zhang and Zhuoshu Li and Zihui Gu and Zijia Zhu and Zijun Liu and Zilin Li and Ziwei Xie and Ziyang Song and Ziyi Gao and Zizheng Pan},
      year={2025},
      eprint={2412.19437},
      archivePrefix={arXiv},
      primaryClass={cs.CL},
      url={https://arxiv.org/abs/2412.19437}, 
}

@inproceedings{nccl_sharp,
author = {Graham, Richard L. and Levi, Lion and Burredy, Devendar and Bloch, Gil and Shainer, Gilad and Cho, David and Elias, George and Klein, Daniel and Ladd, Joshua and Maor, Ophir and Marelli, Ami and Petrov, Valentin and Romlet, Evyatar and Qin, Yong and Zemah, Ido},
title = {Scalable Hierarchical Aggregation and Reduction Protocol (SHARP)TM Streaming-Aggregation Hardware Design and Evaluation},
year = {2020},
isbn = {978-3-030-50742-8},
publisher = {Springer-Verlag},
address = {Berlin, Heidelberg},
url = {https://doi.org/10.1007/978-3-030-50743-5_3},
doi = {10.1007/978-3-030-50743-5_3},
booktitle = {High Performance Computing: 35th International Conference, ISC High Performance 2020, Frankfurt/Main, Germany, June 22–25, 2020, Proceedings},
pages = {41–59},
numpages = {19},
location = {Frankfurt am Main, Germany}
}

@misc{cuda_checkpoint,
  title        = {CUDA Checkpoint and Restore Utility},
  author       = {{NVIDIA Corporation}},
  howpublished = {\url{https://github.com/NVIDIA/cuda-checkpoint}},
  year         = {2025},
  note         = {Accessed: 2026-05}
}

@misc{criu,
  title        = {CRIU: Checkpoint/Restore In Userspace},
  author       = {{OpenVZ Project}},
  howpublished = {\url{https://criu.org/}},
  year         = {2012},
  note         = {Accessed: 2026-05}
}

\newpage
\clearpage

\appendix
\begin{figure}[tb]
    \centering
    \begin{subfigure}{\linewidth}
        \centering
        \ifdefined\ArxivVersion
            \includegraphics[width=0.85\linewidth]{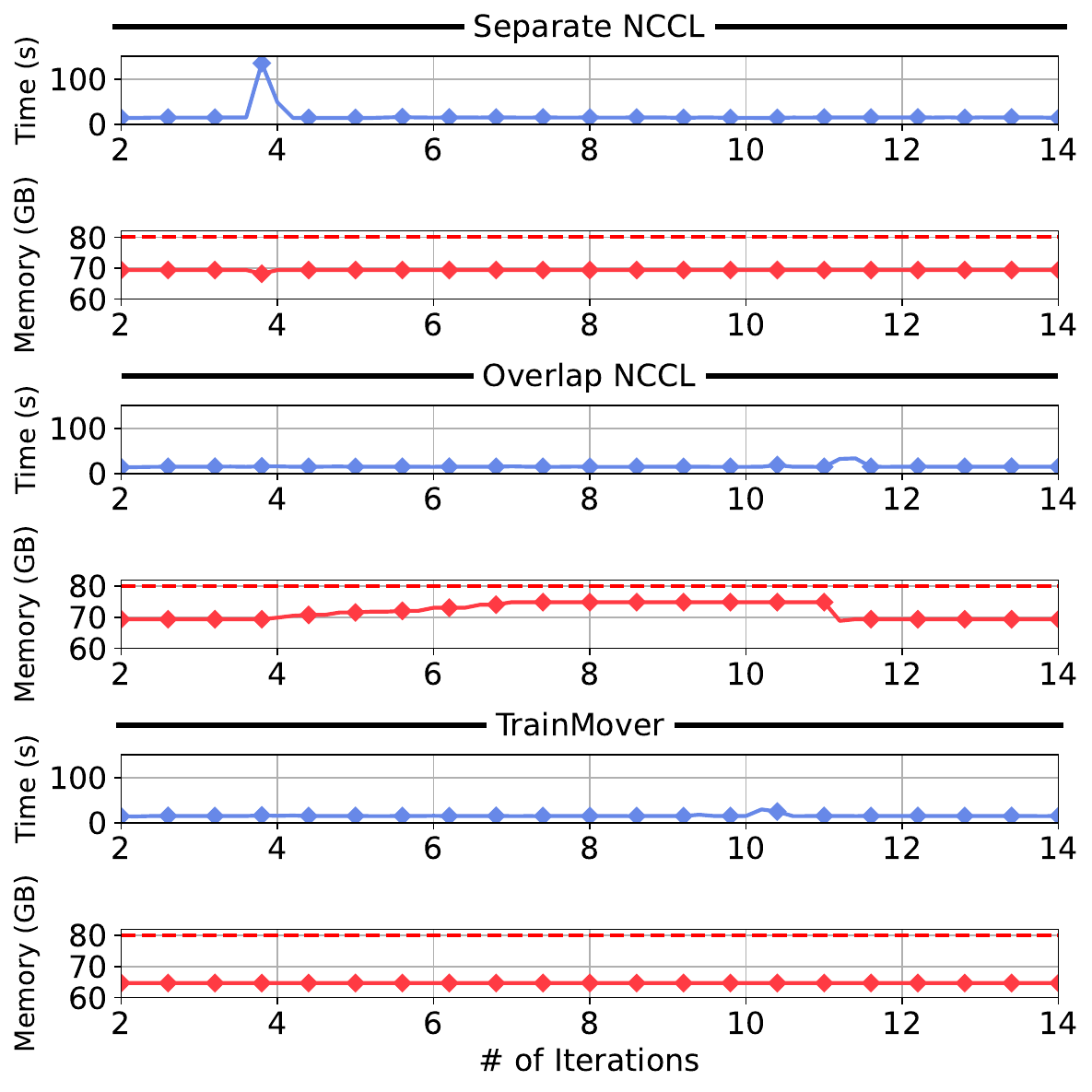}
        \else
            \includegraphics[width=0.85\linewidth]{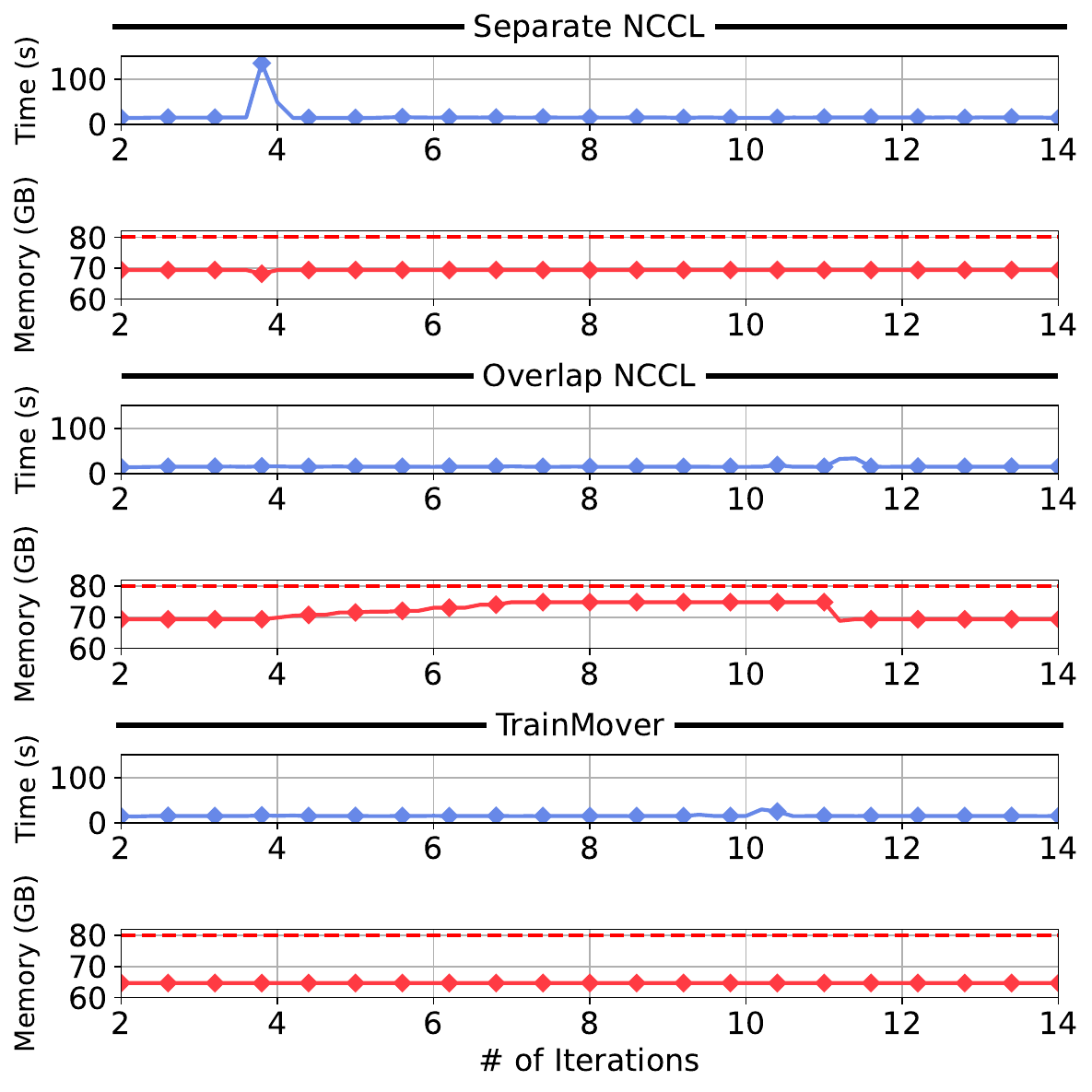}
        \fi
        \caption{Timeline}
        \label{fig:nccl_breakdown}
    \end{subfigure}

    \vspace{0.5em} 

    \begin{subfigure}{\linewidth}
        \centering
        \ifdefined\ArxivVersion
            \includegraphics[width=0.6\linewidth]{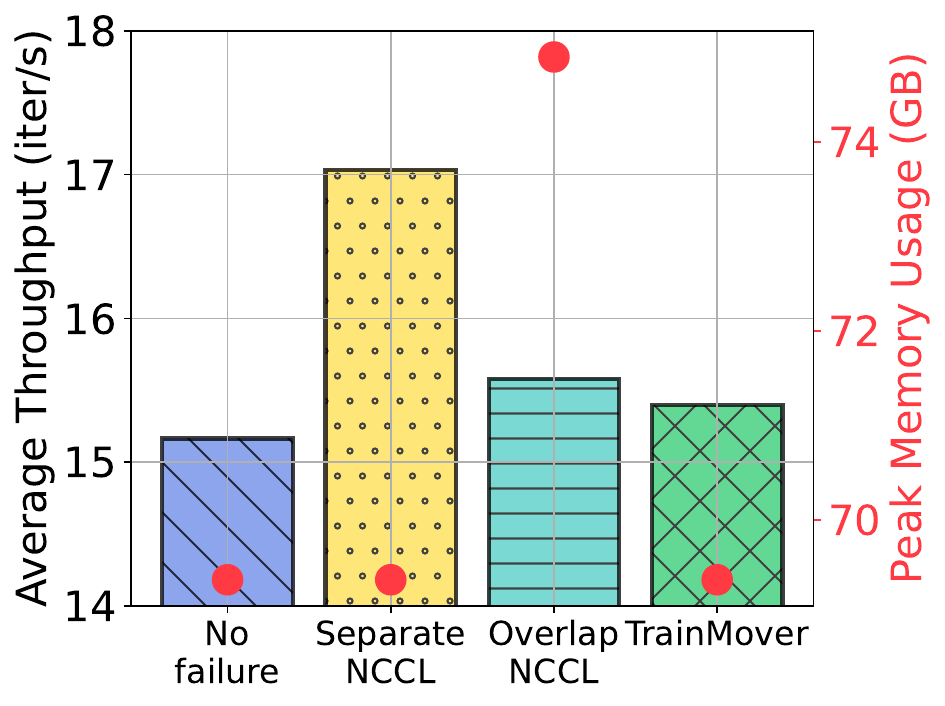}
        \else
            \includegraphics[width=0.6\linewidth]{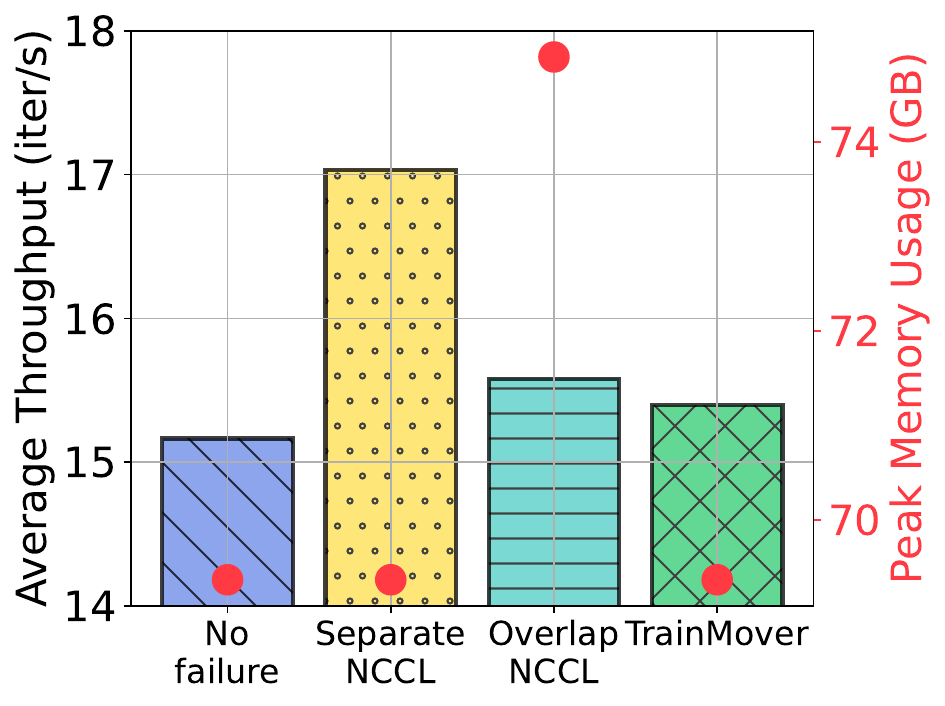}
        \fi
        \caption{Throughput and Peak Memory}
        \label{fig:nccl_breakdown_summary}
    \end{subfigure}

    \caption{Time and memory cost timeline for different NCCL design decisions.}
    \label{fig:nccl_analysis}
\end{figure}

\section{NCCL Decision Choice}
\label{eval:ncclbreakdown}
Our NCCL design achieves zero memory and minimal downtime overhead by overlapping all non-critical path establishment costs. In Figure~\ref{fig:nccl_breakdown}, we show the iteration time (including NCCL re-setup time) and GPU memory usage across three different ways of handling changes in NCCL group participants during migration. (1) \textit{Separate NCCL} involves completely destroying and re-instantiating NCCL groups whenever the group composition changes. This approach is used by Oobleck, Parcae, Bamboo, and all existing systems that rely on native PyTorch support. (2) \textit{Overlap NCCL} is another baseline where we modify PyTorch to allow multiple global groups to exist simultaneously, enabling new members to join within a separate NCCL context. (3) \sysname NCCL is our design, which incorporates a two-stage approach and reuses existing primitives.

For \textit{Separate NCCL}, downtime increases to approximately 8$\times$ the duration of normal training around the 4th iteration, when migration occurs and participant changes take place. However, GPU memory usage remains unchanged, as new groups are initialized only after the old groups are destroyed.
In contrast, \textit{Overlap NCCL} experiences a small downtime around the 11th iteration, when the final migration completes and the old NCCL groups are removed (migration spans from the 4th to the 12th iteration in the backend). However, this design incurs a high memory overhead, increasing from 71GB to 77GB, because many new NCCL groups (e.g., DP/TP/PP groups) must be initialized, and there are two sets of groups (frontend and backend) existing simultaneously.

\sysname's NCCL design achieves zero memory overhead and only a small downtime around the 10th iteration, when migration completes. This is because the second stage of NCCL instantiation (inter-machine connections) occurs on the critical path, requiring the old inter-machine NCCL connections to be destroyed and rebuilt with new participants within a short period. The NCCL reuse mechanism ensures zero memory overhead during this process.

In summary, as shown in Figure~\ref{fig:nccl_breakdown_summary}, our approach performs closely to the no-failure case, without introducing any additional memory overhead. In contrast, both the separate and overlapping NCCL approaches fall short, either compromising throughput (0.88$\times$) or incurring additional memory overhead.

\end{document}